# LegalScore: Development of a Benchmark for Evaluating AI Models in Legal Career Exams in Brazil

Roberto Caparroz [†], Marcelo Roitman[*], Beatriz Graziano Chow[*], Caroline Giusti[*], Larissa Torhacs[*], Pedro Aurélio Sola[‡], João Henrique M. Diogo[*], Luiza Balby[*], Carolina Dolabela L. Vasconcelos[*], Leonardo Roberti Caparroz[*], Albano Prado Franco[*]

## Abstract

This research introduces 'LegalScore', a specialized index for assessing how generative artificial intelligence models perform in a selected range of career exams that require a legal background in Brazil. The index evaluates fourteen different types of artificial intelligence models' performance, from proprietary to open-source models, in answering objective questions applied to these exams. The research uncovers the response of the models when applying English-trained large language models to Brazilian legal contexts, leading us to reflect on the importance and the need for Brazil-specific training data in generative artificial intelligence models. Performance analysis shows that while proprietary and most known models achieved better results overall, local and smaller models indicated promising performances due to their Brazilian context alignment in training. By establishing an evaluation framework with metrics including accuracy, confidence intervals, and normalized scoring, 'LegalScore' enables systematic assessment of artificial intelligence performance in legal examinations in Brazil. While the study demonstrates artificial intelligence's potential value for exam preparation and question development, it concludes that significant improvements are needed before AI can match human performance in advanced legal assessments. The benchmark creates a foundation for continued research, highlighting the importance of local adaptation in artificial intelligence development.

## 1. Introduction

Brazil has many undergraduate law programs, especially considering a significant increase since 2000 (OAB, 2020). In 2022, it was estimated that there would be approximately four million Bachelors of Laws (OAB, 2022). These professionals have several career options after graduating with a LLB degree, including the public service careers, which is extremely traditional and attractive amongst these graduates.

The competition to be able to perform as a civil servant is not easy, since there is a wide range of positions opened every year, varying according to each exams' specifics, including requirements and levels of difficulty. Among Brazil's most traditional and prestigious exams are meant for judicial positions, public prosecution, and other important Federal bodies, such as the Federal Revenue Service. To be able to fill a public position in Brazil aside from the fact that it brings a reasonable status in society, it also ensures good benefits and pay.

---

[†] Visiting Professor at Massachusetts Institute of Technology (MIT) and Professor at Fundação Getúlio Vargas (FGV)
[*] Researcher at Fundação Getúlio Vargas (FGV)
[‡] Student at Universidade Estadual Paulista "Júlio de Mesquita Filho" (UNESP)

In addition to these factors, there is another element of attractiveness in competitive examinations: the job security provided by the very nature of public careers. Investing in public jobs and positions ensures consolidated rights and guarantees for those approved (Rodrigues & Caron, 2016), making public service attractive to those seeking professional stability and higher-than-average pay.

The interest in legal careers is no coincidence. The Brazilian Constitution promotes and regulates the institute of public examination in art. 37, II, establishing it as the main means of access to public service through specific examinations that assess candidates' performance.

As a side effect, the premise of public competitions is creating an institutionalized and meritocratic access system (Araujo, 2020), which has consolidated them as a model deeply rooted in Brazilian culture.

Considering the advance of generative artificial intelligence (Generative AI) models, the debate on public exams is taking on new contours, especially regarding the effectiveness of exams as a means of access to coveted careers. One pertinent question concerns the performance of different generative AI models when subjected to the same exams as human candidates. What would be the result obtained by the main models currently available when submitted to Brazil's federal jobs, considering a legal background? Is it feasible to state that AI has the capability to outperform humans in this context?

The research proposed in this article summarizes the results of a study conducted over approximately three months. The research analyzed the performance of the leading AI models in various public exams, including the OAB (Brazilian Bar Association) exam, which qualifies Bachelor of Laws to duly practice private law.

The aim of the research, in addition to answering the proposed questions, was based on the following aspects: (i) assessing whether the main AI models, both proprietary and open source, are aligned with the Brazilian legal landscape; (ii) investigating whether the so-called language gap (Caparroz, 2024), resulting from the predominant training of models in the English language, plays a significant role in the performance of AI models in competitive exams applied to Brazil federal offices; and (iii) to propose an accuracy index to evaluate the Generative AI models' performance in this context.

The core empirical research was based on the performance evaluation of public exams and the Bar Exam of a carefully selected group of Large Language Models (Fan et al., 2023). Generative AI models have attracted the general public's attention in the last two years, especially after the launch of ChatGPT (Tang, 2023).

One relevant side note of this project is the potential of these models to serve as complementary tools for studying and solving questions, offering immediate feedback, personalized support, and help in understanding complex concepts (Hadi et al., 2023).

Another central point of the research was the analysis of the limitations of LLMs (Wolf et al., 2023). Factors such as the mismatch between the training data, predominantly done so in the English language, and the Brazilian legal landscape were considered, as well as the ability of these models to comprehend contextual issues and apply specific knowledge demanded in the exams. These limitations highlight the need to adapt models to local singularities and use them as support without replacing traditional learning and practice methods.

In addition, these restrictions are directly related to the scarcity of financial resources and the difficulty of obtaining quality datasets trained in languages other than English (Caparroz, 2024). Finally, the study sought to understand the role of LLMs in terms of a performance index. Although it is not yet possible to affirm the total superiority of AI models over human candidates, considering the state of the art and the sample selected, it is plausible that, as these models evolve, which has occurred exponentially (Villalobos et al., 2022) the impasse will be revisited frequently.

Therefore, this research proposes the introduction of an index capable of evaluating the performance of generative AI in the public career and documenting improvements in the performance of each model, firmly in a context in which LLMs are proliferating rapidly and the training process is becoming increasingly robust.

Given the issues raised, this is the first extensive study based on detailed analysis, culminating in the development of a specific benchmark to evaluate the performance of AI models in public exams for legal careers. This work, therefore, not only sets out an authentic and innovative approach on the field of Law & Technology but also represent a thorough representation of Brazil's culture and society in the age of AI, exploring the native language of

Portuguese and Brazilian public career exams lineage as central actors to be investigated.

## 2. Existing benchmarks for language models

The aim of this article is to compare various language models and evaluate their performance in human tasks that do not involve merely logical, ready-made, or, in some cases, mathematical answers. This is because, when performing their tasks, language models access a previously trained data set (Zhao et al., 2023), which serves as the basis for generating the answers presented to the user.

In order to improve the performance of models in responses that require understanding and interpretative reasoning, benchmarks have been developed to enable comparative performance analysis (Wang et al., 2024). This data makes it possible not only to assess the reliability of the answers provided but also to identify points for improvement that can be implemented by the developers.

### 2.1. Legal and non-legal benchmarks

AI models have often been evaluated using multiple-choice questions, which, in theory, offer an objective criterion for measuring their capabilities. However, several recent studies have demonstrated inconsistencies in this approach (Zhou & Duan, 2024). It is claimed that multiple-choice questions may not be a reliable way of assessing the real capabilities of generative AI models when compared to the production of long texts (Li et al., 2024). Despite this warning, for this phase of the research, the analysis of multiple-choice questions applied in legal exams in Brazil proved to be an adequate criterion for comparing the performance of the AI models tested with the results obtained by human candidates.

In the second stage, the ability of AI models to answer discursive questions from the same competitions analyzed in the first phase will be evaluated to obtain an overview of the results in all stages of the main competitions.

Several studies have analyzed the performance of AI models on multiple-choice questions applied in various contexts. A study with UBENCH (Benchmarking Uncertainty in Large Language Models) revealed the limitations faced by large-scale language models in solving and executing tasks involving multiple-choice questions. These difficulties are mainly due to the need to carry out semantic analysis and understanding in order to solve questions that are ambiguous or require more elaborate reasoning in order to carry out the requested task (Wang et al., 2024).

Another benchmark analyzed was multimodal models, i.e., those that require the interpretation of text, images, and possibly other document formats. In this field, SEED-Bench stands out, which includes more than twenty-four thousand multiple-choice questions on various subjects and areas of knowledge. The aim was to assess how language models promote integration between these modalities (Li et al., 2024).

The need to evaluate AI models in specific domains of human knowledge has led to legal benchmarks, such as LawBench (Fei et al., 2023) and LegalBench (Guha et al., 2023).

LawBench tested 51 AI models in various tasks, such as single-label classification (SLC), multi-label classification (MLC),

regression, extraction, and content generation. The results indicated that ChatGPT 4 obtained the best performance in the legal context evaluated, outperforming the others by a wide margin (Fei et al., 2023). LegalBench, on the other hand, tested the legal reasoning capacity of 20 AI models based on tasks prepared by professionals in the field (Guha et al., 2023).

The analysis of these and other studies raised a significant concern because these investigations use something other than Portuguese parameters as a data source. In the cases mentioned, the AI models used databases in English and Chinese, which restricts the analysis to the context of the respective countries. In the case of Brazil, it can be seen in various scenarios that even simple questions related to local history or geography, common to the repertoire of elementary school children, are not answered accurately by various AI models due to the lack of datasets representative of the country's reality (Caparroz, 2024).

## 2.2. Overview of Brazilian civil service examination

Unlike many countries, public careers are the dream of a significant part of the Brazilian population for the reasons previously discussed.

According to the census published by the newspaper Folha Dirigida, in cooperation with the website QConcursos (More than 10,000 public exams held in 2023 | Folha Dirigida, [n.d.]). In 2023 alone, more than 10,000 vacancies were offered for public positions in the most diverse sectors of public administration, covering candidates with training from high school to higher education, at all three levels of the federation. It is estimated that in October 2023 alone, more than 2,000 competitive examinations were held.

The National Confederation of Industry (CNI), in its 15th Economic Note, entitled "The weight of the civil service in Brazil compared to other countries" (CNI, [n.d.]), concluded that although Brazil does not have a high number of civil servants in relation to the proportion of the total population and workforce - representing 5.6% - this rate is higher than the average for Latin American countries. However, it is significantly lower than the average for OECD countries, which is 9.6%.

Given this scenario, it is undeniable that public examinations are increasingly attracting candidates of all profiles and for the most varied activities, which is also boosting related activities, such as the publishing and preparatory course markets.

## 3. Methodology

To make up the initial object of research, the main AI models available to the public were selected in October 2024, including proprietary models from large technology companies and open-source models with more flexible licenses. It should be noted that only multilingual models capable of responding to prompts and questions written in Portuguese were considered.

The models tested, in the latest versions available in October 2024, were:

- ChatGPT 3.5 Turbo, ChatGPT o1-mini and ChatGPT 4o, from OpenAI:
  https://www.openai.com/chatgpt
- Claude Sonnet 3.5, from Anthropic:
  https://www.anthropic.com/claude
- Llama 3.1 405B, from Meta:
  https://ai.meta.com/llama

- Gemini 1.5 Pro, from Google:
  https://ai.google.com/gemini
- Nemotron 340B from Nvidia:
  https://www.nvidia.com/en-us/research/ai/nemotron
- Mistral Large-2, from Mistral AI:
  https://mistral.ai/models/mistral-large-2
- Command-R-Plus from Cohere:
  https://cohere.ai/models/command-r-plus
- Reka Core 67B, from Reka AI:
  https://www.reka.ai/core67b
- Perplexity AI:
  https://www.perplexity.ai
- Grok-2, from X:
  https://x.com/grok-2

For comparison purposes, two models were added to the list that were trained with data in Portuguese and information about Brazil:

- Sabiá-3, developed by Maritaca AI:
  https://www.maritaca.ai/
- Amazônia AI, developed by WideLabs:
  https://amazoniaia.com.br/

### 3.1. Access to chosen models

The same prompt was used for all the models, and whenever possible, the tests were carried out on the official platforms, such as the ChatGPT versions, Command-R-Plus (https://coral.cohere.com), Reka Core 67B (https://www.reka.ai/), Amazonia IA (https://amazoniaia.com.br) and Sabiá-3 (https://www.maritaca.ai/). For Grok-2 and Nemotron 340B, the direct chat function on the https://lmarena.ai/ platform was used. For the Nemotron 340B, direct access to the platform was also used in some cases (https://build.nvidia.com/nvidia/nemotron-4-340b-instruct).

Another widely used tool was Poe, launched by the company Quora (www.poe.com), which allows access to various AI models. Due to the availability of access, we chose to test some of the selected models via the Poe platform, such as Claude Sonnet 3.5, Llama 3.1 405B, Gemini 1.5 Pro and Mistral Large-2.

Although these models are available on the Poe platform, in some cases, to optimize the experiment, the Poe platform and the respective official platform were used concurrently to assess their performance on the selected tests.

In all cases, new chats were always created to analyze a specific test, to avoid any kind of influence on the answers.

## 3.2. Selection of tests and data preparation

The tests were selected based on their relevance to the legal exam scene in Brazil for higher education. We chose national exams, with many candidates and public careers considered more important, with a higher degree of difficulty in the questions, in addition to the national exam for law practice.

The exams tested were the following: Brazilian Bar Association (OAB), Labor Prosecutor's Office, Federal Judge, Federal Prosecutor's Office, Federal Revenue Service, and the National Magistrates' Examination (ENAM).

As a selection criterion, the study chose the most recent exams applied in the last five years, except for the Public Prosecutor's exam, where the last two exams were used (from 2021 and 2017). For the Brazilian Bar Association (OAB) exam, the ten most recent exams, taken between June 2021 and July 2024 (Exam 41), were tested. Although OAB Exam 42 was completed before the study concluded, it was excluded from consideration because the official results had not yet been released.

The tests were taken from the official websites of the examining bodies: FGV (https://conhecimento.fgv.br/), FCC (https://www.concursosfcc.com.br) and CESPE/Cebraspe (https://www.cebraspe.org.br).

For all the tests, spreadsheets have been drawn up, which, in addition to including the questions, indicate the difficulty level, the official templates, and the answers provided by each model.

The questions were classified as "easy" or "difficult". To use an objective criterion, minimally anchored in the reality of public examination candidates, the statistics available on the website https://www.qconcursos.com were used. This site makes the tests available so any user can answer the questions. Considering the total number of respondents, the percentage of correct answers for each question was used as the distinguishing criterion. Questions with a score of less than 50% were considered "difficult," and questions with 50% or more correct were considered "easy".

Some tests had questions annulled due to errors in the templates or the impossibility of answering them. In this case, regardless of the alternative indicated by the models, the answer was considered "correct," the same criterion used for human candidates who took the tests in person.

The various exams analyzed include all the leading legal subjects taught in Brazil and some specific subjects, such as Portuguese, English, Economics, Statistics, Accounting, Data Fluency, and Public Finance, which are all part of the Federal Revenue Auditor exam.

The results of all models, categorized by test and subject, are available in the Annexes, which also include explanations of the competitions and their specific characteristics.

## 4. Test protocol

A team of nine researchers tested the models in all the races between October and the beginning of December 2024. All the models were tested using the most up-to-date versions available at the time.

All tests were carried out in zero-shot mode (Kojima, 2022) without any prior training or examples. The zero-shot approach is widely used to evaluate the performance of AI models, as there is a significant effect of the few-shot thought chain stimulus on the essential zero-shot stimulus (Martínez, 2024).

Among the possible research techniques, zero-shot was chosen due to the greater reliability of the answers presented. The approach guarantees greater assertiveness in relation to the option chosen, given that the AI model has never, in theory, had contact with the format of the question presented or with the expected answer (Orsini, 2023).

All the tests were carried out in prompt-question format. The prompt-question pairs were submitted individually, one by one, with the full text and without any changes to the test being evaluated. The option of repeating the prompt in each question aimed to cancel out the "memory" effect that some larger models have, which could affect the results of smaller models if there was only one original prompt at the start of the test and only the reproduction of the questions afterward.

After some preliminary tests, the following prompt in Portuguese was adopted as the standard: "Consider that you are a candidate in a higher education competition. Answer the multiple-choice questions I am about to give you, knowing that only one of the alternatives is correct. I want you to give me the percentage confidence interval for each answer. Question: ...".

The confidence index was used to gauge whether the model, according to its knowledge base, was confident in the answer provided. With the prompt used, practically all the models indicated the alternative considered correct, with comments on why it was chosen and why the other alternatives were considered incorrect.

All the answers provided by the AI models tested were compared with the official template and considered correct or incorrect according to the alternative indicated. In almost all the public exams tested, the questions have five alternatives. Only in the Brazilian Bar Association exam did the questions have four alternatives.

In public examinations, many of the questions require candidates to use associative reasoning, indicating, for example, which assertions would be correct or what would be the appropriate sequence for the proposed assertions (true, false, true, for example).

The method used to analyze the performance of each model was predominantly quantitative, based on the number of correct answers. Also, by way of "research findings," in some cases, a qualitative analysis of the data found was carried out according to, for example, the degree of difficulty of each question. (Herrmann-Werner et al., 2024).

As explained, the tests were carried out without prior training of the models, and the first answer submitted was always considered, even when it was necessary to resubmit the question or complete the prompt, such as in cases where the confidence index was not submitted.

## 4.1 Quality controls

At various times, the models failed to provide a correct alternative. In these cases, the initial standard prompt was supplemented by inserting additional commands such as "as a candidate you are required to submit an answer" or "as a candidate you are required to indicate an alternative (a, b, c, d or e) as an answer" to force the AI model to submit an alternative. This scenario occurred frequently in Gemini 1.5 Pro, probably because the model understood that the question, which referred to a hypothetical case, was a real legal problem. In some rare situations, the models presented more than one alternative as correct, which required a complementary prompt, determining that the model presented only one correct alternative.

The confidence index was almost always given as a percentage, in line with what was requested. In some cases, instead of the percentage, the models indicated "high," "medium," or "low" as the confidence index.

## 4.2 Limitations

Legal studies have several limitations, especially in Brazil, where there are continuous legislative changes and significant fluctuations in case law due to the enormous complexity of the system. For this reason, it was decided to evaluate only the most recent competitions, usually held in the last five years, as any changes in legislation may have impacted some of the answers provided by the models. The parameters used were designed to minimize this impact.

Another limitation to consider is the possibility that some models had already been trained with tested questions. This is the case with Sabiá-3, who was probably trained for the Brazilian Bar Association and the National Magistrates' Exam (Abonizio, Hugo et al.,2024). In this specific case, the possibility of contamination in the zero-shot approach cannot be ruled out.

## 5. LegalScore development

The index, which will be updated periodically with new tests and AI models, aims to determine the effectiveness and alignment of responses in various scenarios.

In this first stage, which only included multiple choice questions, according to the methodology presented, the LegalScore index involved analyzing the correct answers of each model in the various tests, with subsequent standardization of the data, performance calculations, and normalization for a final scale capable of comparing the results.

The procedures used in all the exams, with variations for each set of tests, are described below. For example, a reference was adopted in the Brazilian Bar Association exams, which have 80 questions, to convert the number of correct questions into percentages. The data was organized in a matrix where the rows represent the models evaluated, and the columns correspond to the editions of the exam.

The number of correct answers for each model per test was converted into a performance percentage using the following formula, which transforms absolute values into a standardized metric from 0 to 100%:

$$Percentage = \left(\frac{Total\ Hits}{80}\right) \times 100$$

To consolidate the performance of the models, the Z-score was used, a standardized scoring method that measures the deviation of each value from the mean, adjusted by the standard deviation of the distribution, according to the formula:

$$Z = \frac{X - \mu}{\sigma}$$

Where:

X: Percentage of model hits in a specific test;

μ: Average of the percentages of all the models in the same test;

σ: Standard deviation of the percentages in the test.

The Z-scores were calculated, by model, for all the Brazilian Bar Exam tests and then added together to generate the General Performance Index, which represents the relative performance of each AI model compared to the others.

To make it easier to interpret the results and avoid negative values, the General Performance Index was normalized to a scale of 0 to 10 using a linear transformation formula:

$$Score = \left(\frac{X - Xmin}{Xmax - Xmin}\right) \times 10$$

Where:

X: General Index of the model;

Xmin: Lowest General Index calculated;

Xmax: Highest General Index calculated.

As a result, the best-performing model will get the maximum score (10), while the worst-performing model will get zero.

To ensure that the methodology adopted allows for the inclusion of new AI models in the future, standardization and normalization will be recalculated based on the new distributions so that the performance index can evolve coherently, allowing for adequate comparisons between current models and those that will be evaluated.

Finally, to visualize the results more easily, a bar chart was used to show the overall and relative performance of the models evaluated.

The procedure described was adopted for all the tests in the competitions evaluated.

## 6. Overall results of the competitions evaluated

## 6.1 Exam of the Brazilian Bar Association (OAB)

| | OAB 32 | OAB 33 | OAB 34 | OAB 35 | OAB 36 | OAB 37 | OAB 38 | OAB 39 | OAB 40 | OAB 41 |
|---|---|---|---|---|---|---|---|---|---|---|
| | Total Correct | Total Correct | Total Correct | Total Correct | Total Correct | Total Correct | Total Correct | Total Correct | Total Correct | Total Correct |
| ChatGPT 3.5 Turbo | 41 | 47 | 41 | 43 | 41 | 46 | 34 | 48 | 45 | 51 |
| ChatGPT 4o | 62 | 64 | 71 | 63 | 67 | 65 | 67 | 68 | 68 | 69 |
| ChatGPT oi-mini | 47 | 43 | 41 | 43 | 49 | 53 | 50 | 50 | 62 | 52 |
| Claude Sonnet 3.5 | 63 | 64 | 67 | 70 | 75 | 66 | 69 | 66 | 68 | 68 |
| LLama 3.1 405B | 49 | 52 | 59 | 61 | 66 | 61 | 57 | 59 | 64 | 65 |
| Gemini 1.5 Pro | 60 | 58 | 66 | 63 | 70 | 61 | 58 | 58 | 59 | 58 |
| Command-R-Plus | 43 | 45 | 49 | 54 | 51 | 52 | 47 | 52 | 50 | 47 |
| Mistral Large-2 | 48 | 52 | 57 | 54 | 57 | 49 | 56 | 54 | 55 | 56 |
| Amazonia AI | 59 | 56 | 61 | 64 | 63 | 52 | 50 | 58 | 47 | 66 |
| Reka Core 67B | 48 | 40 | 37 | 50 | 52 | 56 | 49 | 45 | 55 | 50 |
| Perplexity AI | 51 | 53 | 58 | 53 | 53 | 55 | 52 | 57 | 58 | 53 |
| Sabiá 3 | 37 | 58 | 66 | 64 | 67 | 61 | 53 | 58 | 62 | 63 |
| GROK 2 | 51 | 58 | 60 | 43 | 67 | 55 | 59 | 56 | 58 | 57 |
| Nemotron 340B | 52 | 57 | 59 | 64 | 66 | 62 | 59 | 51 | 59 | 56 |

Table 1: Overall performance of the models by hits

| Models | General Index |
|---|---|
| Claude Sonnet 3.5 | 14,66 |
| ChatGPT 4th | 13,25 |
| Gemini 1.5 Pro | 5,85 |
| LLama 3.1 405B | 4,25 |
| Sabiá 3 | 3,60 |
| Nemotron 340B | 2,95 |
| Amazonia AI | 1,68 |
| GROK 2 | 0,56 |
| Perplexity AI | -1,96 |
| Mistral Large-2 | -3,06 |
| ChatGPT oi-mini | -8,34 |
| Command-R-Plus | -8,86 |
| Reka Core 67B | -9,46 |
| ChatGPT 3.5 Turbo | -15,13 |

Table 2: OAB General Normalized Index

| Model | Note |
|---|---|
| Claude Sonnet 3.5 | 10 |
| ChatGPT 4th | 9,53 |
| Gemini 1.5 Pro | 7,04 |
| LLama 3.1 405B | 6,51 |
| Sabiá 3 | 6,29 |
| Nemotron 340B | 6,07 |
| Amazonia AI | 5,64 |
| GROK 2 | 5,27 |
| Perplexity AI | 4,42 |
| Mistral Large-2 | 4,05 |
| ChatGPT oi-mini | 2,28 |
| Command-R-Plus | 2,11 |
| Reka Core 67B | 1,9 |
| ChatGPT 3.5 Turbo | 0 |

Table 3: Score given to each model

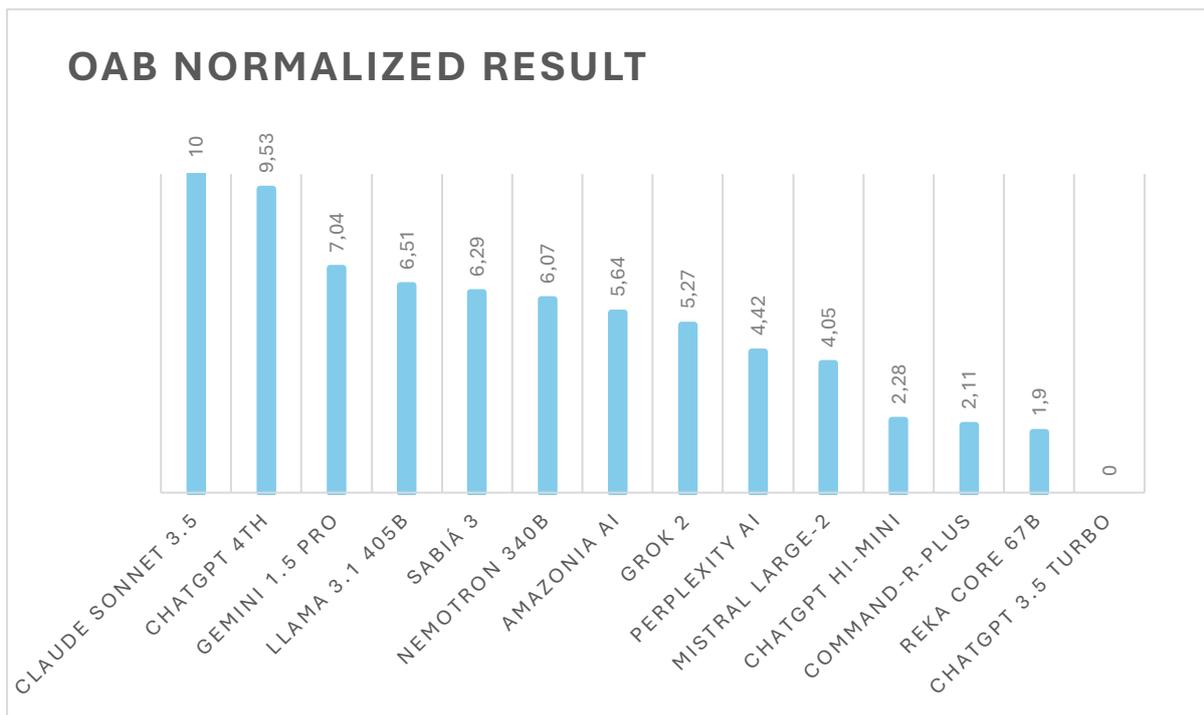

Graph 1: Normalized result

## 6.2 Federal Judge (TRF)

| Model | TRF-3 2022 Total Correct | TRF-4 2022 Total Correct | TRF-1 2023 Total Correct |
|---|---|---|---|
| ChatGPT 3.5 Turbo | 36 | 37 | 37 |
| ChatGPT 4th | 76 | 67 | 62 |
| ChatGPT oi-mini | 32 | 35 | 35 |
| Claude Sonnet 3.5 | 53 | 76 | 66 |
| LLama 3.1 405B | 58 | 57 | 57 |
| Gemini 1.5 Pro | 56 | 60 | 58 |
| Command-R-Plus | 34 | 37 | 29 |
| Mistral Large-2 | 50 | 52 | 49 |
| Amazonia AI | 45 | 47 | 55 |
| Reka Core 67B | 31 | 29 | 45 |
| Perplexity AI | 43 | 31 | 49 |
| Sabiá 3 | 34 | 46 | 59 |
| GROK 2 | 22 | 35 | 53 |
| Nemotron 340B | 27 | 27 | 58 |

Table 4: Overall performance of the models by hits

| Model | General Index | Note |
|---|---|---|
| ChatGPT 4th | 4,76 | 10 |
| Claude Sonnet 3.5 | 4,15 | 9,24 |

| Gemini 1.5 Pro | 2,5 | 7,15 |
|---|---|---|
| LLama 3.1 405B | 2,34 | 6,96 |
| Mistral Large-2 | 0,73 | 4,93 |
| Amazonia AI | 0,68 | 4,87 |
| Sabiá 3 | 0,16 | 4,21 |
| Perplexity AI | -1,13 | 2,58 |
| Nemotron 340B | -1,6 | 2 |
| GROK 2 | -1,87 | 1,66 |
| ChatGPT 3.5 Turbo | -2,25 | 1,18 |
| Reka Core 67B | -2,38 | 1,01 |
| ChatGPT oi-mini | -2,91 | 0,35 |
| Command-R-Plus | -3,19 | 0 |

Table 5: Score given to each model

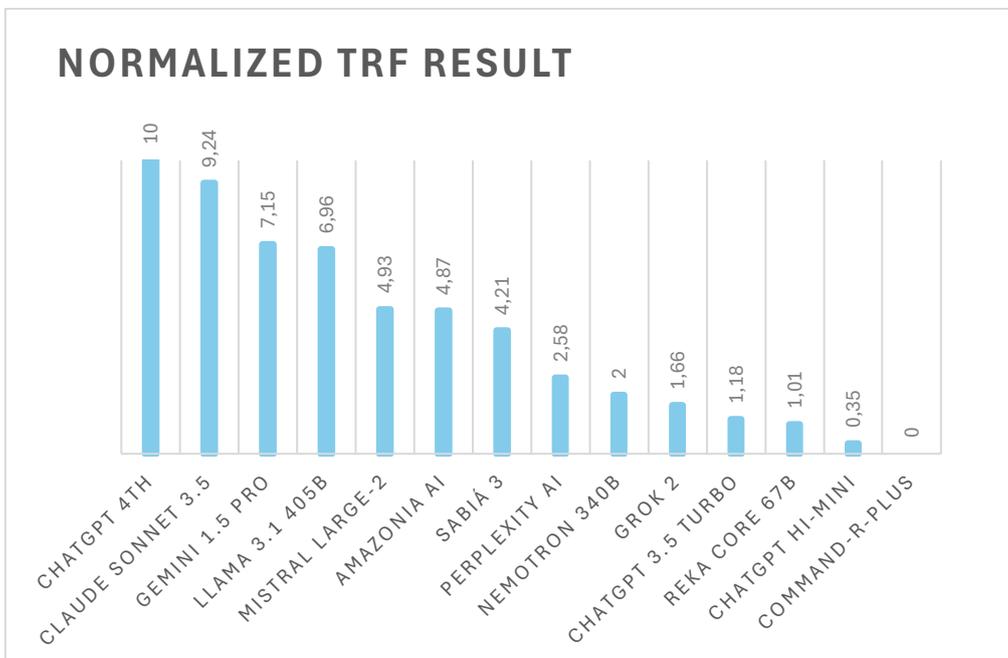

Graph 2: Normalized result

## 6.3 Public Prosecutor (MPF)

| | 29th Competition | 30th competition |
|---|---|---|
| **Model** | **Total Correct** | **Total Correct** |
| ChatGPT 3.5 Turbo | 45 | 50 |
| ChatGPT 4th | 70 | 72 |
| ChatGPT oi-mini | 58 | 57 |
| Claude Sonnet 3.5 | 92 | 92 |
| LLama 3.1 405B | 75 | 73 |
| Gemini 1.5 Pro | 70 | 88 |
| Command-R-Plus | 56 | 39 |
| Mistral Large-2 | 64 | 66 |
| Amazonia AI | 70 | 67 |
| Reka Core 67B | 56 | 55 |
| Perplexity AI | 55 | 65 |
| Sabiá 3 | 79 | 77 |
| GROK 2 | 72 | 75 |
| Nemotron 340B | 73 | 72 |

Table 6: Overall model performance by hits

| Model | General Index | Note |
|---|---|---|
| Claude Sonnet 3.5 | 3,8 | 10 |
| Gemini 1.5 Pro | 1,69 | 6,92 |
| Sabiá 3 | 1,67 | 6,89 |
| LLama 3.1 405B | 1,05 | 5,99 |
| GROK 2 | 0,95 | 5,83 |
| Nemotron 340B | 0,82 | 5,65 |
| ChatGPT 4th | 0,57 | 5,28 |
| Amazonia AI | 0,22 | 4,77 |
| Mistral Large-2 | -0,35 | 3,94 |
| Perplexity AI | -1,17 | 2,75 |
| ChatGPT oi-mini | -1,48 | 2,29 |
| Reka Core 67B | -1,79 | 1,85 |
| Command-R-Plus | -2,91 | 0,21 |
| ChatGPT 3.5 Turbo | -3,06 | 0 |

Table 7: Score given to each model

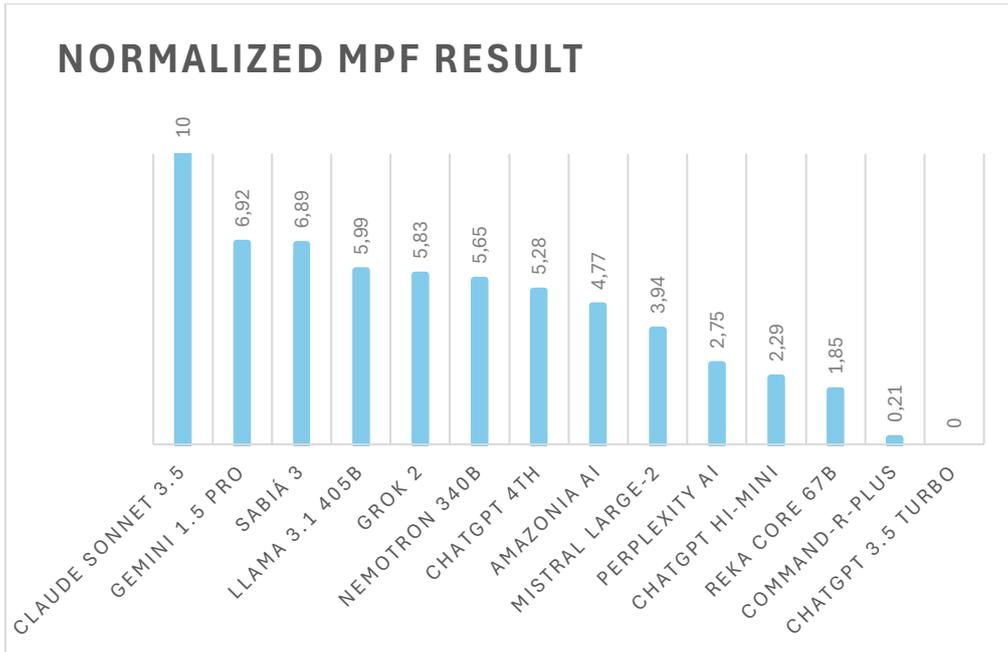

Graph 3: Normalized result

## 6.4 National Judicial Exam (ENAM)

| | ENAM 1 | ENAM 2 |
|---|---|---|
| **Model** | **Total Correct** | **Total Correct** |
| ChatGPT 3.5 Turbo | 33 | 33 |
| ChatGPT 4th | 46 | 62 |
| ChatGPT oi-mini | 30 | 47 |
| Claude Sonnet 3.5 | 61 | 64 |
| LLama 3.1 405B | 41 | 54 |
| Gemini 1.5 Pro | 48 | 58 |
| Command-R-Plus | 34 | 40 |
| Mistral Large-2 | 33 | 18 |
| Amazonia AI | 44 | 50 |
| Reka Core 67B | 35 | 36 |
| Perplexity AI | 37 | 41 |
| Sabiá 3 | 51 | 23 |
| GROK 2 | 45 | 52 |
| Nemotron 340B | 43 | 51 |

Table 8: Overall performance of the models by hits

| Model | General Index | Note |
|---|---|---|
| ChatGPT 4th | 4,76 | 10 |
| Claude Sonnet 3.5 | 4,15 | 9,24 |
| Gemini 1.5 Pro | 2,5 | 7,15 |
| LLama 3.1 405B | 2,34 | 6,96 |
| Mistral Large-2 | 0,73 | 4,93 |
| Amazonia AI | 0,68 | 4,87 |
| Sabiá 3 | 0,16 | 4,21 |
| Perplexity AI | -1,13 | 2,58 |
| Nemotron 340B | -1,6 | 2 |
| GROK 2 | -1,87 | 1,66 |
| ChatGPT 3.5 Turbo | -2,25 | 1,18 |
| Reka Core 67B | -2,38 | 1,01 |
| ChatGPT oi-mini | -2,91 | 0,35 |
| Command-R-Plus | -3,19 | 0 |

Table 9: Score given to each model

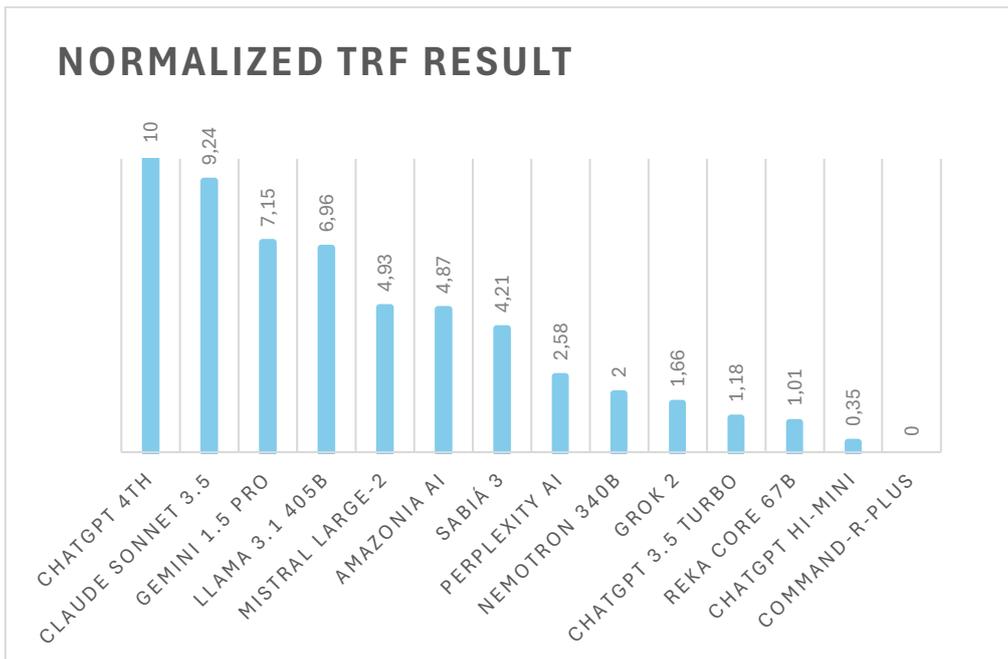

Graph 4: Normalized result

## 6.5 Federal Revenue Tax Auditor (AFRFB)

| Model | Total Correct | Total Incorrect |
|---|---|---|
| ChatGPT 3.5 Turbo | 64 | 76 |
| ChatGPT 4th | 112 | 28 |
| ChatGPT oi-mini | 100 | 40 |
| Claude Sonnet 3.5 | 118 | 22 |
| LLama 3.1 405B | 94 | 46 |
| Gemini 1.5 Pro | 103 | 37 |
| Command-R-Plus | 73 | 67 |
| Mistral Large-2 | 93 | 47 |
| Amazonia AI | 94 | 46 |
| Reka Core 67B | 82 | 58 |
| Perplexity AI | 105 | 35 |
| Sabiá 3 | 103 | 37 |
| GROK 2 | 105 | 35 |
| Nemotron 340B | 89 | 51 |

Table 10: Overall model performance by hits

| | General Index | Note |
|---|---|---|
| Claude Sonnet 3.5 | 1,52 | 10 |
| ChatGPT 4th | 1,13 | 8,95 |
| Perplexity AI | 0,66 | 7,63 |
| GROK 2 | 0,66 | 7,63 |
| Gemini 1.5 Pro | 0,56 | 7,37 |
| Sabiá 3 | 0,56 | 7,37 |
| ChatGPT oi-mini | 0,27 | 6,58 |
| LLama 3.1 405B | -0,11 | 5,53 |
| Amazonia AI | -0,11 | 5,53 |
| Mistral Large-2 | -0,2 | 5,26 |
| Nemotron 340B | -0,4 | 4,74 |
| Reka Core 67B | -0,87 | 3,42 |
| Command-R-Plus | -1,54 | 1,58 |
| ChatGPT 3.5 Turbo | -2,12 | 0 |

Table 11: Score given to each model

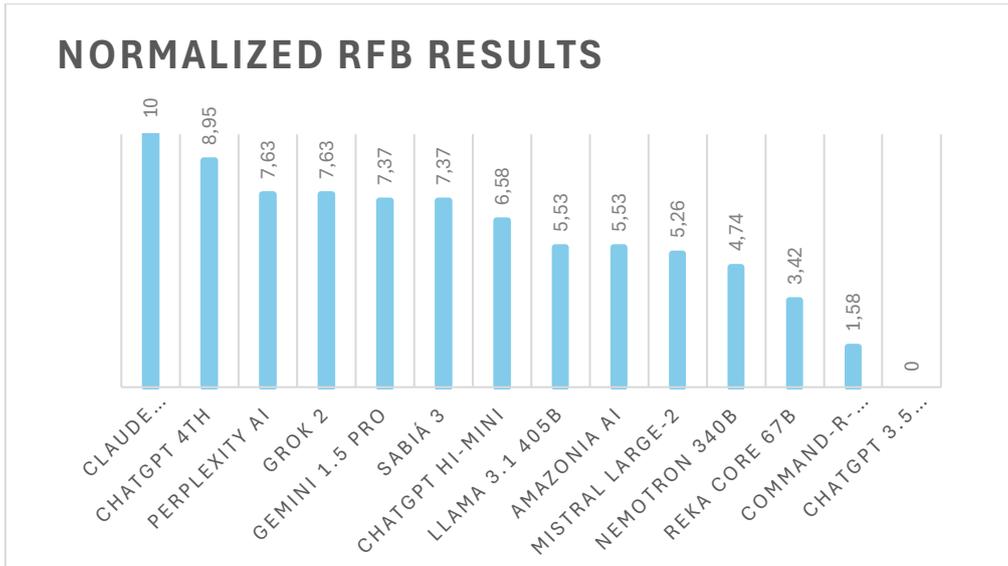

Graph 5: Normalized result

## 6.6 Public Ministry of Labor (MPT)

| | 22ⁿᵈ Competition | 23ʳᵈ competition |
|---|---|---|
| **Model** | **Total Correct** | **Total Correct** |
| ChatGPT 3.5 Turbo | 27 | 36 |
| ChatGPT 4th | 50 | 58 |
| ChatGPT oi-mini | 43 | 39 |
| Claude Sonnet 3.5 | 63 | 66 |
| LLama 3.1 405B | 50 | 56 |
| Gemini 1.5 Pro | 42 | 55 |
| Command-R-Plus | 37 | 34 |
| Mistral Large-2 | 42 | 46 |
| Amazonia AI | 44 | 60 |
| Reka Core 67B | 39 | 48 |
| Perplexity AI | 39 | 45 |
| Sabiá 3 | 59 | 65 |
| GROK 2 | 49 | 59 |
| Nemotron 340B | 50 | 60 |

Table 12: Overall performance of the models by hits

| **Model** | **General Index** | **Note** |
|---|---|---|
| Claude Sonnet 3.5 | 3,26 | 10 |
| Sabiá 3 | 2,73 | 9,22 |
| Nemotron 340B | 1,28 | 7,06 |
| ChatGPT 4th | 1,09 | 6,79 |
| GROK 2 | 1,07 | 6,76 |
| LLama 3.1 405B | 0,9 | 6,51 |
| Amazonia AI | 0,62 | 6,1 |
| Gemini 1.5 Pro | -0,07 | 5,08 |
| Mistral Large-2 | -0,92 | 3,82 |
| Reka Core 67B | -1,06 | 3,61 |
| Perplexity AI | -1,34 | 3,19 |
| ChatGPT oi-mini | -1,47 | 3 |
| Command-R-Plus | -2,6 | 1,33 |
| ChatGPT 3.5 Turbo | -3,5 | 0 |

Table 13: Score given to each model

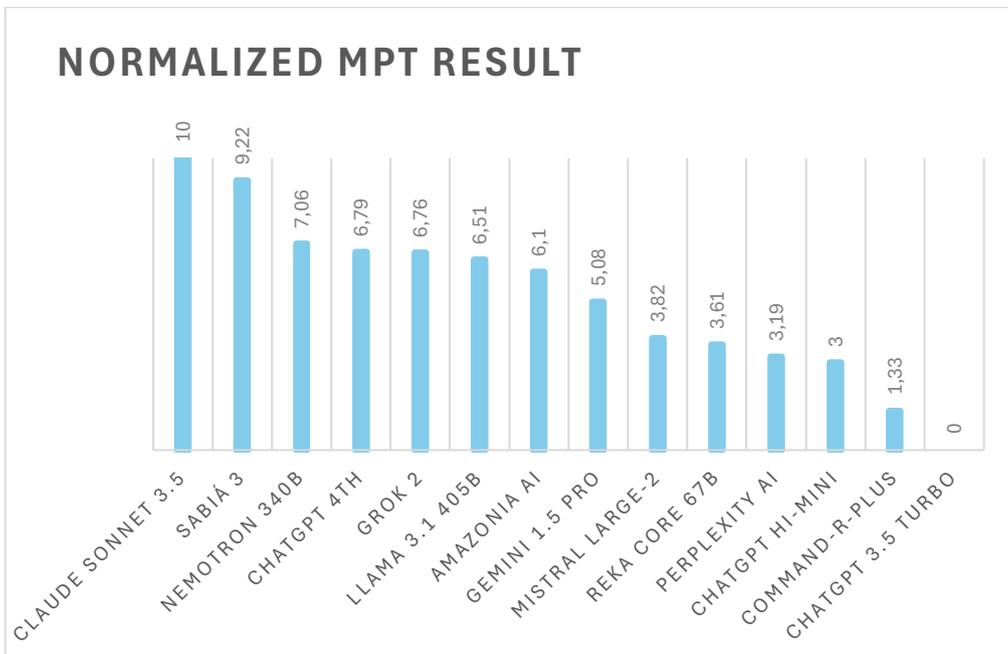

Graph 6: Normalized result

## 7. General Classification

To determine which models obtained the best results, 14 AI models were evaluated, with 14 points awarded to the first-placed model in each contest, 13 points to the second-placed model, and so on.

The final ranking, with the total scores for each model, is shown in the table below.

| Model | Final ranking |
|---|---|
| Claude Sonnet 3.5 | 83 |
| ChatGPT 4th | 72 |
| Gemini 1.5 Pro | 66 |
| Sabiá 3 | 59 |
| LLama 3.1 405B | 57 |
| GROK 2 | 54 |
| Nemotron 340B | 49 |
| Amazonia AI | 48 |
| Perplexity AI | 40 |
| Mistral Large-2 | 33 |
| ChatGPT oi-mini | 26 |
| Reka Core 67B | 19 |
| Command-R-Plus | 14 |
| ChatGPT 3.5 Turbo | 10 |

Table 14: Overall ranking

Therefore, according to the proposed system, the best AI model evaluated was Claude Sonnet 3.5, followed by ChatGPT 4o, Gemini 1.5 Pro, and Sabiá 3, a model trained with datasets specific to Brazil.

Claude's position is in line with the original perception of the research leader. Practical experience in various legal scenarios over several months indicated that Anthropic's model showed better results than its two main competitors and the research confirmed this.

## 8. Discussion

### 8.1 Interpretation of results

The 14 AI models selected in this study showed varying performance in the different tests of each competition.

The most consistent result was obtained in the Brazilian Bar Association exam, which requires a minimum pass rate of 50% of the 80 questions in each exam. Virtually all the models evaluated would have passed any of the last ten OAB exams, except ChatGPT 3.5 in exam 37 and Sabiá 3 in exam 32.

This result can be used as a starting point for an in-depth reflection on the model for assessing future lawyers in Brazil. While the AI models have been relatively successful in the OAB's multiple-choice exams (see annexes), considering the tests evaluated, the percentage of human candidates who definitely pass the exams is around 15%.

An important observation needs to be made here. The OAB exams consist of two phases: the first, with 80 multiple-choice questions, and the second, with discursive questions, according to the area of practice chosen by the candidate, as well as the drafting of a legal paper. This research, in Phase 1, tested the 14 AI models only in relation to the multiple-choice questions, as shown. The next research stage will check the performance of all the models on discursive questions.

The definitive result, for comparison purposes, still needs to be investigated. However, it is reasonable to imagine that the performance of the models, in general, will be superior to that of the human candidates.

Even so, it is understood that legal education in the country, which is known to be poor, needs to be rethought, as well as how university graduates are assessed.

On the other hand, the most important legal exams in Brazil were more challenging for the AI models evaluated. Most of them did not achieve the minimum scores required to pass the first stage, multiple choice, of the exams analyzed.

A notable exception was the competition for the position of Federal Revenue Auditor, which has long been considered one of the most difficult in the country, not only because of the level of the questions but also because it requires candidates to have, in addition to legal knowledge, expertise in exact subjects, such as logical reasoning, economics, statistics, finance, and data fluency, as well as mastery of the Portuguese and English languages.

As expected, the models obtained excellent results in the language tests (which basically deal with interpretative questions) and in the mathematical subjects, although they found it difficult to solve more extensive and complex questions, such as accounting, which require reasoning and development.

The point outside the curve was the performance of the Claude Sonnet 3.5 and ChatgGPT 4o, which would not only have passed the contest but would have been the first in the overall standings, as the following graph shows.

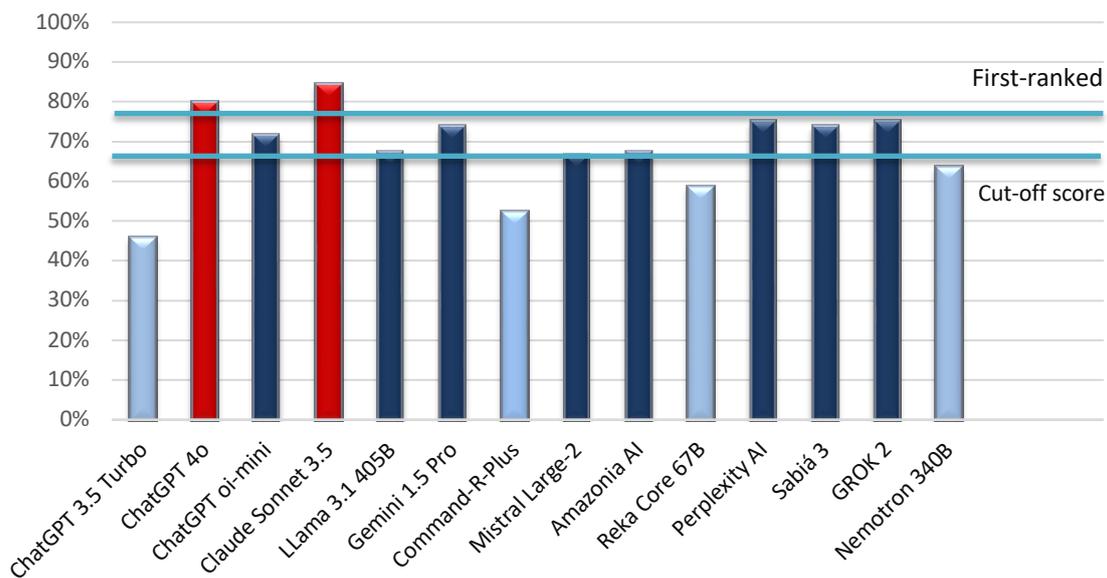

On the other hand, the poor performance in strictly legal competitions can largely be attributed to the fact that most of the models were trained with datasets predominantly in English, which impaired their understanding of specific legal terminology or language nuances typical of the Portuguese language (Caparroz, 2024).

Another challenge identified was the difficulty in addressing ambiguous questions requiring more in-depth reasoning. The models also needed to be improved in their ability to correlate information when faced with interdisciplinary questions.

In general terms, the models struggled most in areas such as Tax Law, Labor Law, and Criminal Law, particularly with questions that required analysis of case law or more specific knowledge, such as the content of regulations and regional laws. Indeed, the content required in the competitions was outside the data used, which is more generic and restricted to national and better-known laws, such as the codes of the primary disciplines.

This finding is particularly significant, as it underscores that the more specific the question, the worse the performance of all evaluated AI models in nearly every instance. This highlights the critical need to develop datasets tailored to the Brazilian context (Caparroz, 2024).

As things stand, AI models offer different forms of support to candidates:

- Solving questions: understanding the format of the questions and training for quick answers.
- Explanations: providing feedback and explanations on complex topics, helping with the understanding of concepts.
- Simulated tests: the creation of simulated tests, allowing both the training of candidates and the measurement of their performance in a controlled environment.
- Personalization of learning: identifying patterns in candidates' mistakes and recommending specific study topics to improve performance.

The study also indicates that the bodies that conduct the tests can benefit from using AI models to create and revise questions, as well as to parameterize the difficulty of the tests, provided, of course, that various precautions are taken regarding the possible leakage of questions.

## 8.2 Summary of Findings

The study evaluated 14 AI models, categorized into four groups for comparison. The first group comprises large proprietary models such as ChatGPT4o, Gemini 1.5 Pro, and Claude Sonnet 3.5. As expected, this group performed best, which can be attributed to two factors: a) the huge number of parameters and b) the fact that, as they are commercial models, they were probably trained with datasets more representative of the Brazilian reality.

The second group comprises models considered large but substantially smaller than the three leaders, such as LLama 3.1 405B, GROK 2, Nemotron 340B, and Perplexity AI, which uses a combination of large models. These models were in an intermediate position in the overall ranking. The third group is made up of smaller models, such as Mistral Large-2, ChatGPT oi-mini, Reka Core 67B, Command-R-Plus, and ChatGPT 3.5 Turbo (in this case, a relatively large model, with 175 billion parameters, but old compared to the others). These models performed the worst, coming last in the overall ranking.

Finally, the fourth group comprises the two models trained for the Brazilian reality, Sabiá 3 and Amazonia IA. Unfortunately, Brazilian companies do not provide information on the models' size or the datasets they were trained in. It is assumed that they are small/medium-sized models in terms of parameters but that they obtained good rankings (4th place and eighth place, respectively), which shows that, for regional and language alignment, the datasets tend to be as important if not more important, than the raw capacity of the models.

The relevance of the datasets in the context evaluated can be confirmed by the fact that, in several questions, practically all the models showed errors when indicating the same alternative as the answer. This allows us to conclude that, regardless of size, most of these models were trained on the same *datasets*, which proved insufficient to provide adequate answers.

## 8.3 Contributions to the field

The findings suggest that while AI models have potential as complementary tools, they are still far from matching human performance in more complex public exams. The linguistic, cultural, and contextual limitations observed in the research highlight the need for public investment in developing datasets adapted to the reality and characteristics of the complex Brazilian legal system.

LegalScore contributes to future research in artificial intelligence applied to Brazilian law by creating a reference. The index could be developed as a tool for improving models, making it possible to objectively assess the improvements incorporated into the new versions tested and contributing to the development of more efficient educational solutions.

The benchmark model developed for LegalScore can be replicated in other countries, which probably have peculiarities that the AI models available on the market have not yet considered.

Future studies should deepen the analysis in other cultural contexts, which also require specific datasets aligned with the expectations of non-Anglophone countries, such as Brazil.

LegalScore presents a new paradigm for evaluating the performance of AI models in complex human tasks, reinforcing the importance of aligning emerging technologies with local and sectoral demands, especially in critical educational and professional contexts, such as legal careers.

## Conclusion

This research developed an objective analysis of the performance of various AI models in Brazilian public exams aimed at selecting candidates for relevant legal positions, such as the Judiciary (in the Federal and State spheres), the Labor Prosecutor's Office, the Public Prosecutor's Office, the Tax Auditor of the Federal Revenue Service of Brazil, and the aptitude test for the practice of law.

The work resulted in a pioneering benchmark, called LegalScore, used to measure the accuracy and efficiency of the AI models tested in their most recent versions. LegalScore's permanent creation will make it possible to evaluate new models as they become available to the public.

The research findings show that AI models face difficulties in multiple-choice questions that require contextual interpretation, especially in questions that demand an understanding of specific laws, infra-legal norms, case law, and local particularities.

Despite these limitations, LegalScore is believed to offer a reliable metric for evaluating the evolution of models and identifying possible points for improvement, especially regarding the need to create datasets aligned with the Brazilian legal landscape and reality.

The index is also intended to serve as a reference for the debate on legal education in Brazil, especially considering that the models evaluated are quite recent and that their performance could soon surpass that of human candidates if the current format of the tests is maintained.

With the exponential advance of generative AI, all legal professionals will need to develop new skills, which must be identified and disseminated on a large scale.

## Appendices

### Brazilian Federal Revenue Service

The competition is considered one of the most difficult and competitive in Brazil, as anyone with a university degree can participate. The AI models were tested based on the last competition, held in 2023. There were two multiple-choice tests, with 140 questions in total, on the following subjects: Portuguese Language, Foreign Language (English or Spanish), Logical and Quantitative Reasoning, Civil, Criminal, and Business Law, Constitutional and Administrative Law, General and Public Administration and Data Fluency, as well as specific knowledge of Tax Law, General and Advanced Accounting, Tax and Customs Legislation, International Trade and Customs Legislation, Auditing, Economics and Public Finance.

One of the competition's characteristics is the variety of subjects, in addition to legal knowledge, which often creates difficulties for candidates. However, AI models perform very well in language questions (especially English, due to their language training) and maths-related subjects.

In the competition evaluated, AI models had difficulty, above all, in two subjects: a) Accounting, due to long questions that require the development of concatenated calculations, and b) Public Finance, due to the specificities of Brazil legislation.

Candidates who pass the multiple-choice test have their discursive exams corrected, and the best-placed candidates within the number of vacancies offered pass.

### Analysing the cut-off marks for the position of Tax Auditor

The research analysed the distribution of cut-off scores for the Tax Auditor positions, considering both the broad competition and the vacancies reserved for blacks and people with disabilities (PwD). The analysis covers the maximum scores, the percentages achieved in relation to the total points, and the minimum scores required for the discursive tests to be corrected.

The Auditor exam had 140 multiple-choice questions, each worth 1 point.

### Cut-off Point: Federal Revenue Tax Auditor

| Candidate | Available positions | Cut-off Point | % of Cut-off Point |
|---|---|---|---|
| Open competition | 172 | 93 points | 66,43% |
| Black | 46 | 86 points | 61,43% |
| People with disabilities | 12 | 85 points | 60,71% |

### Performance of the Best-ranked in the Federal Revenue Auditor Contest

| Candidate | First place score | % of Maximum Score |
|---|---|---|
| Open competition | 111 points | 79,29% |
| Black | 97 points | 69,29% |
| People with disabilities | 94 points | 67,14% |

### Comparison and Reflections

The difference between the scores of the top candidates and the cut-off marks reinforces the idea that the Internal Revenue Service competitions remain highly competitive, even with the reduction in the number of vacancies in recent years. Regarding the position of Tax Auditor, the difference between the first-placed candidate (111 points) and the cut-off mark (93 points) was 18 points, or around 12.86 percent of the maximum score.

Analyzing the cut-off marks and the scores of the first-placed candidates in the Federal Revenue Service competition shows a rigorous selection process with extensive and challenging questions.

The cut-off marks were relatively low, which shows the degree of difficulty of the tests faced by human candidates.

**General conclusion**

**Analysis of the Federal Revenue Auditor Contest**

| Model | Total Correct | Total Incorrect | % Accuracy (Total) | % Accuracy (Easy Questions) | % Accuracy (Difficult Questions) |
|---|---|---|---|---|---|
| **ChatGPT 3.5 Turbo** | 64 | 76 | 46% | 54% | 34% |
| **ChatGPT 4o** | 112 | 28 | 80% | 93% | 61% |
| **ChatGPT oi-mini** | 100 | 40 | 71% | 76% | 60% |
| **Claude Sonnet 3.5** | 118 | 22 | 84% | 89% | 77% |
| **LLama 3.1 405B** | 94 | 46 | 67% | 73% | 59% |
| **Gemini 1.5 Pro** | 103 | 37 | 74% | 79% | 64% |
| **Command-R-Plus** | 73 | 67 | 52% | 62% | 38% |
| **Mistral Large-2** | 93 | 47 | 66% | 71% | 56% |
| **Amazonia AI** | 94 | 46 | 67% | 73% | 58% |
| **Reka Core 67B** | 82 | 58 | 59% | 65% | 50% |
| **Perplexity AI** | 105 | 35 | 75% | 82% | 63% |
| **Sabiá 3** | 103 | 37 | 74% | 79% | 63% |
| **GROK 2** | 105 | 35 | 75% | 82% | 63% |
| **Nemotron 340B** | 89 | 51 | 64% | 66% | 60% |

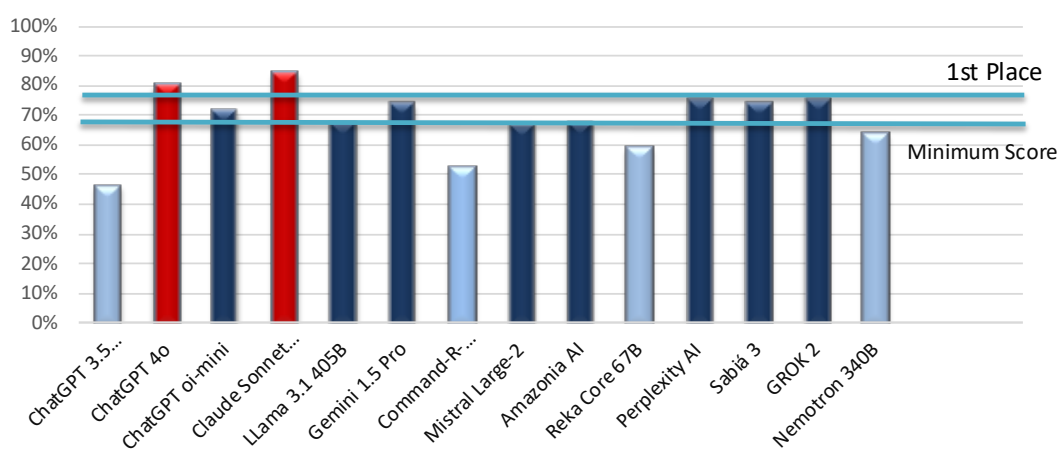

**Consolidated Grade RFB**

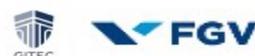

**Graphical Analysis of the Performance of Artificial Intelligence Models Tested in the First Exam, conducted during the morning**

| Model | Total Correct | Total Incorrect | % Accuracy (Total) | % Accuracy (Easy Questions) | % Accuracy (Difficult Questions) |
|---|---|---|---|---|---|
| ChatGPT 3.5 Turbo | 38 | 42 | 48% | 52% | 41% |
| ChatGPT 4o | 66 | 14 | 83% | 96% | 65% |
| ChatGPT oi-mini | 69 | 11 | 86% | 96% | 74% |
| Claude Sonnet 3.5 | 70 | 10 | 88% | 98% | 74% |
| LLama 3.1 405B | 54 | 26 | 68% | 76% | 56% |
| Gemini 1.5 Pro | 64 | 16 | 80% | 87% | 71% |
| Command-R-Plus | 44 | 36 | 55% | 65% | 41% |
| Mistral Large-2 | 60 | 20 | 75% | 78% | 71% |
| Amazonia AI | 58 | 22 | 73% | 80% | 62% |
| Reka Core 67B | 49 | 31 | 61% | 67% | 53% |
| Perplexity AI | 65 | 15 | 81% | 91% | 68% |
| Maritaca AI (Sabiá 3) | 65 | 15 | 81% | 91% | 68% |
| GROK 2 | 65 | 15 | 81% | 91% | 68% |
| Nemotron 340B | 52 | 28 | 65% | 67% | 62% |

## % Accuracy **Exam 1 - Morning**

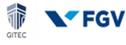
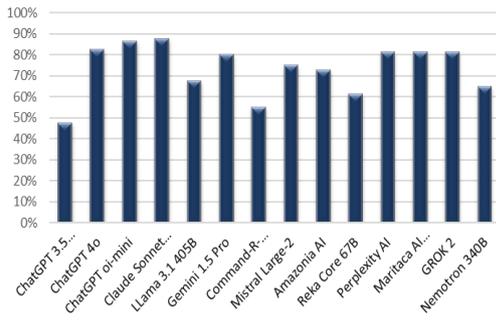

## % Accuracy **(English)**

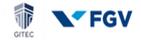
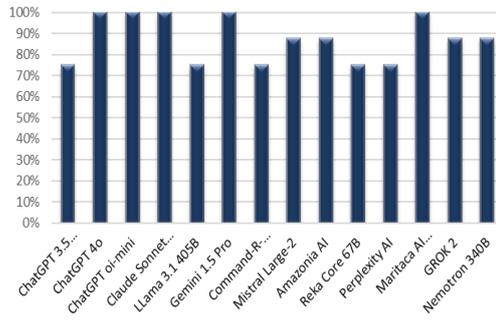

## % Accuracy **(Easy Questions)**

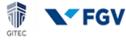
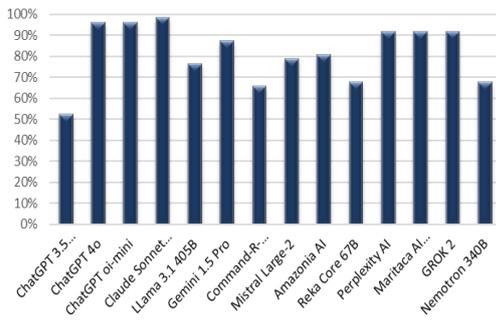

## % Accuracy **(Logical Reasoning)**

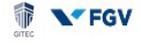
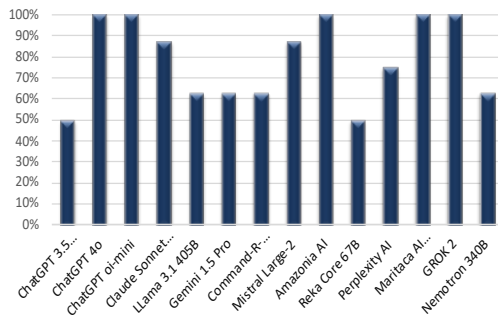

## % Accuracy **(Difficult Questions)**

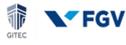
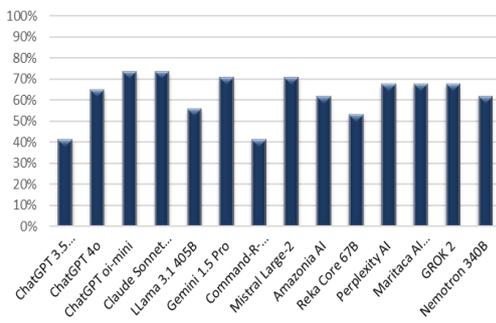

## % Accuracy **(Statistics)**

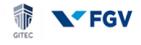
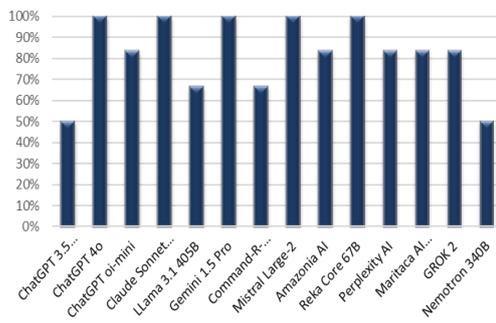

## % Accuracy **(Portuguese)**

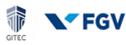
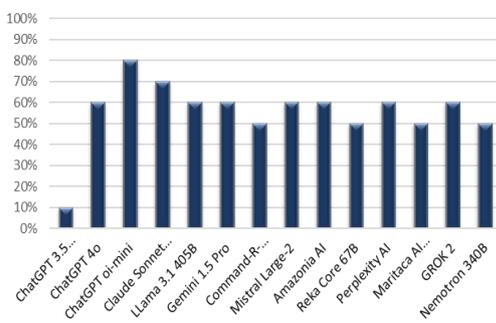

## % Accuracy **(Economy)**

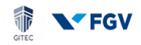
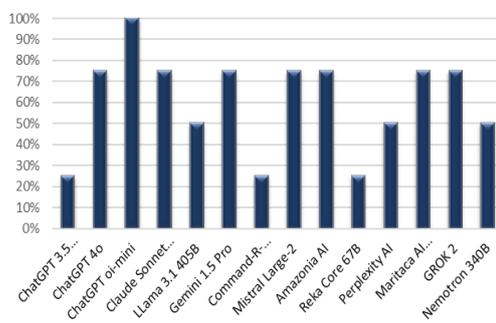

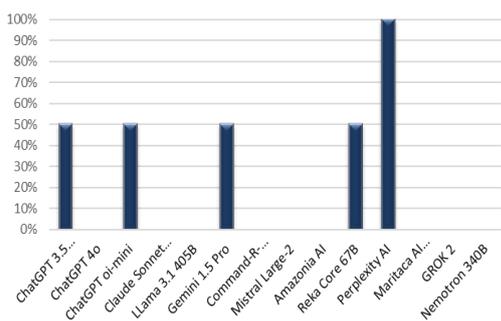

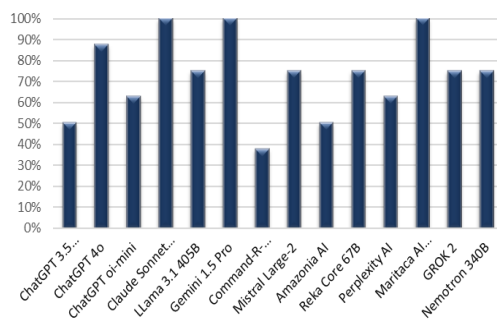

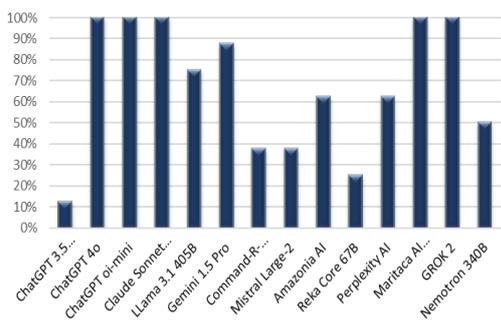

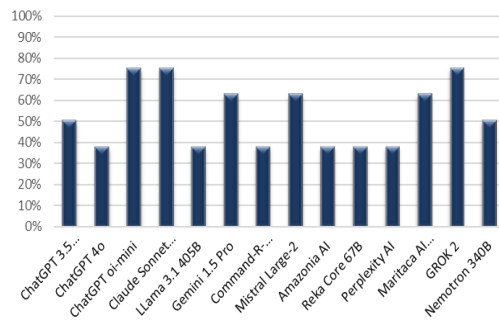

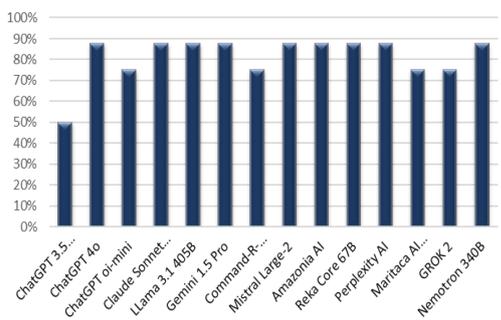

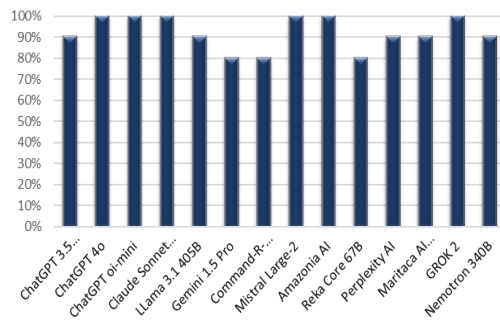

**Graphical Analysis of the Performance of Artificial Intelligence Models Tested in the Second Exam, conducted during the afternoon**

| Model | Total Correct | Total Incorrect | % Accuracy (Total) | % Accuracy (Easy Questions) | % Accuracy (Difficult Questions) |
|---|---|---|---|---|---|
| **ChatGPT 3.5 Turbo** | 26 | 34 | 43% | 56% | 27% |
| **ChatGPT 4o** | 46 | 14 | 77% | 91% | 58% |
| **ChatGPT oi-mini** | 31 | 29 | 52% | 56% | 46% |
| **Claude Sonnet 3.5** | 48 | 12 | 80% | 79% | 81% |
| **LLama 3.1 405B** | 40 | 20 | 67% | 71% | 62% |
| **Gemini 1.5 Pro** | 39 | 21 | 65% | 71% | 58% |
| **Command-R-Plus** | 29 | 31 | 48% | 59% | 35% |
| **Mistral Large-2** | 33 | 27 | 55% | 65% | 42% |
| **Amazonia AI** | 36 | 24 | 60% | 65% | 54% |
| **Reka Core 67B** | 33 | 27 | 55% | 62% | 46% |
| **Perplexity AI** | 40 | 20 | 67% | 74% | 58% |
| **Sabiá 3** | 38 | 22 | 63% | 68% | 58% |
| **GROK 2** | 40 | 20 | 67% | 74% | 58% |
| **Nemotron 340B** | 37 | 23 | 62% | 65% | 58% |

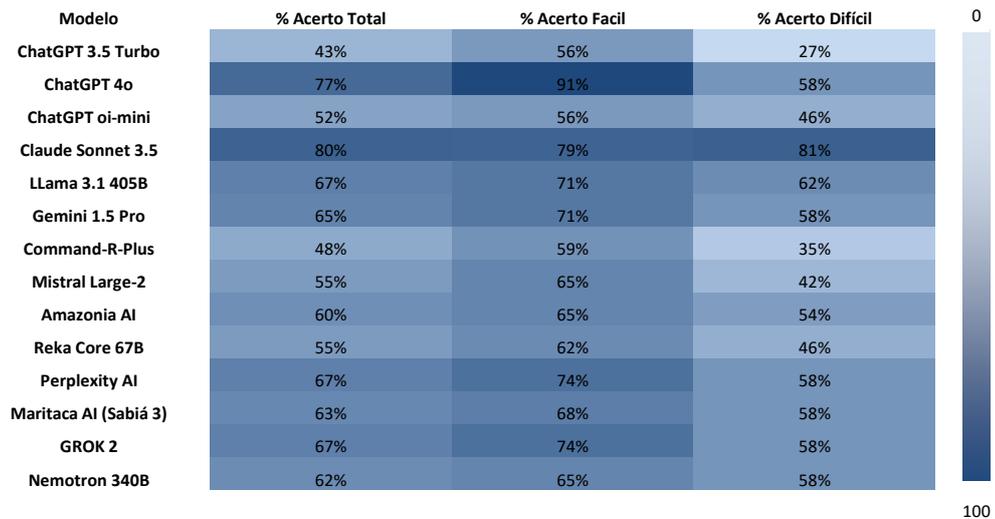

| Modelo | % Acerto Total | % Acerto Facil | % Acerto Difícil |
|---|---|---|---|
| ChatGPT 3.5 Turbo | 43% | 56% | 27% |
| ChatGPT 4o | 77% | 91% | 58% |
| ChatGPT oi-mini | 52% | 56% | 46% |
| Claude Sonnet 3.5 | 80% | 79% | 81% |
| LLama 3.1 405B | 67% | 71% | 62% |
| Gemini 1.5 Pro | 65% | 71% | 58% |
| Command-R-Plus | 48% | 59% | 35% |
| Mistral Large-2 | 55% | 65% | 42% |
| Amazonia AI | 60% | 65% | 54% |
| Reka Core 67B | 55% | 62% | 46% |
| Perplexity AI | 67% | 74% | 58% |
| Maritaca AI (Sabiá 3) | 63% | 68% | 58% |
| GROK 2 | 67% | 74% | 58% |
| Nemotron 340B | 62% | 65% | 58% |

## % Accuracy **Exam 2 - Afternoon**

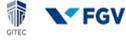

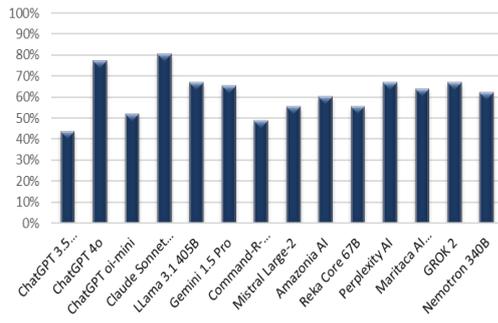

## % Accuracy **(Constitutional Law)**

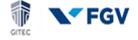

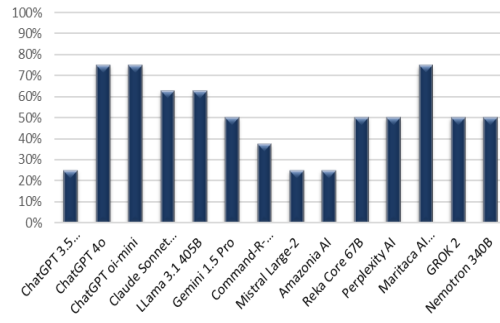

## % Accuracy **(Easy Questions)**

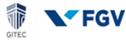

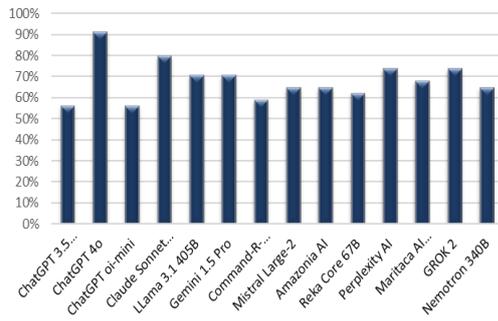

## % Accuracy **(Social Security Law)**

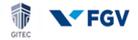

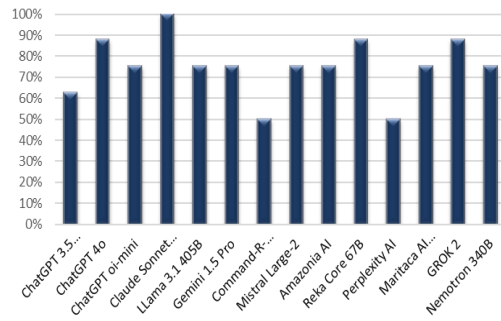

## % Accuracy **(Difficult Questions)**

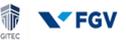

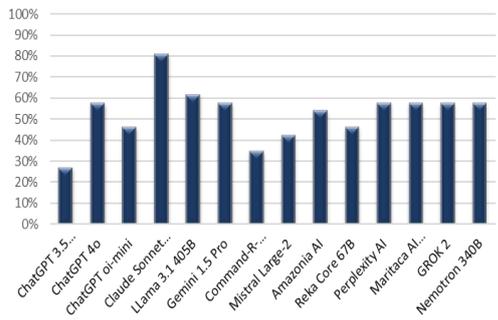

## % Accuracy **(Tax Law)**

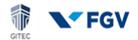

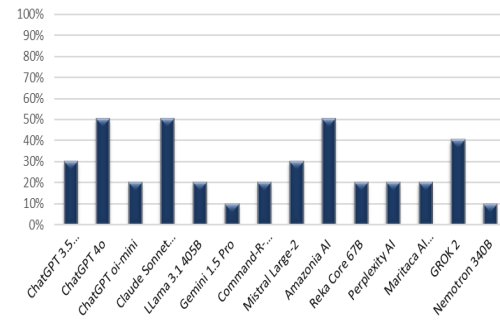

## % Accuracy **(Administrative Law)**

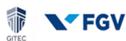

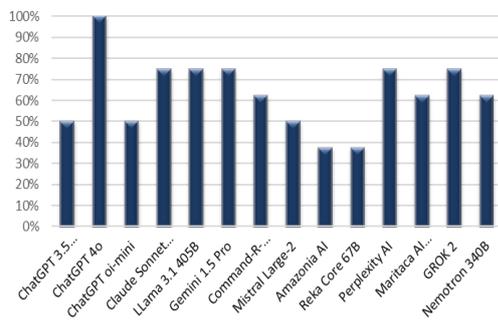

## % Accuracy **(Tax Legislation)**

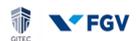

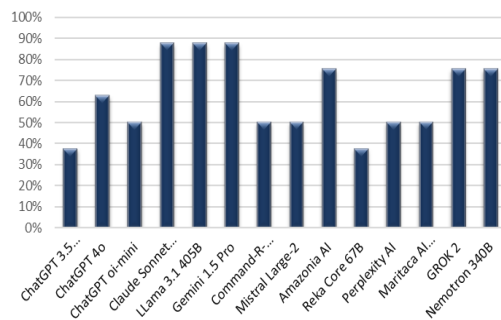

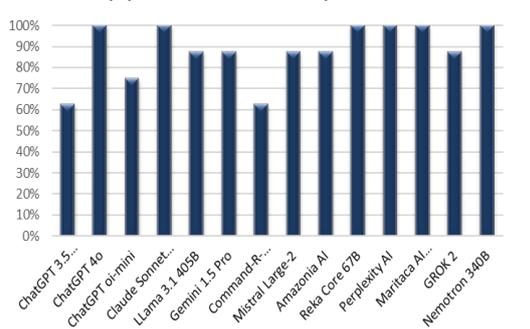

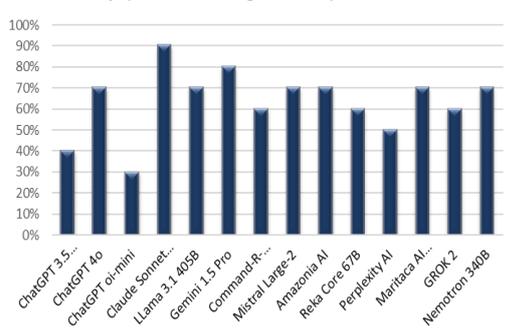

## Federal Prosecutor

This is a high-level civil service examination for the position of Federal Prosecutor in Brazil. The exam is considered one of the most challenging in the legal field in the country. Eligibility is restricted to law graduates with at least three years of legal experience.

The examination questions are prepared and graded by the institution's own Examination Committee, chaired by the Prosecutor General of the Republic. The committee also includes two additional members of the Federal Public Ministry (MPF), a legal scholar appointed by the Superior Council of the MPF, and a lawyer appointed by the Federal Council of the Brazilian Bar Association (OAB).

The selection process consists of five written exams, including a general multiple-choice exam, four essay-based exams covering specific groups of subjects, oral examinations for each subject, and an assessment of professional credentials.

The multiple-choice exam includes 120 questions that cover the following areas of law: Constitutional Law, Legal Methodology, International Human Rights Protection, Electoral Law, Administrative Law, Environmental Law, Tax Law, Financial Law, Public International Law, Private International Law, Economic Law, Consumer Law, Civil Law, Civil Procedure, Criminal Law, and Criminal Procedure.

The subjects are divided into the following groups:

GROUP I

Constitutional Law, Legal Methodology, International Human Rights Protection, Electoral Law

GROUP II

Administrative Law, Environmental Law, Tax Law, Financial Law, Public International Law, and Private International Law

GROUP III

Economic Law, Consumer Law, Civil Law, and Civil Procedure

GROUP IV

Criminal Law and Criminal Procedure

The multiple-choice exams from the last two Federal Prosecutor selection processes were submitted to the AI models for evaluation.

The exam for the 29[th] Selection Process was administered on March 12, 2017. This selection offered 83 positions, with 10% of the vacancies reserved for people with disabilities. A total of 13,772 candidates applied, resulting in a competition ratio of 165.9 candidates per vacancy (Edital 14/2016, 2016).

The exam for the 30[th] Selection Process was administered on November 27, 2022. This selection offered 13 positions, with 10% of the vacancies reserved for people with disabilities, 20% for Black candidates, and 5% for Indigenous candidates. A total of 5,303 candidates applied, resulting in a competition ratio of 407.9 candidates per vacancy (Edital PGR/MPF Nº 6, 2022).

## Performance of the Top Scorer In The Federal Prosecutor Examinations

| Selection Process | Group I | Group II | Group III | Group IV | Average Score | % of the maximum score |
|---|---|---|---|---|---|---|
| 29th | 60.00 | 90.00 | 83.33 | 50.00 | 70.83 | 70.83 % |
| 30th | 96.66 | 96.66 | 96.66 | 83.33 | 93.33 | 93.33% |

## Cutoff Score

| Selection Process | Group I | Group II | Group III | Group IV | Average Score |
|---|---|---|---|---|---|
| 29th | 50.00 | 50.00 | 53.33 | 50.00 | 50.83 |
| 30th General Category | 83.33 | 63.33 | 80.00 | 56.66 | 70.83 |
| 30th PwD | 50.00 | 73.33 | 63.33 | 50.00 | 59.16 |
| 30th Black candidates | 56.66 | 56.66 | 53.33 | 50.00 | 54.16 |

The difference between the scores of the top-ranked candidates and the cutoff scores shows the high level of competitiveness in the Federal Prosecutor selection process. In the case of the 30[th] Selection Process, the gap between the top scorer in the general category (average score of 93.33) and the cutoff score (average of 70.83 points) was 22.5 points, equivalent to approximately 22.5% of the maximum score.

This margin indicates that, while reaching the cutoff score is challenging, there is significant room for standout performance, particularly in the general category.

Additionally, a considerable improvement can be observed in the performance of the top-ranked candidates between the 29[th] and 30[th] selection processes. While the top scorer in the 29[th] Selection Process achieved 70.83% of the maximum score, the best-performing candidate in the 30[th] Selection Process reached 93.33%. This could reflect both a higher qualification of candidates and a potentially lower degree of difficulty in the most recent examination.

Conversely, the highest score in the 29[th] Selection Process (70.83) corresponds to the lowest score in the 30[th] Exam.

## Analysis of The 29[th] Federal Prosecutor Selection Process, conducted in 2017

| Model | Total Correct | Total Incorrect | % Accuracy (Total) | % Accuracy (Easy Questions) | % Accuracy (Difficult Questions) |
|---|---|---|---|---|---|
| ChatGPT 3.5 Turbo | 45 | 75 | 38% | 41% | 30% |
| ChatGPT 4o | 70 | 50 | 58% | 63% | 50% |
| ChatGPT oi-mini | 58 | 62 | 48% | 55% | 35% |

| Model | | | | | |
|---|---|---|---|---|---|
| **Claude Sonnet 3.5** | 92 | 28 | 77% | 84% | 63% |
| **LLama 3.1 405B** | 75 | 45 | 63% | 70% | 48% |
| **Gemini 1.5 Pro** | 70 | 50 | 58% | 69% | 38% |
| **Command-R-Plus** | 56 | 63 | 47% | 53% | 35% |
| **Mistral Large-2** | 64 | 56 | 53% | 60% | 40% |
| **Amazonia AI** | 70 | 50 | 58% | 65% | 45% |
| **Reka Core 67B** | 56 | 64 | 47% | 49% | 43% |
| **Perplexity AI** | 55 | 65 | 46% | 55% | 28% |
| **Maritaca AI (Sabiá 3)** | 79 | 41 | 66% | 71% | 55% |
| **GROK 2** | 72 | 48 | 60% | 69% | 43% |
| **Nemotron 340B** | 73 | 47 | 61% | 71% | 40% |

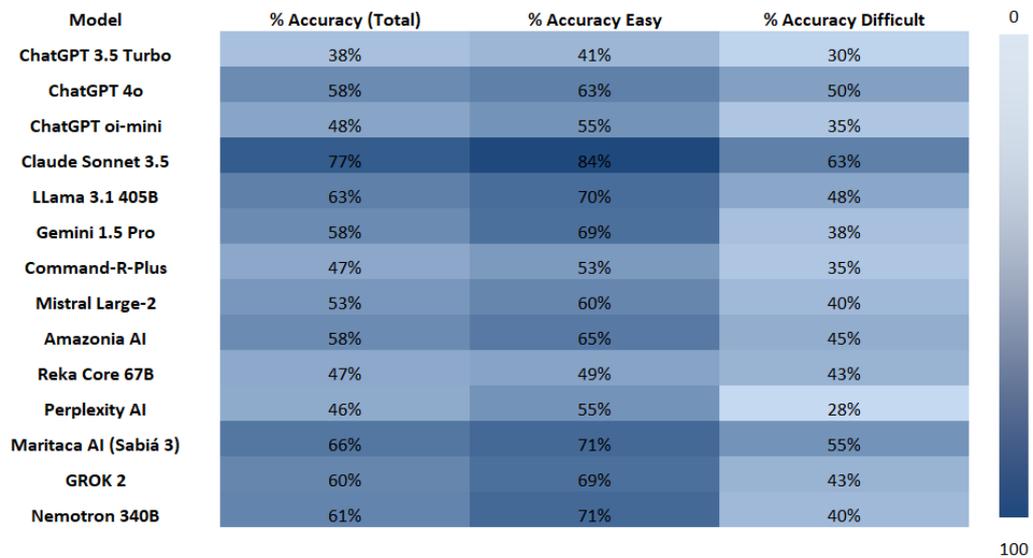

### % Accuracy (Total)

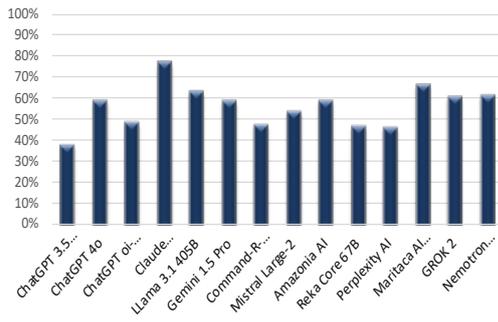

### Philosophy of Law

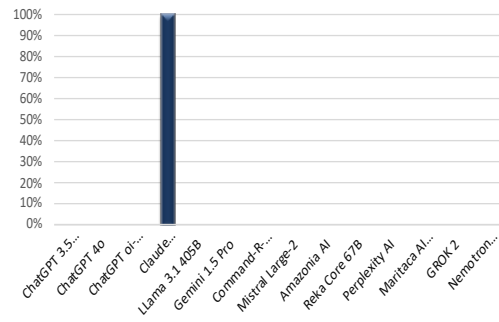

### % Accuracy (Easy Questions)

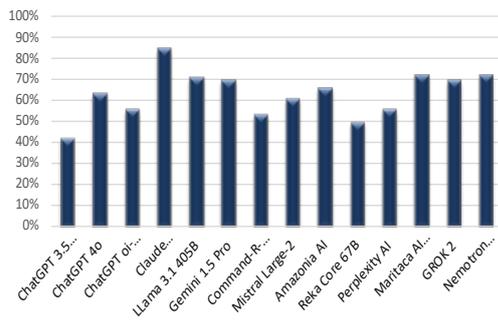

### Human Rights

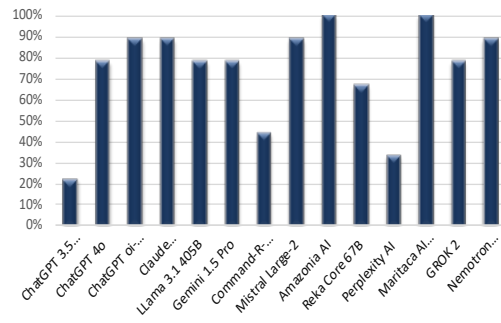

### % Accuracy (Difficult Questions)

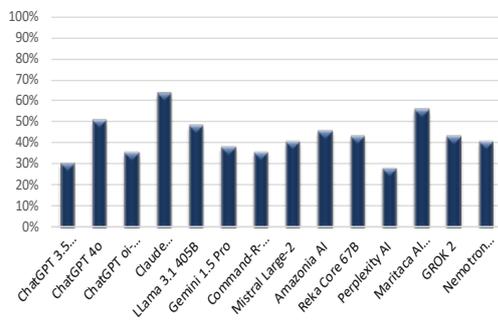

### Electoral Law

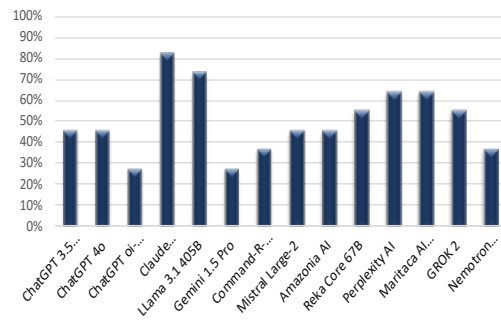

### Constitutional Law

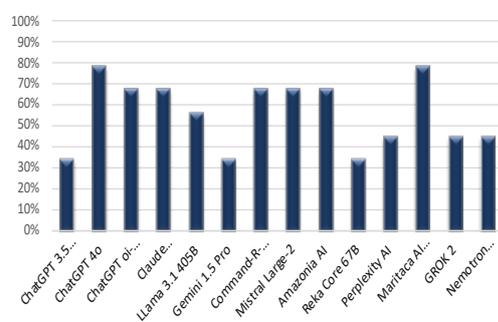

### Administrative Law

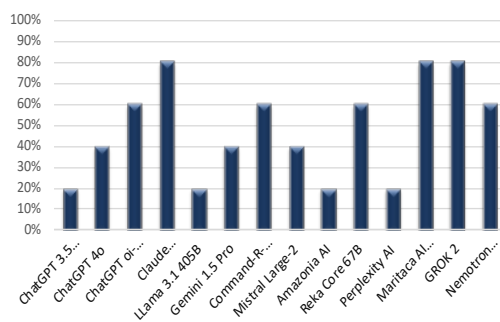

## Environmental Law

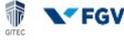

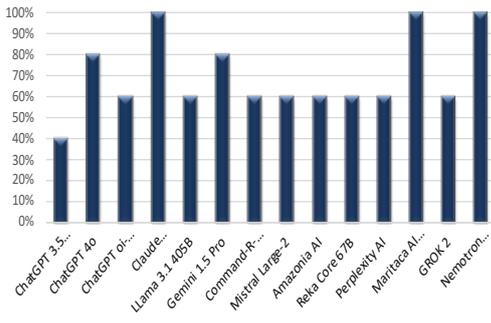

## Economic Law

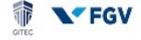

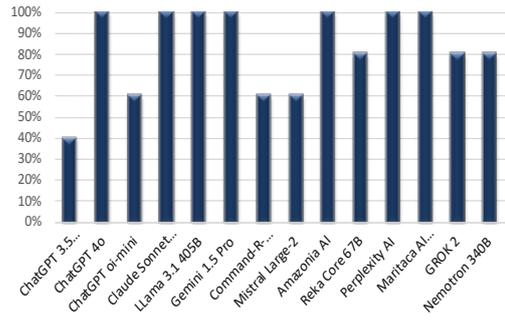

## Tax Law

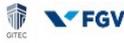

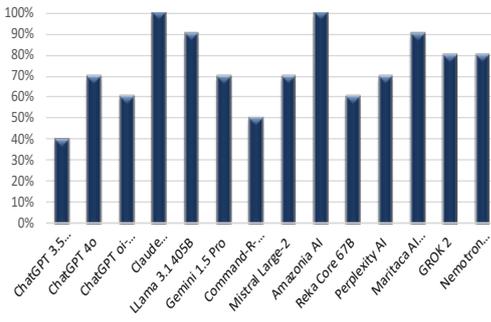

## Consumer Law

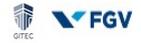

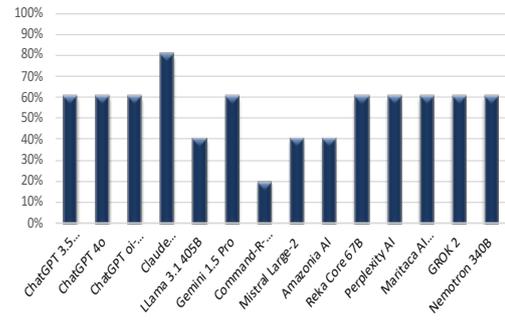

## Public International Law

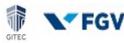

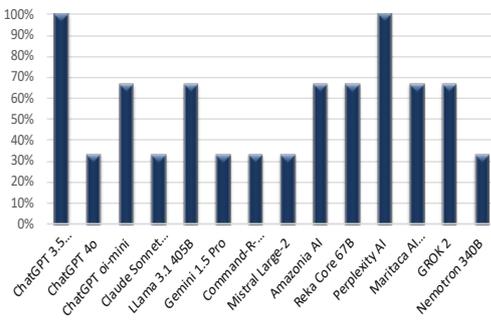

## Civil Law

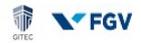

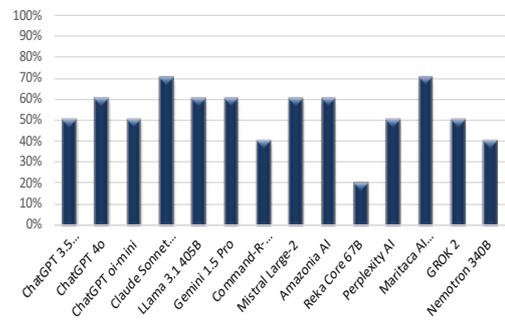

## Private International Law

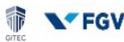

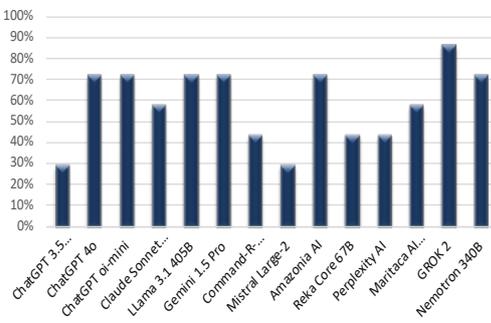

## Civil Procedure Law

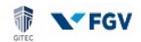

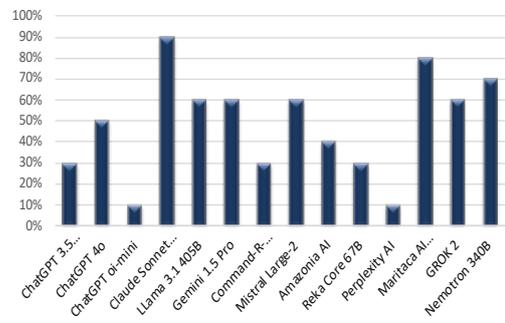

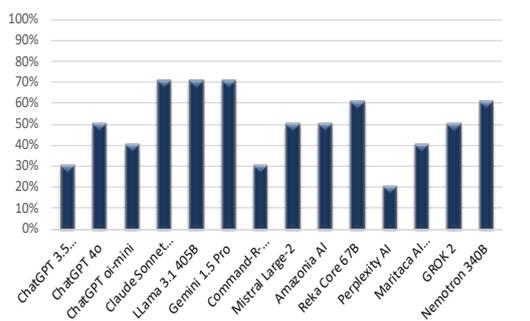

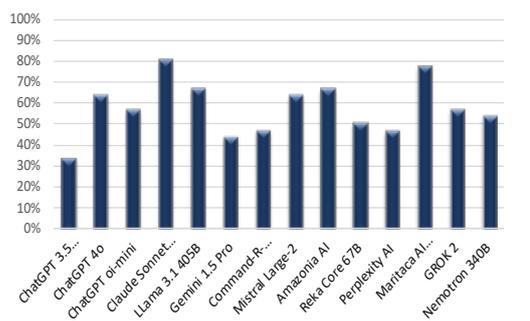

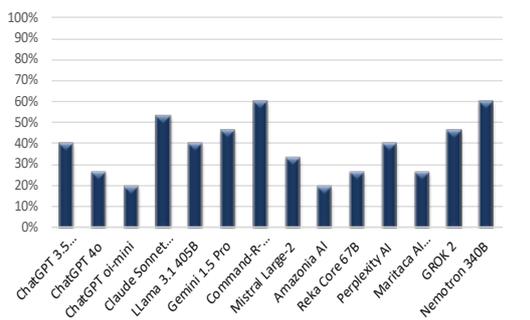

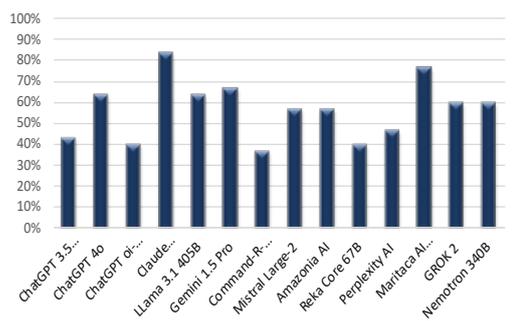

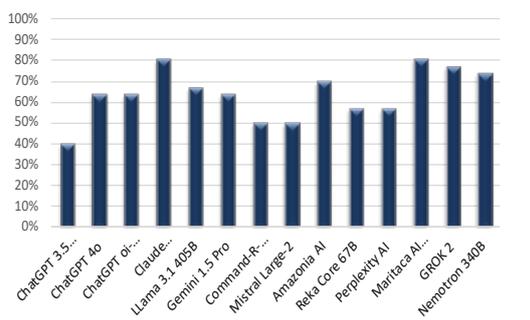

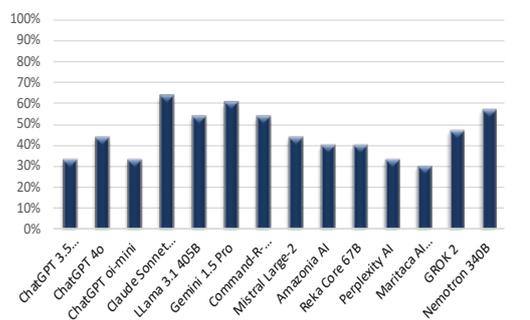

**Analysis of The 30th Federal Prosecutor Selection Process, conducted In 2022**

| Model | Total Correct | Total Incorrect | % Accuracy (Total) | % Accuracy (Easy Questions) | % Accuracy (Difficult Questions) |
|---|---|---|---|---|---|
| **ChatGPT 3.5 Turbo** | 50 | 70 | 42% | 45% | 31% |
| **ChatGPT 4o** | 72 | 48 | 60% | 67% | 38% |
| **ChatGPT oi-mini** | 57 | 63 | 48% | 49% | 41% |
| **Claude Sonnet 3.5** | 92 | 28 | 77% | 81% | 62% |
| **LLama 3.1 405B** | 73 | 47 | 61% | 65% | 48% |
| **Gemini 1.5 Pro** | 88 | 32 | 73% | 78% | 59% |
| **Command-R-Plus** | 39 | 81 | 33% | 34% | 28% |

| Mistral Large-2 | 66 | 54 | 55% | 62% | 34% |
|---|---|---|---|---|---|
| Amazonia AI | 67 | 53 | 56% | 60% | 41% |
| Reka Core 67B | 55 | 65 | 46% | 51% | 31% |
| Perplexity AI | 65 | 55 | 54% | 57% | 45% |
| Maritaca AI (Sabiá 3) | 77 | 43 | 64% | 69% | 48% |
| GROK 2 | 75 | 45 | 63% | 67% | 48% |
| Nemotron 340B | 72 | 48 | 60% | 65% | 45% |

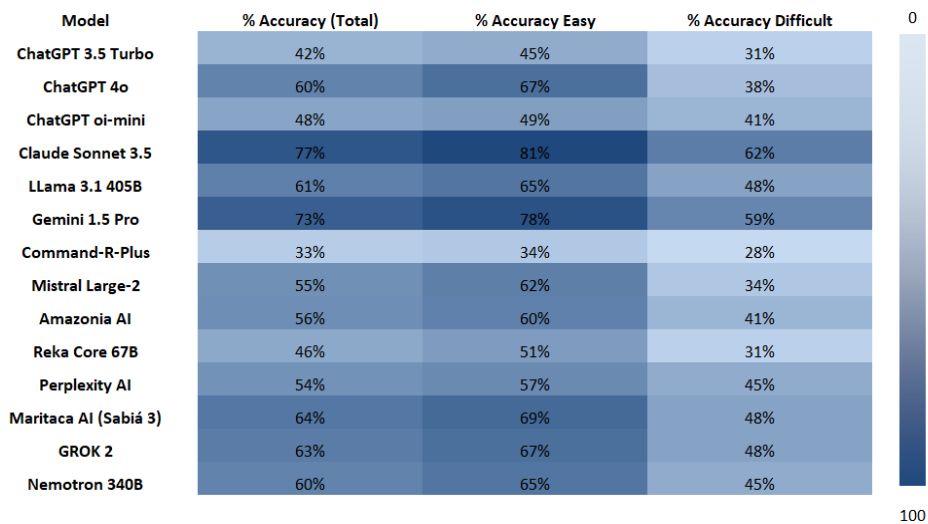

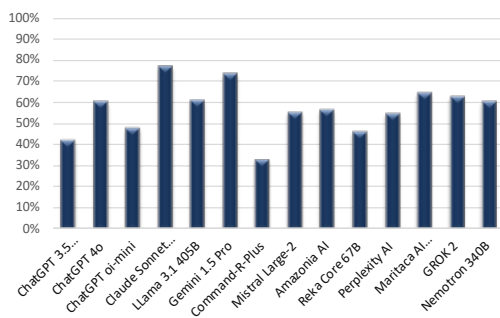

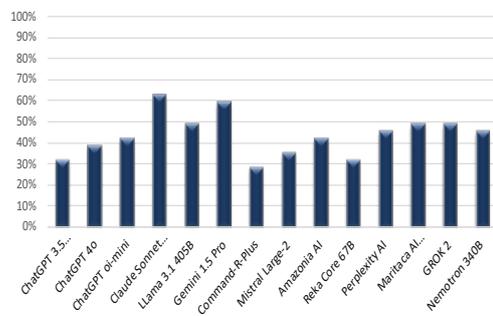

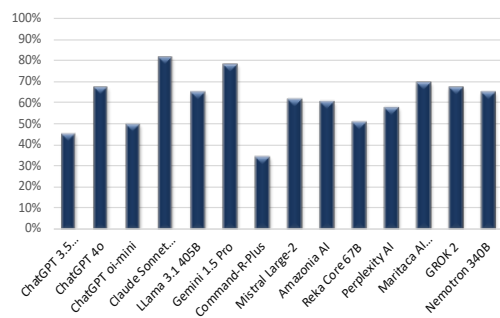

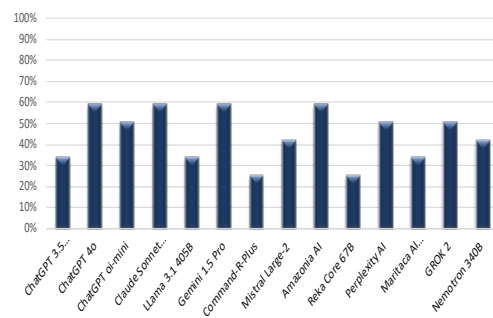

## Human Rights

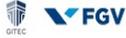
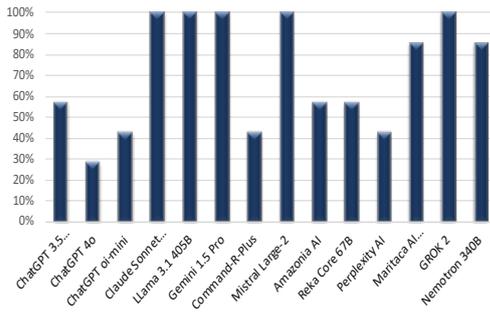

## Tax Law

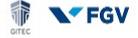
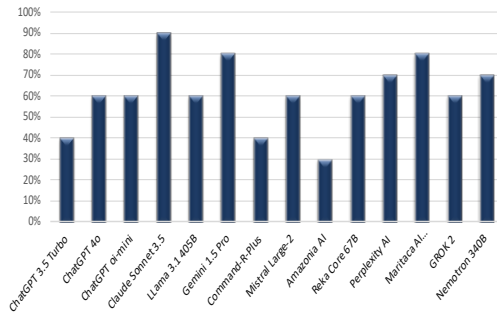

## Electoral Law

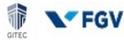
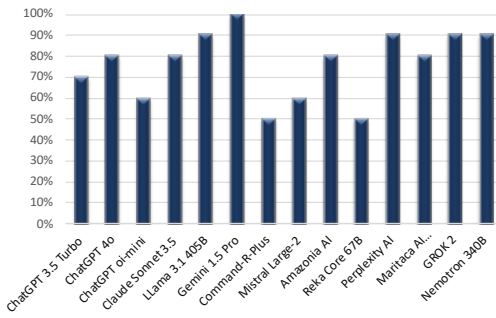

## Public International Law

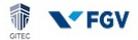
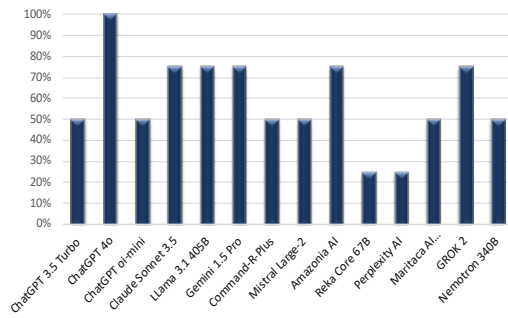

## Administrative Law

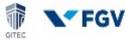
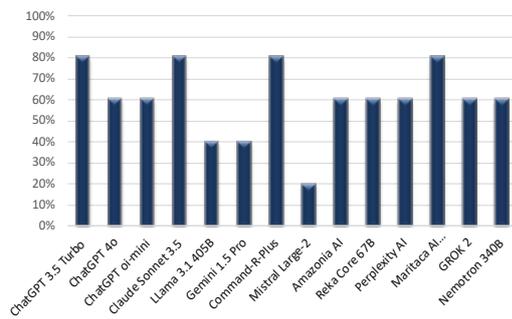

## Private International Law

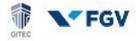
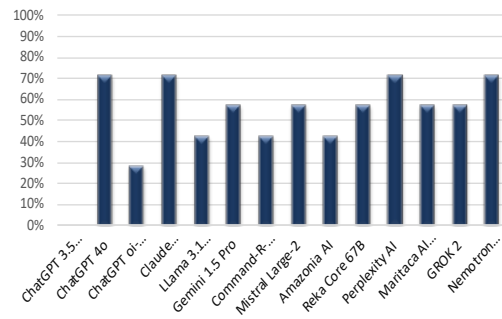

## Environmental Law

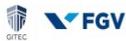
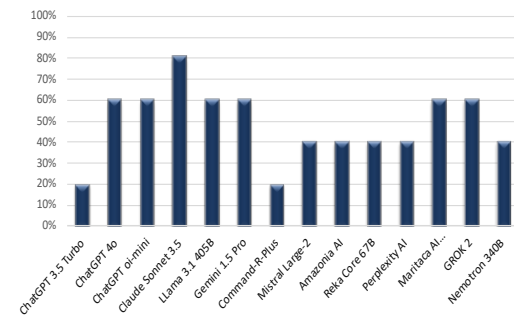

## Economic Law

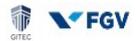
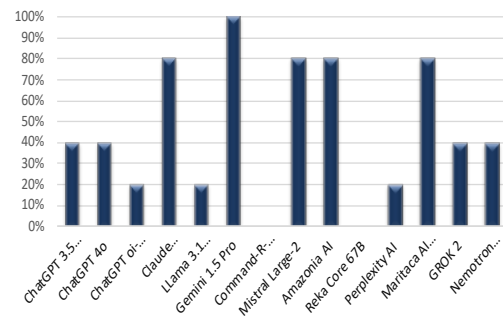

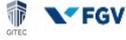

### Consumer law

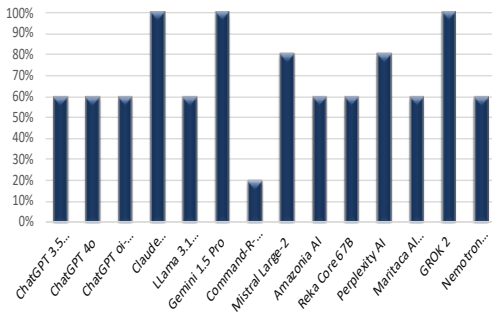

### Criminal Procedure Law

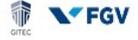
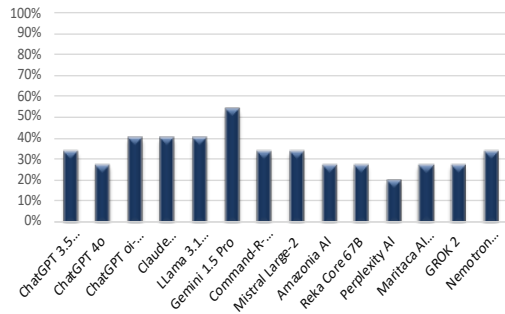

### Civil Law

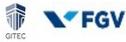
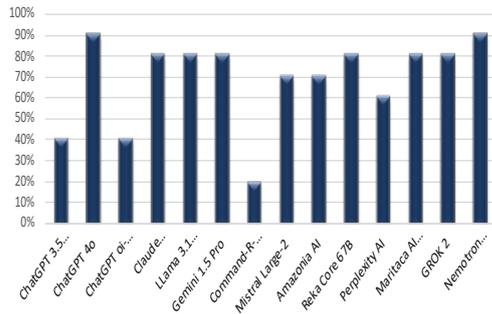

### Group 1

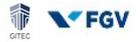
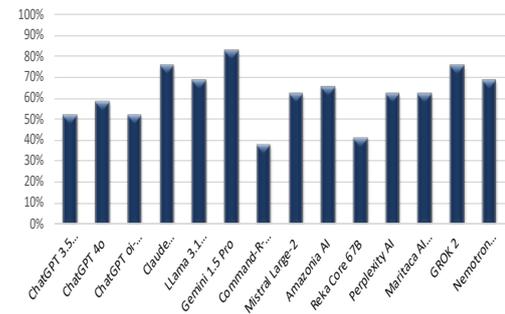

### Civil Procedure Law

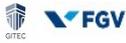
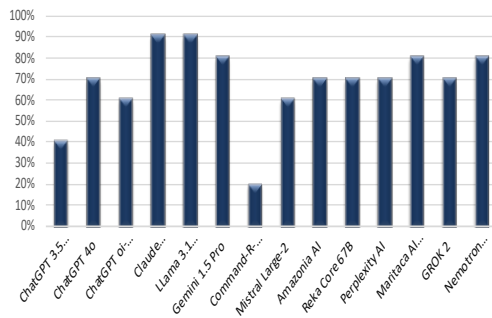

### Group 2

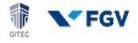
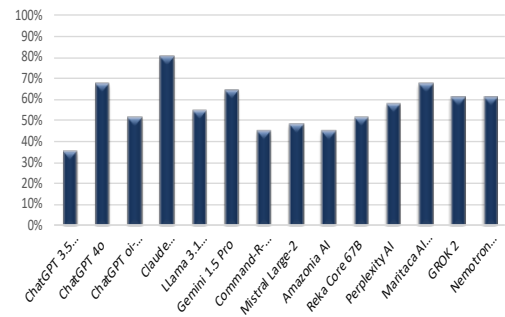

### Criminal Law

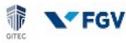
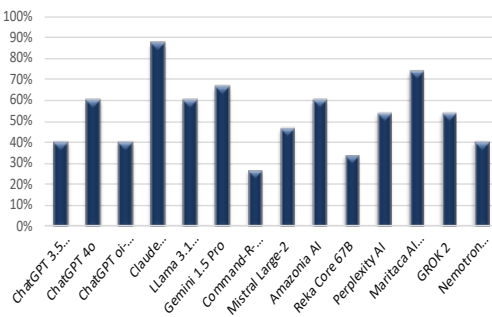

### Group 3

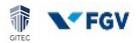
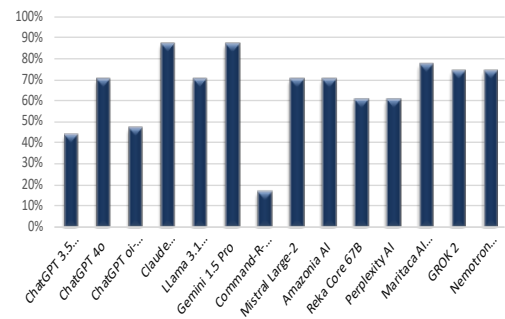

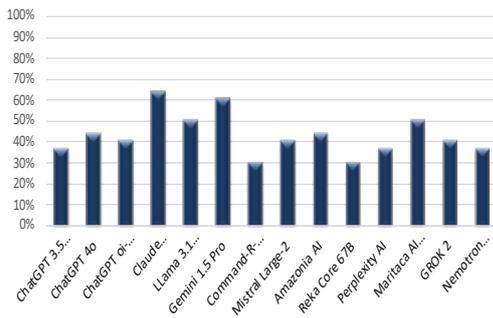

Group 4

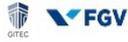

**Comparison between the two Federal Prosecutor Exams**

| Model | 29th Selection Process<br>Total Correct | 30th Selection Process<br>Total Incorrect |
|---|---|---|
| **ChatGPT 3.5 Turbo** | 45 | 50 |
| **ChatGPT 4o** | 70 | 72 |
| **ChatGPT oi-mini** | 58 | 57 |
| **Claude Sonnet 3.5** | 92 | 92 |
| **LLama 3.1 405B** | 75 | 73 |
| **Gemini 1.5 Pro** | 70 | 88 |
| **Command-R-Plus** | 56 | 39 |
| **Mistral Large-2** | 64 | 66 |
| **Amazonia AI** | 70 | 67 |
| **Reka Core 67B** | 56 | 55 |
| **Perplexity AI** | 55 | 65 |
| **Maritaca AI (Sabiá 3)** | 79 | 77 |
| **GROK 2** | 72 | 75 |
| **Nemotron 340B** | 73 | 72 |

## Federal Judge

The public selection process for the position of Federal Judge in the Federal Judiciary of Brazil follows a strict structure, regulated and 'organized by the National School of Judicial Education (Enfam), under the supervision of the National Council of Justice (CNJ), with support from the National School for the Training and Improvement of Labor Judges (Res. 75/2009).

Currently, the Federal Judiciary is divided into six Regional Federal Courts ("TRFs"), as follows:

TRF1: Headquartered in Brasília, it has jurisdiction over the Federal District (DF) and 13 states: Acre (AC), Amapá (AP), Amazonas (AM), Bahia (BA), Goiás (GO), Maranhão (MA), Mato Grosso (MT), Pará (PA), Piauí (PI), Rondônia (RO), Roraima (RR), and Tocantins (TO).

TRF2: Headquartered in Rio de Janeiro, it has jurisdiction over the states of Rio de Janeiro (RJ) and Espírito Santo (ES).

TRF3: Headquartered in São Paulo, it has jurisdiction over the states of São Paulo (SP) and Mato Grosso do Sul (MS).

TRF4: Headquartered in Porto Alegre, it has jurisdiction over the states of Rio Grande do Sul (RS), Santa Catarina (SC), and Paraná (PR).

TRF5: Headquartered in Recife, it has jurisdiction over the states of Ceará (CE), Rio Grande do Norte (RN), Paraíba (PB), Pernambuco (PE), Alagoas (AL), and Sergipe (SE).

TRF6: Headquartered in Belo Horizonte, it has jurisdiction over the state of Minas Gerais (MG).

The public selection process is conducted in the "exams and credentials" modality, as stipulated in Articles 93(I) and 96(I)(c) of the Brazilian Federal Constitution. The examination includes an objective test with a minimum of 50 questions designed to assess reasoning skills, problem-solving ability, and aptitude for the judiciary. The test covers the following areas of knowledge: I – Constitutional Law; II – Administrative Law; III – General Legal Knowledge and Humanistic Education; IV – Human Rights; V – Civil Procedure; VI – Civil Law (6 questions); VII – Business Law; and VIII – Criminal Law.

The selection process is divided into five main stages, as outlined in CNJ Resolution No. 75/2009:

I - First Stage: A multiple-choice exam with both eliminatory and classificatory purposes.

II - Second Stage: Two written exams, also eliminatory and classificatory. During these tests, candidates may consult legislation that contains no annotations or commentary, but consulting doctrinal works, jurisprudence summaries, or interpretative guidance is prohibited.

The first written exam is discursive and consists of questions on any topics from the specific curriculum of the judiciary branch in question.

The second written exam is practical judgment writing, addressing legal topics from the program. It requires I – in Federal and State Courts, the drafting of two judgments, one civil and one criminal, on successive days; II – In Labor Courts, the drafting of one labor judgment; III – In Military Courts, the drafting of a criminal judgment. For the judgment-writing test, if more than one judgment is required, a minimum score of 6 (out of 10) must be achieved on each for approval.

III - Third Stage: An eliminatory phase comprising the following: a) Investigation of the candidate's background and social conduct; b) Physical and mental health examination; and c) Psychotechnical evaluation.

IV - Fourth Stage: An oral examination with both eliminatory and classificatory purposes.

V - Fifth Stage: Evaluation of credentials with a classificatory purpose.

Candidates may only proceed to the next stage after being approved in the preceding one.

The selection process includes a social inclusion policy. The cutoff score for general candidates is a minimum of 70% on the multiple-choice exam. For candidates who self-declare as persons with disabilities, Black, or Indigenous, the cutoff score is reduced to a minimum of 50%.

The ranking of qualified candidates will follow a descending order based on the final average score, with the following weight distribution: I – Multiple-choice exam: weight 1; II - First and second written tests: weight 3 for each test; III - Oral examination: weight 2; IV - Evaluation of credentials: weight 1. Under no circumstances will grades be rounded, with fractions beyond the hundredth place disregarded in the evaluations for each stage of the selection process.

The final average will be calculated using a weighted arithmetic mean, considering the weight assigned to each test, and will be expressed with three decimal places.

The multiple-choice exam will consist of three blocks of questions (I, II, and III), detailed in Annexes I, II, III, IV, V, VI, and VII, according to the respective segment of the Brazilian national Judiciary.

The questions in the multiple-choice exam will be formulated so that the answers necessarily reflect either the dominant doctrinal position or the settled jurisprudence of the Superior Courts. A candidate will be considered qualified in the selective objective test if they achieve a minimum of 30% correct answers in each block and a final average of 60% correct answers based on the algebraic sum of the scores from all three blocks.

Candidates will qualify for the second stage as follows: I - In examinations with up to 1,500 applicants, the 200 highest-scoring candidates after appeals will advance; II - In examinations with more than 1,500 applicants, the 300 highest-scoring candidates after appeals will advance; III - In national examinations or those with more than 10,000 applicants, at the discretion of the court, up to 1,500 highest-scoring candidates after appeals may advance. All candidates tied at the cutoff score for the last qualifying position will also advance to the written tests, even if this exceeds the specified limit of vacancies. This limit does not apply to candidates with disabilities, who will advance to the next phase provided they meet the minimum score required of other candidates. This study analyzed Federal Judge examinations conducted over the past five years. During this period, only three selection processes were held: 1) 2022 by TRF3, 2) 2022 by TRF4, and 3) 2023 by TRF1.

## Analysis of The TRF3 Selection Process conducted In 2022

According to the Opening Notice No. 8154853/2021, this selection process offered 106 positions for Substitute Federal Judges. Of this total, 6 positions were reserved for candidates who declared themselves as people with disabilities during preliminary registration, and 21 positions were reserved for candidates who self-identified as Black (Black or Brown) during preliminary registration.

This selection process was divided into five stages, as outlined in CNJ Resolution 75/2009 previously described.

The multiple-choice exam in the first stage consisted of 100 questions, distributed across three blocks as follows: Block I: 35 questions on Constitutional Law, Social Security Law, Criminal Law, Criminal Procedure, Economic Law, and Consumer Protection; Block II: 35 questions on Civil Law, Civil Procedure, Business Law, and Financial and Tax Law;

Block III: 30 questions on Administrative Law, Environmental Law, Public and Private International Law, and General

Legal Knowledge and Humanistic Education. The test lasted five hours, and any form of consultation was strictly prohibited.

**Cutoff Score**: Candidates were required to achieve a minimum of 30% correct answers in each block and a final score of 60% correct answers across the entire test to qualify.

**Cutoff Scores**

| Type of Position | Available Positions | Cutoff Score | % of Cutoff Score |
|---|---|---|---|
| General Category | ** points | **% | **% |
| Black | 6,0 points | **% | **% |
| PwD | 6,0 points | **% | **% |

*Candidates from the Black and Persons with Disabilities (PwD) categories are approved in the first phase with a minimum score of 6.0, as per legal regulations.
** The data regarding the number of registered candidates and the cutoff scores for each phase and candidate category was requested from the TRF3's respective department. However, no response was received by the conclusion of this research.

**Performance of the Top-Scoring Candidates in the Selection Process**

| Type of Position | Multiple-choice | Final Average | % of Maximum Score |
|---|---|---|---|
| General Category | 9.00 | 8.560 | 85.60% |
| Black | 7.30 | 7.739 | 7.74% |
| PwD | 7.30 | 6.474 | 6.47% |

A total of 113 candidates were approved on the general list, including 14 candidates selected for positions reserved for Black candidates and 3 candidates for positions reserved for persons with disabilities, as per Notice 4, approved on May 21, 2024.

**Graphical Analysis of the Performance of Artificial Intelligence Models Tested in Phase I of the TRF3 Selection Process**

| Model | Total Correct | Total Incorrect | % Accuracy (Total) | % Accuracy (Easy Questions) | % Accuracy (Difficult Questions) |
|---|---|---|---|---|---|
| ChatGPT 3.5 Turbo | 36 | 64 | 36% | 33% | 39% |
| ChatGPT 4o | 76 | 24 | 76% | 78% | 75% |
| ChatGPT oi-mini | 31 | 69 | 31% | 31% | 31% |
| Claude Sonnet 3.5 | 53 | 47 | 53% | 55% | 51% |
| LLama 3.1 405B | 57 | 43 | 57% | 55% | 59% |
| Gemini 1.5 Pro | 55 | 45 | 55% | 53% | 57% |
| Command-R-Plus | 33 | 67 | 33% | 31% | 35% |
| Mistral Large-2 | 49 | 51 | 49% | 35% | 63% |
| Amazonia AI | 45 | 55 | 45% | 47% | 43% |
| Reka Core 67B | 31 | 69 | 31% | 39% | 24% |
| Perplexity AI | 42 | 58 | 42% | 49% | 35% |
| Maritaca AI (Sabiá 3) | 33 | 67 | 33% | 31% | 35% |
| GROK 2 | 22 | 78 | 22% | 29% | 16% |
| Nemotron 340B | 27 | 73 | 27% | 24% | 29% |

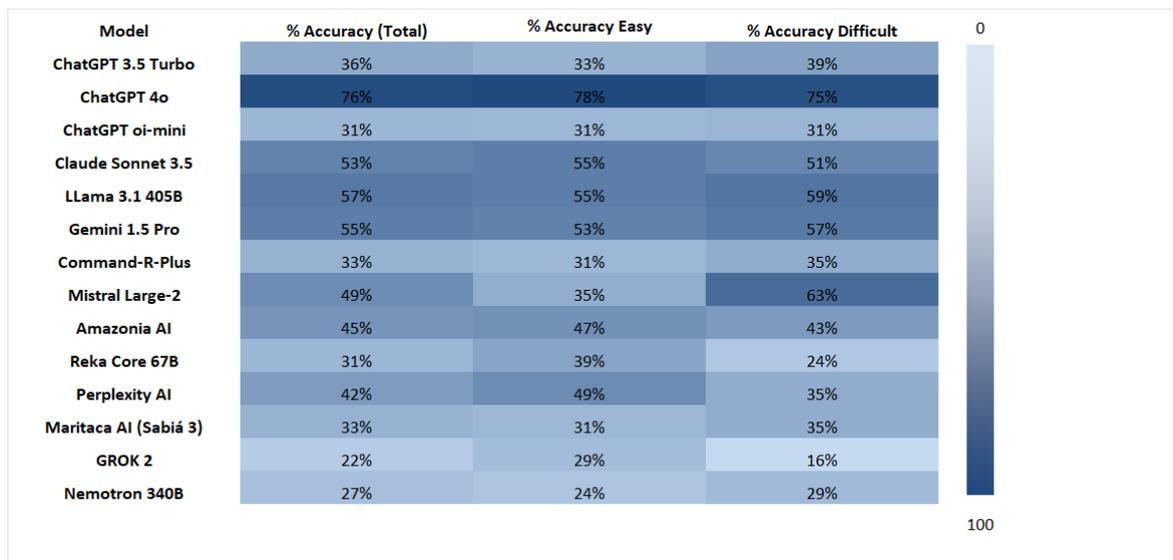

% Accuracy **TRF3 2022** 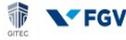

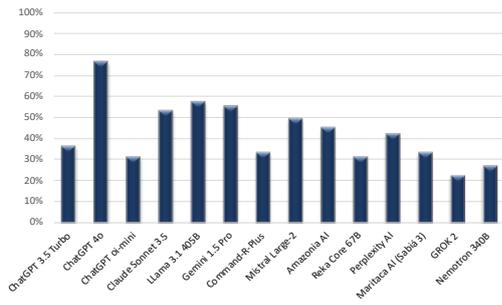

% Accuracy **(Social Security Law)** 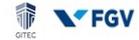

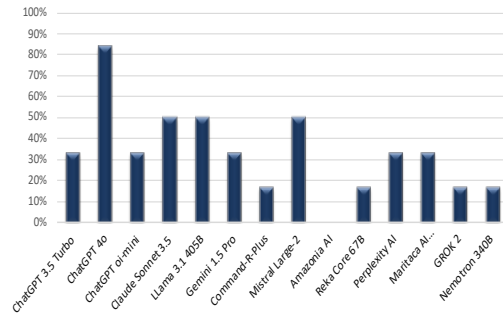

% Accuracy **(Easy Questions)** 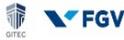

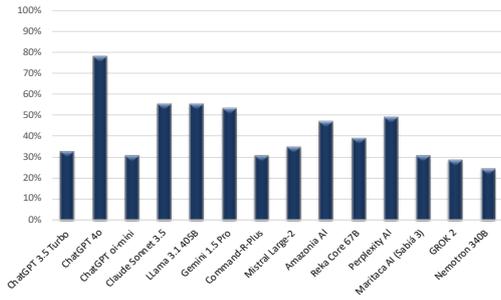

% Accuracy **(Criminal Law)** 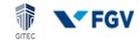

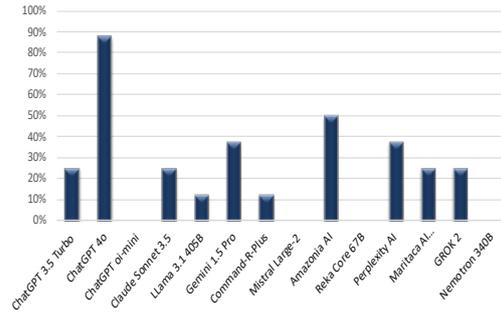

% Accuracy **(Difficult Questions)** 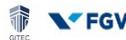

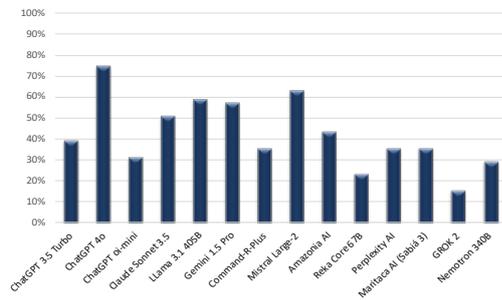

% Accuracy **(Criminal Procedure Law)** 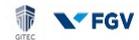

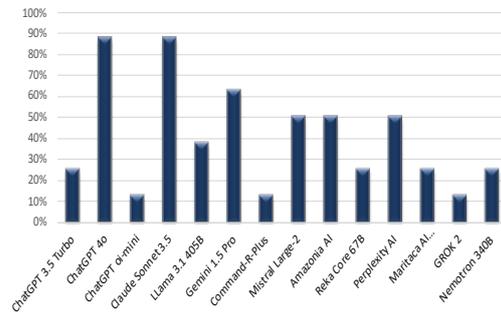

% Accuracy **(Constitutional Law)** 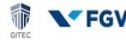

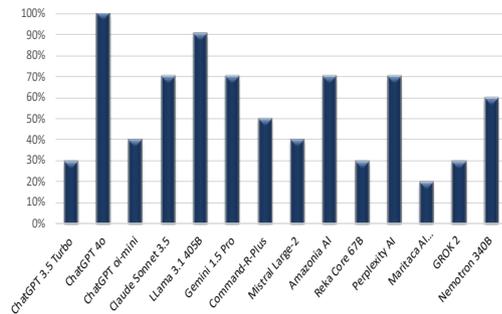

% Accuracy **(Consumer Law)** 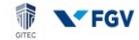

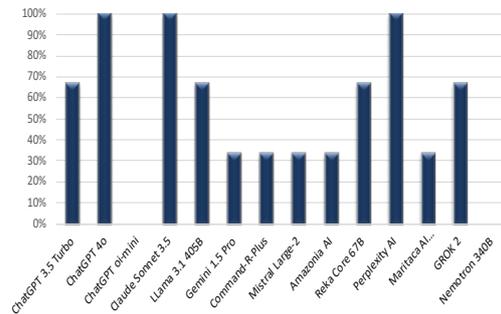

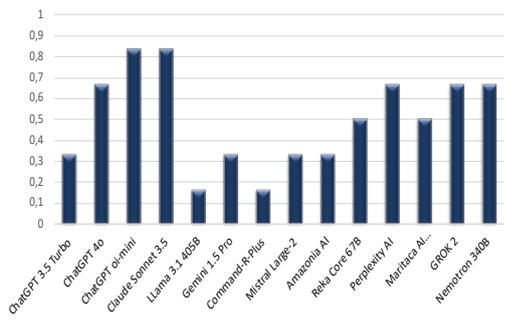

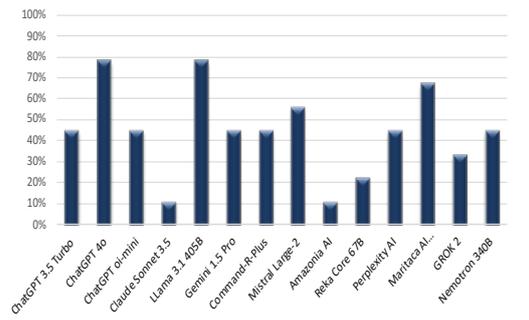

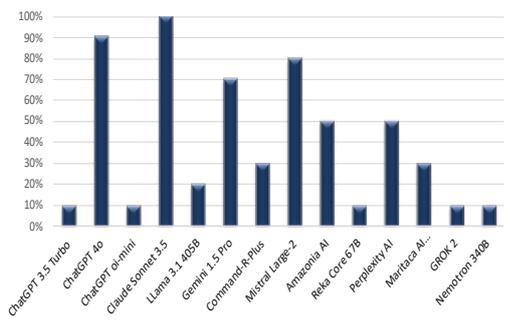

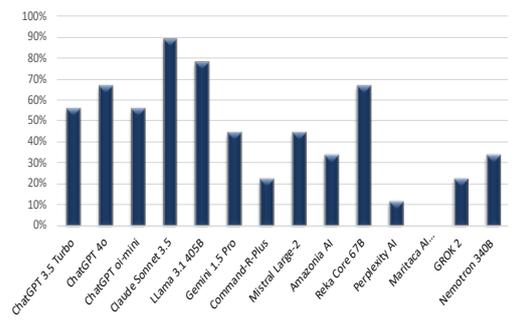

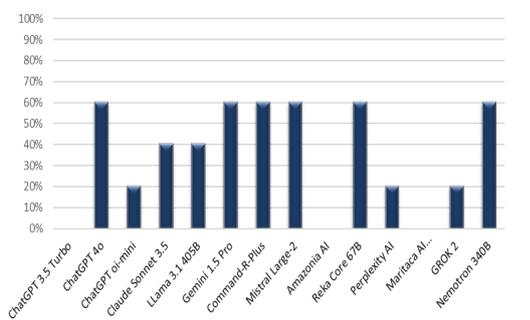

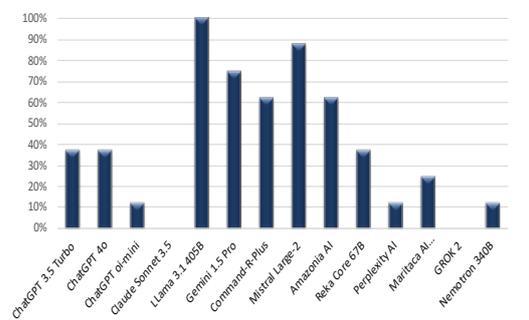

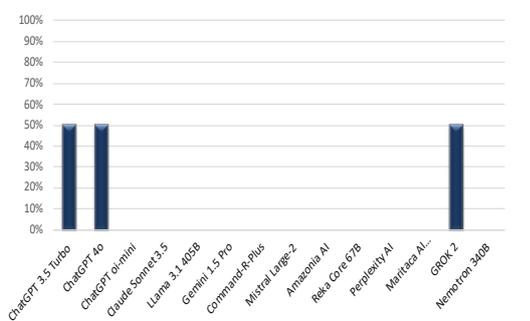

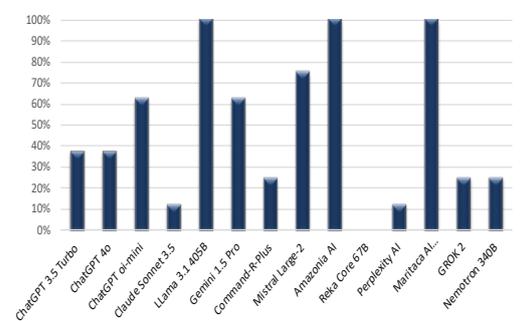

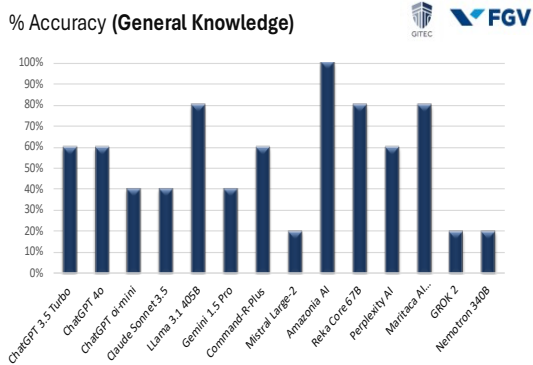

% Accuracy (General Knowledge)

**Analysis of The TRF4 Selection Process conducted in 2022**

According to the Opening Notice, this selection process offered 20 positions for Substitute Federal Judges. Of this total, 1 position was reserved for candidates who declared themselves as people with disabilities during preliminary registration, and 4 positions were reserved for candidates who self-identified as Black (Black or Brown) during preliminary registration. The initial salary is BRL 32,004.65 (thirty-two thousand, four Brazilian reais, and sixty-five centavos).

The multiple-choice exam for the first stage consisted of 100 questions, distributed across three blocks as follows: Block I – 35 questions covering Constitutional Law, Social Security Law, Criminal Law, Criminal Procedure, and Economic Law and Consumer Protection; Block II – 35 questions covering Civil Law, Civil Procedure, Business Law, and Financial and Tax Law; Block III – 30 questions covering Administrative Law, Environmental Law, Public and Private International Law, and General Legal Knowledge and Humanistic Education. The test lasted five hours, and any form of consultation was strictly prohibited. Cutoff Score: A minimum of 30% correct answers was required in each block, along with a final score of 60% correct answers across the total multiple-choice exam.

Registered Candidates: A total of 5,707 candidates were registered, distributed as follows: 1,488 in Rio Grande do Sul (RS), 3,198 in Paraná (PR), and 1,021 in Santa Catarina (SC).

Qualified for the Second Phase: A total of 382 candidates, distributed as follows: 89 in RS, 222 in PR and 71 in SC. Approved Candidates:

A total of 55 candidates were approved, distributed as follows: 17 in RS, 28 in PR,10 in SC, and 1 person with disabilities (PwD).

Candidate-to-Vacancy Ratio: 5707/20 = 285,35

**Cutoff Scores**

| Type of Position | Cutoff Score | % of Cutoff Score |
|---|---|---|
| General Category | 71 points | 71% |
| Black | 60 points | 60% |
| PwD | 60 points | 60% |

*Candidates from the Black and Persons with Disabilities (PwD) categories are approved in the first phase with a minimum score of 6.0, as per legal regulations.

** The data regarding the number of registered candidates and the cutoff scores for each phase and candidate category was requested from the TRF4's respective department. However, no response was received by the conclusion of this research.

**Performance of the Top-Scoring Candidates in the Selection Process**

| General Category | 74.71 points | 74.71 % |
|---|---|---|
| Black | n/c points | % |
| PwD | 65.54 points | 65.54 % |

| Type of Position | Score of the Top-Ranked Candidate | % of Maximum Score |
|---|---|---|
| | | |

**Graphical Analysis of the Performance of Artificial Intelligence Models Tested in Phase I of the TRF4 Selection Process**

| Model | Total Correct | Total Incorrect | % Accuracy (Total) | % Accuracy (Easy Questions) | % Accuracy (Difficult Questions) |
|---|---|---|---|---|---|
| ChatGPT 3.5 Turbo | 37 | 63 | 37% | 35% | 39% |
| ChatGPT 4o | 67 | 33 | 67% | 68% | 64% |
| ChatGPT oi-mini | 35 | 65 | 35% | 35% | 33% |
| Claude Sonnet 3.5 | 76 | 24 | 76% | 82% | 64% |
| LLama 3.1 405B | 57 | 43 | 57% | 59% | 52% |
| Gemini 1.5 Pro | 60 | 40 | 60% | 65% | 48% |
| Command-R-Plus | 37 | 63 | 37% | 39% | 30% |
| Mistral Large-2 | 52 | 48 | 52% | 55% | 45% |
| Amazonia AI | 47 | 53 | 47% | 44% | 55% |
| Reka Core 67B | 29 | 71 | 29% | 26% | 33% |
| Perplexity AI | 31 | 69 | 31% | 27% | 39% |
| Maritaca AI (Sabiá 3) | 46 | 54 | 46% | 50% | 36% |
| GROK 2 | 35 | 65 | 35% | 41% | 21% |
| Nemotron 340B | 27 | 73 | 27% | 23% | 36% |

| Model | % Accuracy Total | % Accuracy Easy | % Accuracy Difficult |
|---|---|---|---|
| ChatGPT 3.5 Turbo | 37% | 35% | 39% |
| ChatGPT 4o | 67% | 68% | 64% |
| ChatGPT oi-mini | 35% | 35% | 33% |
| Claude Sonnet 3.5 | 76% | 82% | 64% |
| LLama 3.1 405B | 57% | 59% | 52% |
| Gemini 1.5 Pro | 60% | 65% | 48% |
| Command-R-Plus | 37% | 39% | 30% |
| Mistral Large-2 | 52% | 55% | 45% |
| Amazonia AI | 47% | 44% | 55% |
| Reka Core 67B | 29% | 26% | 33% |
| Perplexity AI | 31% | 27% | 39% |
| Maritaca AI (Sabiá 3) | 46% | 50% | 36% |
| GROK 2 | 35% | 41% | 21% |
| Nemotron 340B | 27% | 23% | 36% |

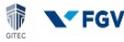

% Accuracy **TRF4 2022**

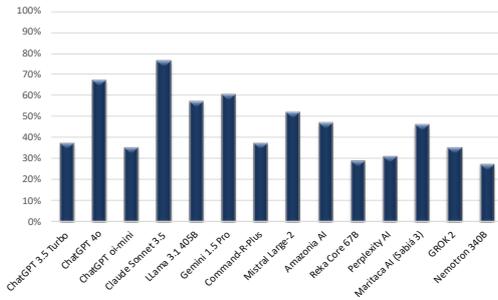

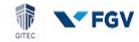

% Accuracy (**Constitutional Law**)

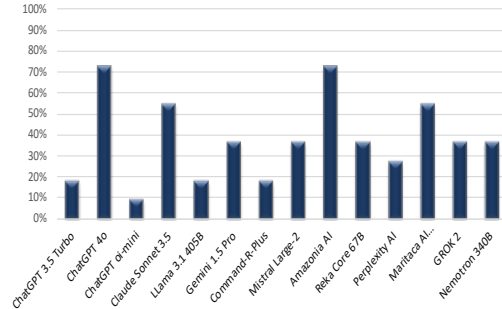

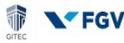

% Accuracy (**Easy Questions**)

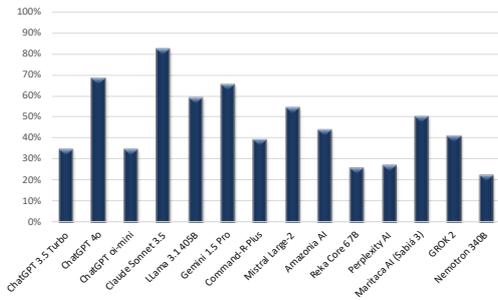

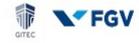

% Accuracy (**Social Security Law**)

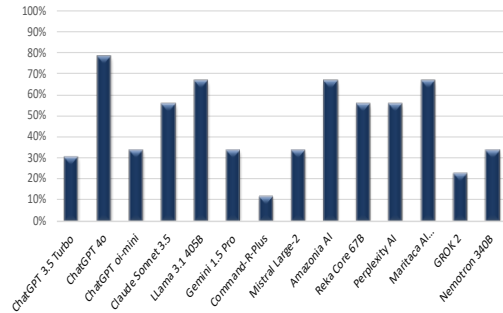

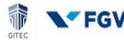

% Accuracy (**Difficult Questions**)

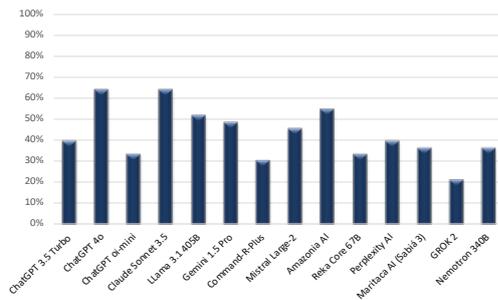

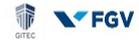

% Accuracy (**Economic Law**)

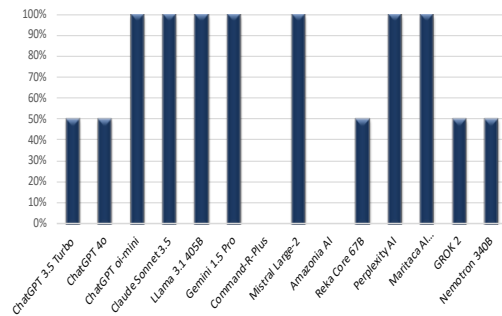

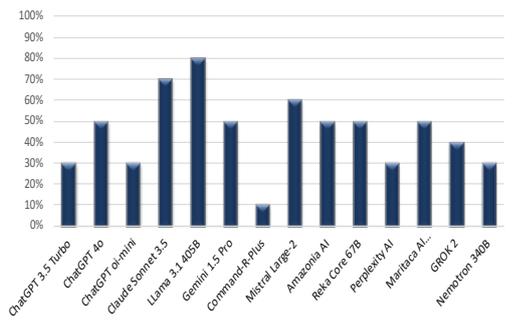

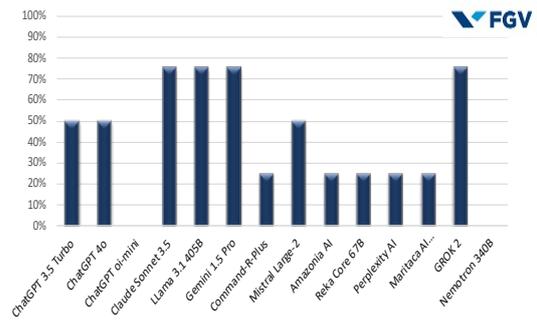

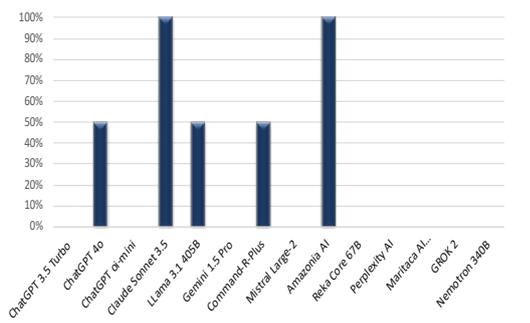

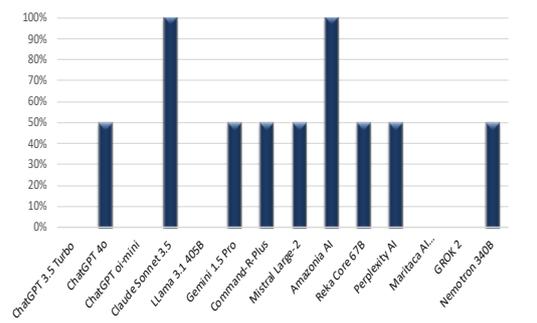

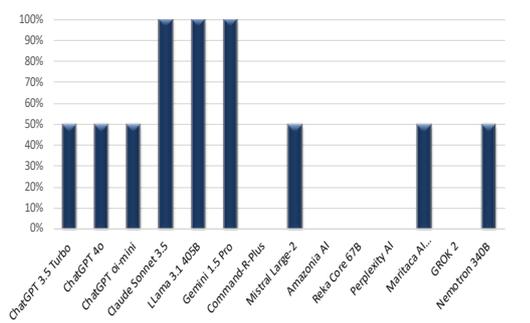

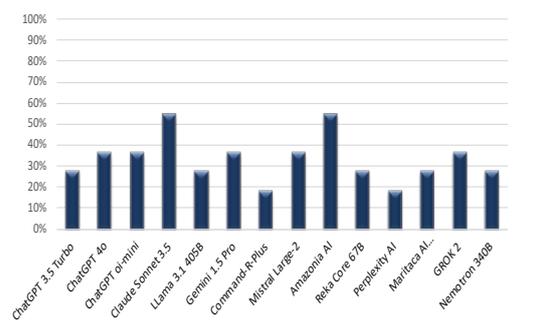

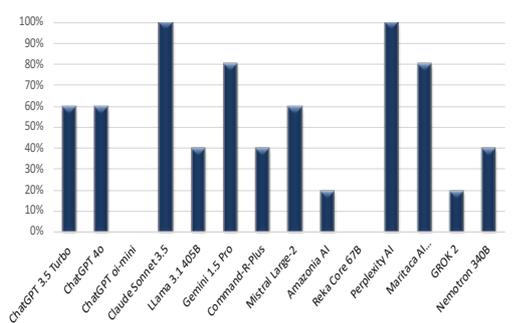

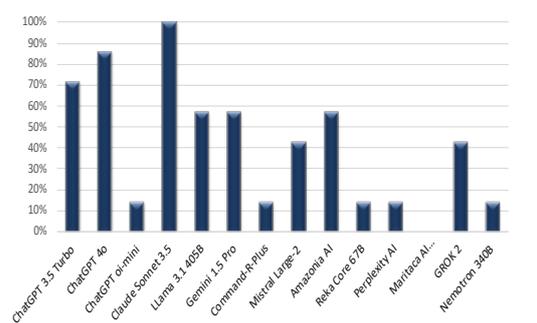

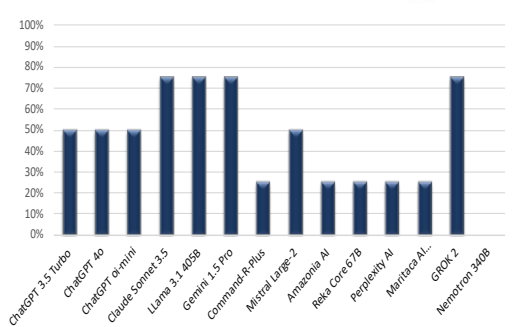

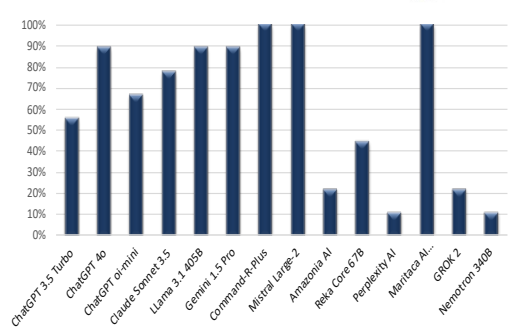

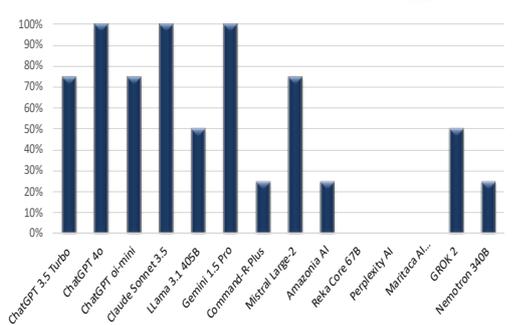

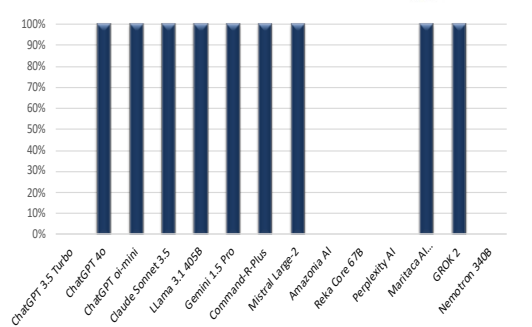

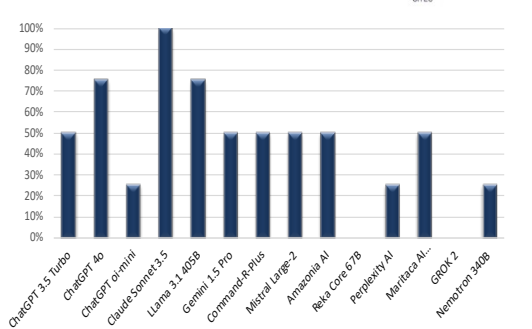

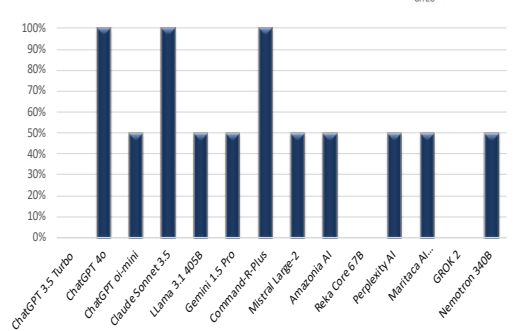

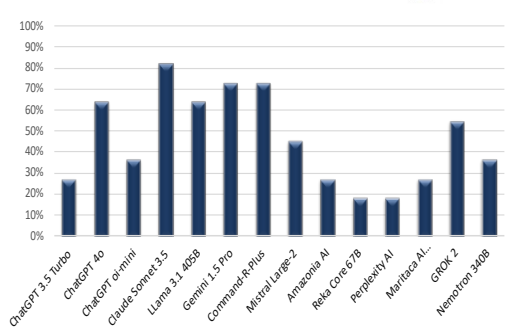

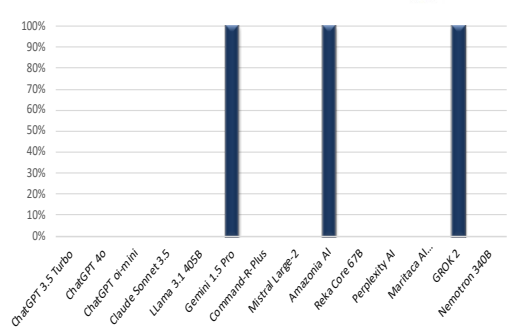

**Analysis of the TRF1 Selection Process conducted in 2023**

According to Opening Notice No. 1/2023, the 17th Public Selection Process for Substitute Federal Judges of the TRF1 offered 1 position for Substitute Federal

Judge, in addition to any positions that may become available. Of the total positions, 5% were reserved for candidates who declared themselves as persons with disabilities during preliminary registration, and 20% for candidates who self-identified as Black during preliminary registration. The initial salary is BRL 32,004.65 (thirty-two thousand, four Brazilian reais, and sixty-five centavos).

This selection process was divided into 5 (five) stages, in accordance with CNJ Resolution 75/2009 described above:

I – multiple choice test;

II – written tests;

a) discursive test;

b) Civil Sentencing practice test;

c) Criminal Sentencing practice test;

III – oral tests

The multiple-choice exam in the first stage consisted of 100 questions, distributed across three blocks as follows: Block I – 35 questions on Constitutional Law, Social Security Law, Criminal Law, Criminal Procedure, and Economic Law and Consumer Protection; Block II – 35 questions on Civil Law, Civil Procedure, Business Law, and Financial and Tax Law; Block III – 30 questions on Administrative Law, Environmental Law, Public and Private International Law, and General Legal Knowledge and Humanistic Education. The exam lasted five hours, and any form of consultation was strictly prohibited.

**Cutoff Score:** Candidates were required to achieve a minimum of 30% correct answers in each block and a final score of 60%

correct answers across the entire multiple-choice exam.

Applicants: A total of 6,916 candidates registered for the exam. Out of these, 58 candidates were approved, including 17 women, 41 men, 1 person with a disability, and 7 self-declared Black candidates. (Res. n. 193/2022; 2022)

Candidate-to-Vacancy Ratio: 6916/1 = 6.916 or 6916/58 = 119,24

**Cutoff Scores:**

| Type of Position | Points | Cutoff Score | % of Cutoff Score |
|---|---|---|---|
| General Category | 6,9 points | 69% | 69% |
| Black | 6,0 points* | 60% | 60% |
| PwD | 6,0 points* | 60% | 60% |

*Candidates from the Black and Persons with Disabilities (PwD) categories are approved in the first phase with a minimum score of 60, as per legal regulations.

**Performance of the Top-Scoring Candidates in the Selection Process**

| Type of Position | Multiple-choice Exam | % of Maximum Score |
|---|---|---|
| General Category | 5,802 | 58,02% |
| Black | 6,119 | 61,19% |
| PwD | 6,272 | 62,72% |

**Graphical Analysis of the Performance of Artificial Intelligence Models Tested in Phase I of the TRF1 Selection Process**

| Model | Total Correct | Total Incorrect | % Accuracy (Total) | % Accuracy (Easy Questions) | % Accuracy (Difficult Questions) |
|---|---|---|---|---|---|
| **ChatGPT 3.5 Turbo** | 39 | 61 | 39% | 37% | 43% |
| **ChatGPT 4o** | 63 | 37 | 63% | 64% | 60% |
| **ChatGPT oi-mini** | 37 | 63 | 37% | 39% | 33% |
| **Claude Sonnet 3.5** | 67 | 33 | 67% | 67% | 67% |
| **LLama 3.1 405B** | 57 | 43 | 57% | 59% | 53% |
| **Gemini 1.5 Pro** | 58 | 42 | 58% | 59% | 57% |
| **Command-R-Plus** | 29 | 71 | 29% | 26% | 37% |
| **Mistral Large-2** | 51 | 49 | 51% | 46% | 63% |
| **Amazonia AI** | 55 | 45 | 55% | 51% | 63% |
| **Reka Core 67B** | 46 | 54 | 46% | 46% | 47% |
| **Perplexity AI** | 49 | 51 | 49% | 47% | 53% |
| **Maritaca AI (Sabiá 3)** | 59 | 41 | 59% | 57% | 63% |
| **GROK 2** | 53 | 47 | 53% | 49% | 63% |
| **Nemotron 340B** | 58 | 42 | 58% | 56% | 63% |

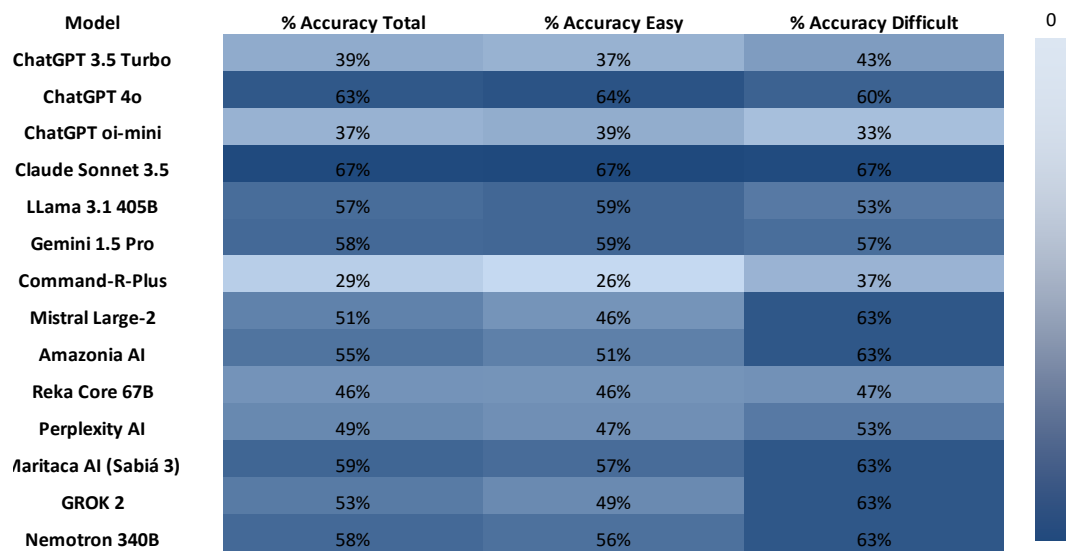

% Accuracy **TRF1 2023 Exam**

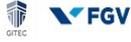
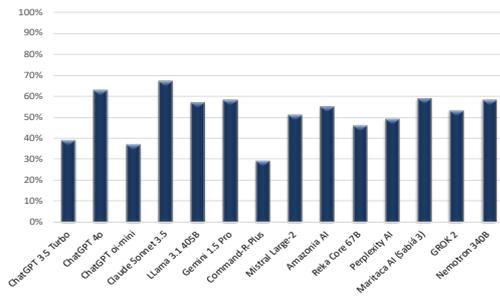

% Accuracy (Environmental Law)

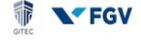
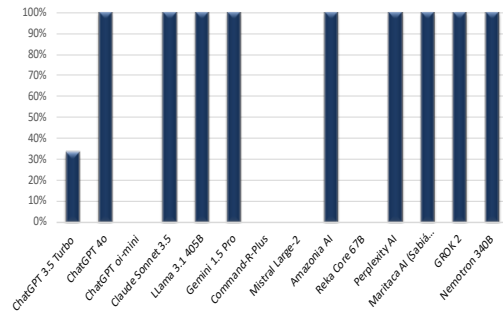

% Accuracy (Easy Questions)

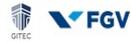
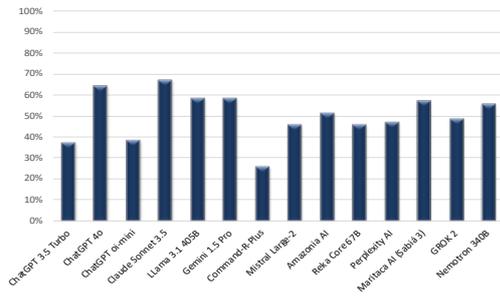

% Accuracy (Civil Procedural Law)

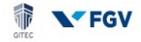
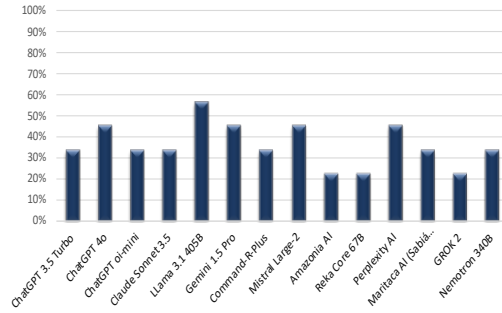

% Accuracy (Difficult Questions)

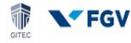
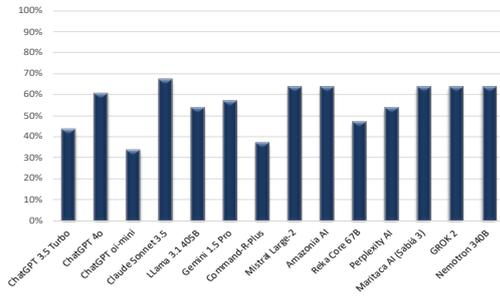

% Accuracy (Criminal Law)

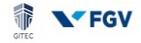
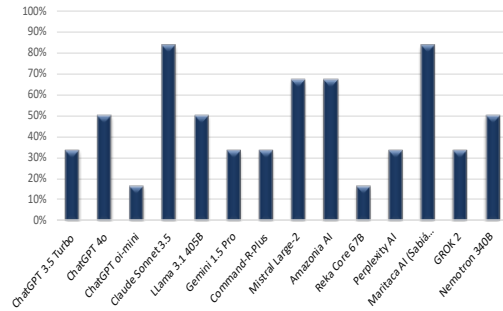

% Accuracy (Constitutional Law )

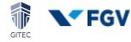
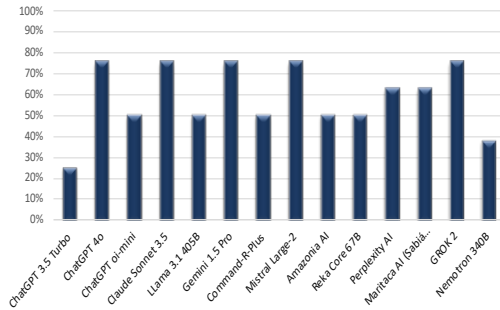

% Accuracy (Current Affairs)

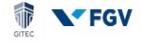
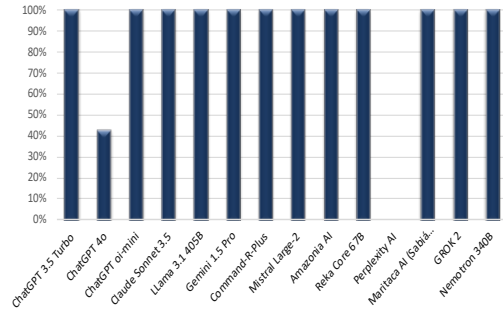

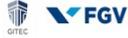

% Accuracy (**Criminal Procedural Law** )

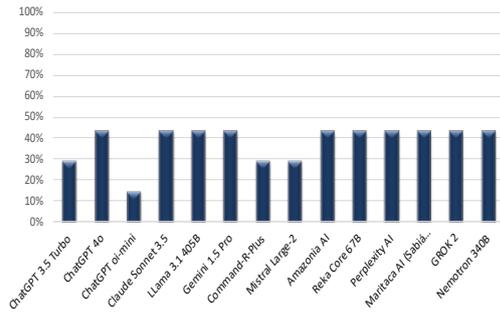

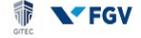

% Accuracy (**Digital Law**)

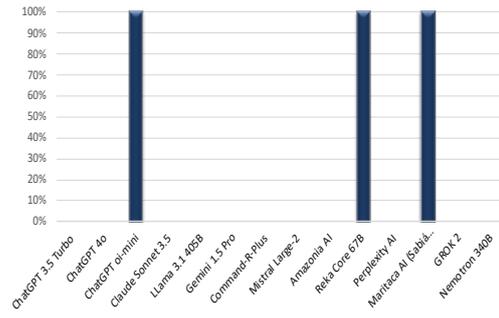

% Accuracy (**Economic Law**)

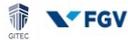

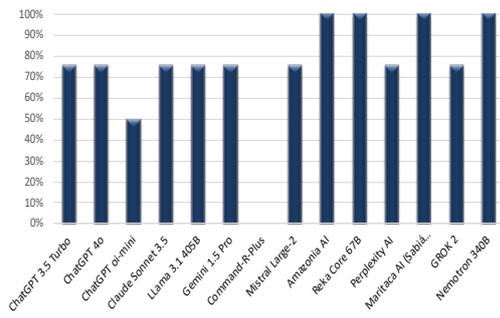

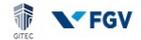

% Accuracy (**Business Law**)

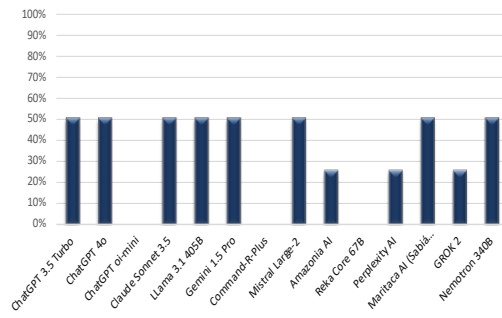

% Accuracy (**Consumer Law**)

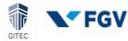

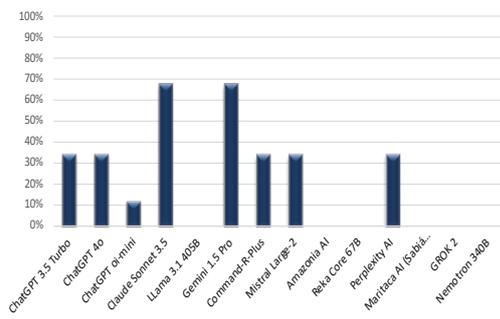

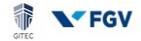

% Accuracy (**Tax Law**)

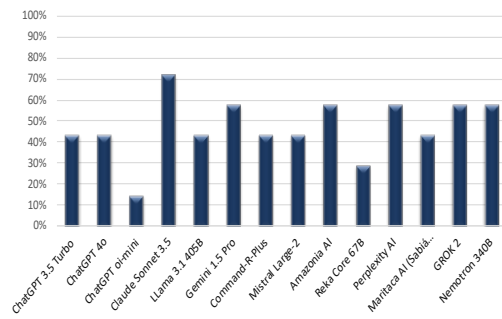

% Accuracy (**Civil Law**)

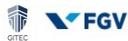

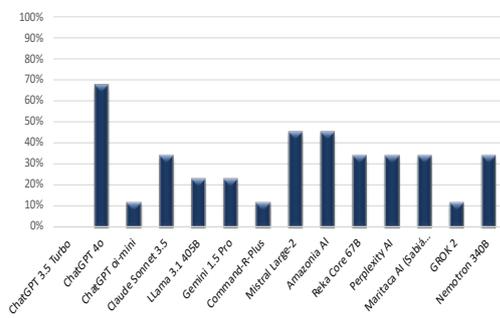

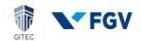

% Accuracy (**Financial Law**)

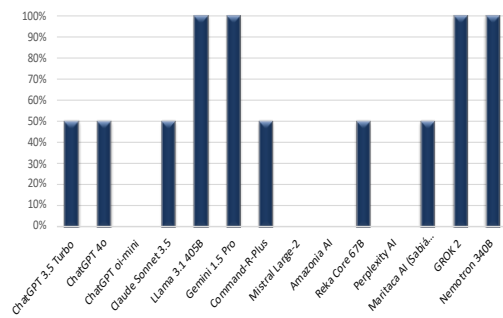

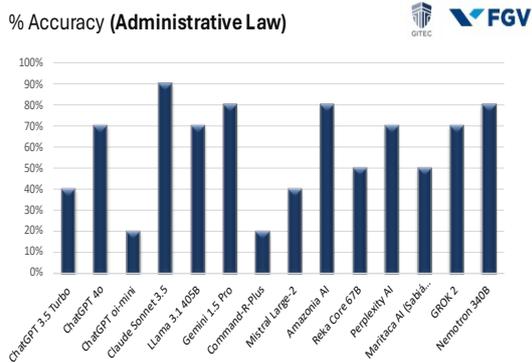

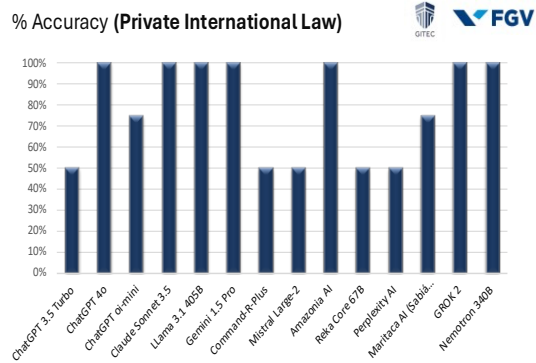

## Comparison Between the Three Selection Processes – TRFs 1, 3, And 4

|  | TRF-3 2022 | TRF-4 2022 | TRF-1 2023 |
|---|---|---|---|
| **Model** | **Total Correct** | **Total Correct** | **Total Correct** |
| **ChatGPT 3.5 Turbo** | 36 | 37 | 39 |
| **ChatGPT 4o** | 76 | 67 | 63 |
| **ChatGPT oi-mini** | 31 | 35 | 37 |
| **Claude Sonnet 3.5** | 53 | 76 | 67 |
| **LLama 3.1 405B** | 57 | 57 | 57 |
| **Gemini 1.5 Pro** | 55 | 60 | 58 |
| **Command-R-Plus** | 33 | 37 | 29 |
| **Mistral Large-2** | 49 | 52 | 51 |
| **Amazonia AI** | 45 | 47 | 55 |
| **Reka Core 67B** | 31 | 29 | 46 |
| **Perplexity AI** | 42 | 31 | 49 |
| **Maritaca AI (Sabiá 3)** | 33 | 46 | 59 |
| **GROK 2** | 22 | 35 | 53 |
| **Nemotron 340B** | 27 | 27 | 58 |

## National Judiciary Exam (ENAM)

The public selection process known as the National Judiciary Exam (ENAM) is conducted to offer positions across various judicial careers, including Regional Federal Courts, Military Courts, and State and Federal District Courts.

The primary objective of this exam is to certify candidates for eligibility to compete for judicial career positions, without specifying a fixed number of vacancies. The certification of qualification is valid for two years and can be automatically renewed once, for an additional two-year period, starting from the date of issuance.

This means that the ENAM serves as an eliminatory, not a classificatory, process. Candidates will be considered qualified if they achieve a score equal to or greater than 70% of correct answers. For candidates who self-identify as Black or Indigenous, the qualification threshold is lowered to 50% of correct answers.

The multiple-choice exam consists of 80 questions, designed to present scenarios requiring resolution by a judge, covering

the following areas of law (ENFAM, 2024a; ENFAM, 2024b):

- Constitutional Law – 16 questions
- Administrative Law – 10 questions
- General Legal Knowledge and Humanistic Education – 6 questions
- Human Rights – 6 questions
- Civil Procedure – 12 questions
- Business Law – 6 questions
- Criminal Law – 12 questions

The Constitutional Law block may include questions related to labor constitutional law, tax law, and criminal procedure.

The key distinction of the ENAM compared to other judiciary career selection processes is its lack of a classificatory nature, functioning instead as a talent pool for vacancies that arise in federal, state, and military judiciary careers.

Both ENAM exams were prepared by the Getulio Vargas Foundation and consisted of 80 multiple-choice questions, each with 5 answer options, of which only one is correct. Candidates were given 5 hours to complete the exam (ENFAM, 2024a; ENFAM, 2024b).

Although it is a national exam, candidates are required to specify the judicial position they are applying for at the time of registration.

## Graphical Analysis of the Performance of Artificial Intelligence Models Tested in ENAM – 1

Considering the cutoff score required for the issuance of the certification of qualification, none of the artificial intelligence models tested in the 1st edition of ENAM were successful. The closest performance was achieved by Claude Sonnet 3.5, with an accuracy rate of 61%, followed by ChatGPT 4o and Amazônia AI, both with 56% accuracy:

| Model | Total Correct | Total Incorrect | % Accuracy (Total) | % Accuracy (Easy Questions) | % Accuracy (Difficult Questions) |
|---|---|---|---|---|---|
| ChatGPT 3.5 Turbo | 33 | 47 | 41% | 52% | 6% |
| ChatGPT 4o | 46 | 34 | 58% | 58% | 56% |
| ChatGPT oi-mini | 30 | 50 | 38% | 42% | 22% |
| Claude Sonnet 3.5 | 61 | 19 | 76% | 81% | 61% |
| LLama 3.1 405B | 41 | 39 | 51% | 56% | 33% |
| Gemini 1.5 Pro | 48 | 32 | 60% | 66% | 39% |
| Command-R-Plus | 34 | 46 | 43% | 48% | 22% |
| Mistral Large-2 | 33 | 47 | 41% | 44% | 33% |
| Amazonia AI | 44 | 36 | 55% | 55% | 56% |
| Reka Core 67B | 35 | 45 | 44% | 48% | 28% |
| Perplexity AI | 37 | 43 | 46% | 50% | 33% |
| Maritaca AI (Sabiá 3) | 51 | 29 | 64% | 69% | 44% |
| **GROK 2** | 45 | 35 | 56% | 63% | 33% |
| **Nemotron 340B** | 43 | 37 | 54% | 55% | 50% |

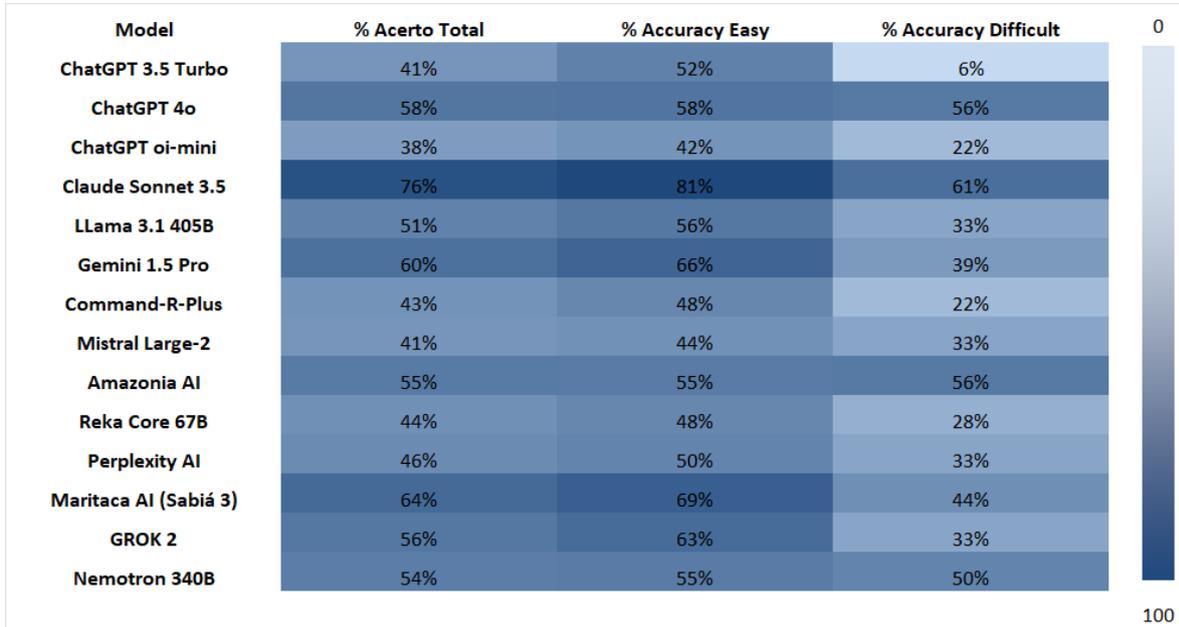

| Model | % Acerto Total | % Accuracy Easy | % Accuracy Difficult |
|---|---|---|---|
| ChatGPT 3.5 Turbo | 41% | 52% | 6% |
| ChatGPT 4o | 58% | 58% | 56% |
| ChatGPT oi-mini | 38% | 42% | 22% |
| Claude Sonnet 3.5 | 76% | 81% | 61% |
| LLama 3.1 405B | 51% | 56% | 33% |
| Gemini 1.5 Pro | 60% | 66% | 39% |
| Command-R-Plus | 43% | 48% | 22% |
| Mistral Large-2 | 41% | 44% | 33% |
| Amazonia AI | 55% | 55% | 56% |
| Reka Core 67B | 44% | 48% | 28% |
| Perplexity AI | 46% | 50% | 33% |
| Maritaca AI (Sabiá 3) | 64% | 69% | 44% |
| GROK 2 | 56% | 63% | 33% |
| Nemotron 340B | 54% | 55% | 50% |

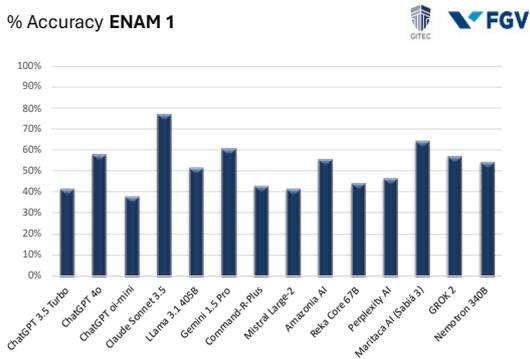

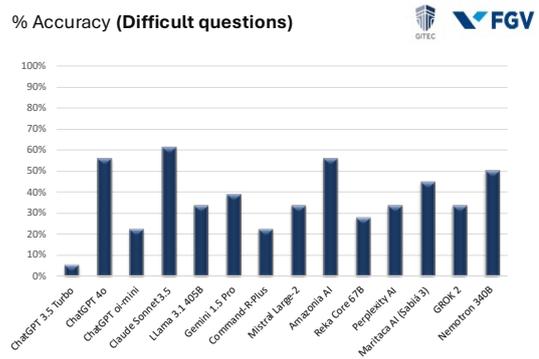

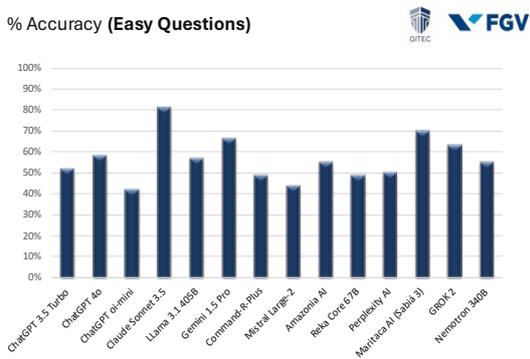

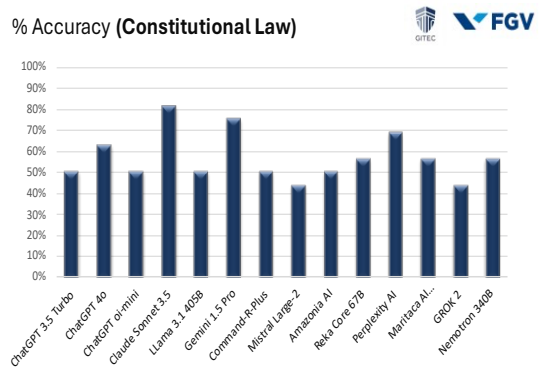

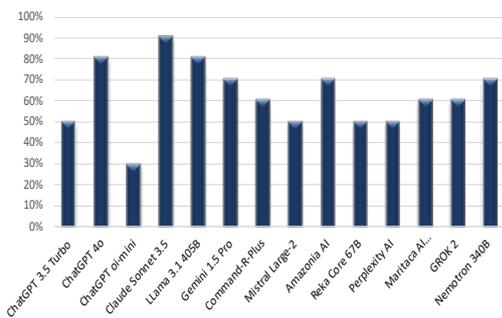

% Accuracy (**Administrative Law**)

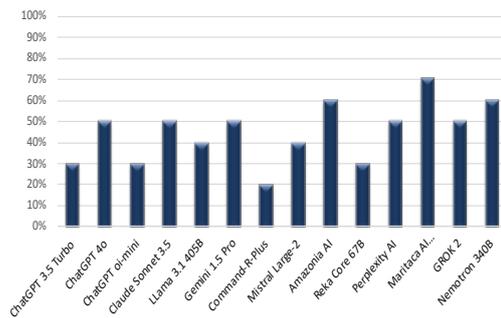

% Accuracy (**Civil Law**)

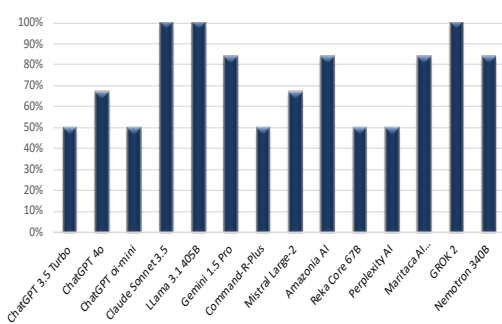

% Accuracy (**General Notions of Law and Humanistic Training**)

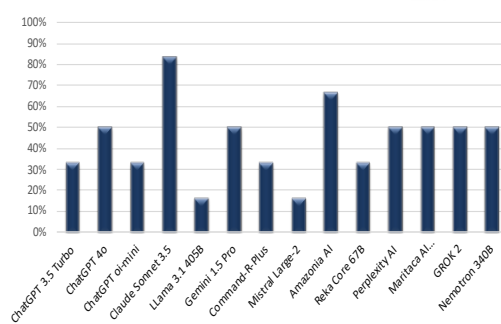

% Accuracy (**Business Law**)

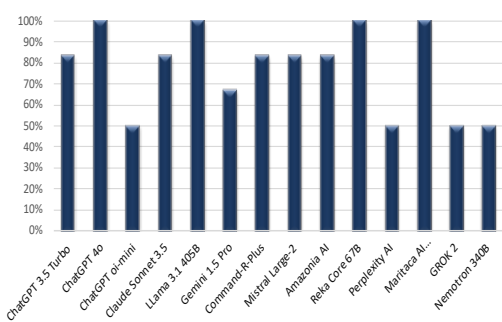

% Accuracy (**Human Rights**)

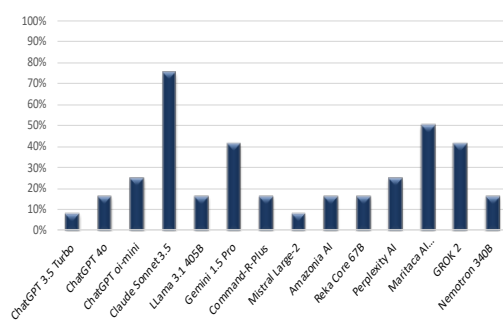

% Accuracy (**Criminal Law**)

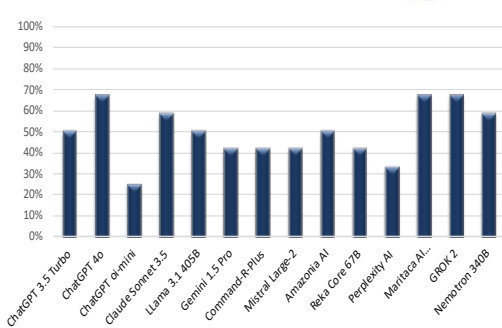

% Accuracy (**Civil Procedure Law**)

## Graphical Analysis of the Performance of Artificial Intelligence Models Tested in ENAM – 2

Once again, it is evident that Claude Sonnet 3.5 was the standout artificial intelligence model in the exam, this time earning the certification of qualification with an 80% accuracy rate. Additionally, ChatGPT 4o and Gemini 1.5 also succeeded in the examination, achieving 78% and 71% accuracy, respectively.

| Model | Total Correct | Total Incorrect | % Accuracy (Total) | % Accuracy (Easy Questions) | % Accuracy (Difficult Questions) |
|---|---|---|---|---|---|
| **ChatGPT 3.5 Turbo** | 33 | 47 | 41% | 46% | 14% |
| **ChatGPT 4o** | 62 | 18 | 78% | 80% | 64% |
| **ChatGPT oi-mini** | 47 | 33 | 59% | 65% | 29% |
| **Claude Sonnet 3.5** | 64 | 16 | 80% | 80% | 79% |
| **LLama 3.1 405B** | 54 | 26 | 68% | 72% | 43% |
| **Gemini 1.5 Pro** | 58 | 22 | 73% | 75% | 57% |
| **Command-R-Plus** | 40 | 40 | 50% | 54% | 29% |
| **Mistral Large-2** | 18 | 62 | 23% | 17% | 43% |
| **Amazonia AI** | 50 | 30 | 63% | 63% | 57% |
| **Reka Core 67B** | 36 | 44 | 45% | 51% | 14% |
| **Perplexity AI** | 41 | 39 | 51% | 58% | 14% |
| **Maritaca AI (Sabiá 3)** | 23 | 57 | 29% | 28% | 29% |
| **GROK 2** | 52 | 28 | 65% | 71% | 36% |
| **Nemotron 340B** | 51 | 29 | 64% | 68% | 43% |

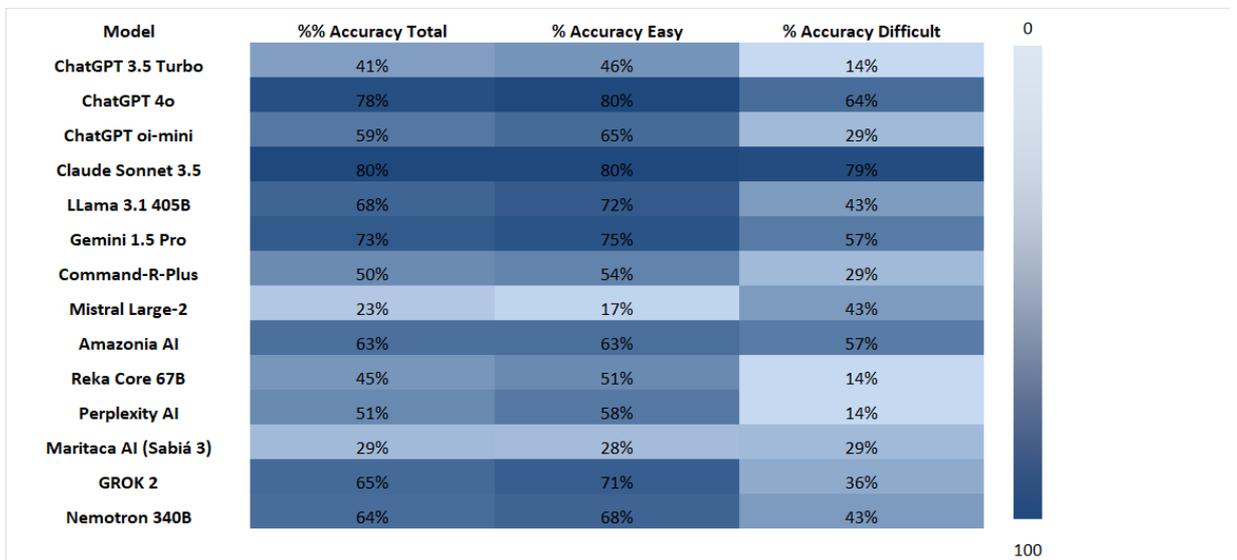

% Accuracy **ENAM 2** 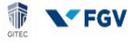

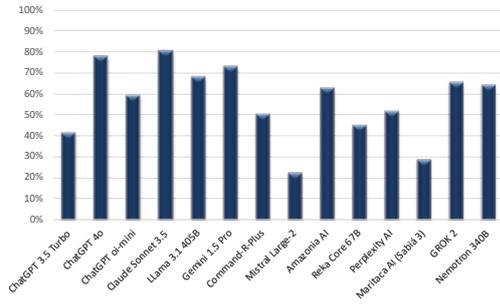

% Accuracy **(Administrative Law)** 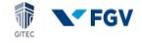

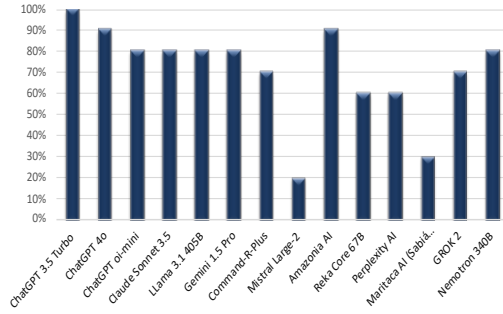

% Accuracy **(Easy Questions)** 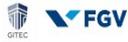

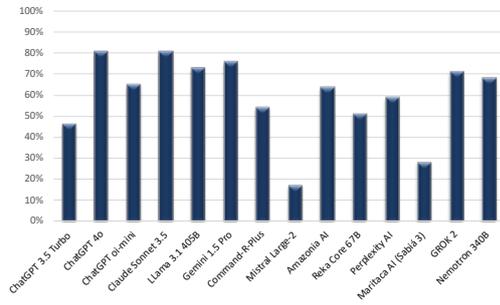

% Accuracy **(General Notions of Law and Humanistic Training)** 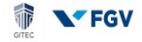

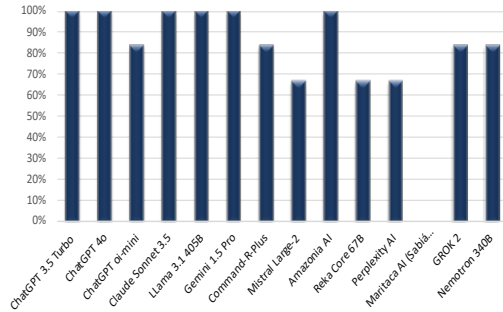

% Accuracy **(Difficult Questions)** 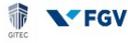

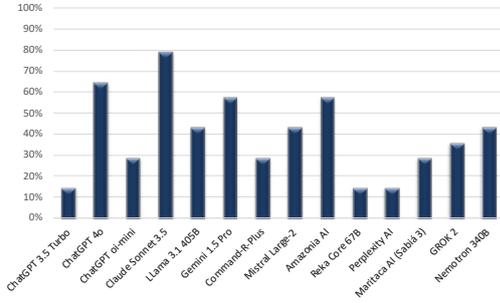

% Accuracy **(Human Rights)** 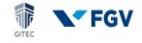

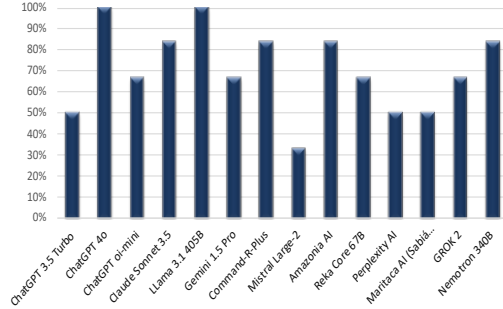

% Accuracy **(Constitutional Law)** 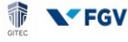

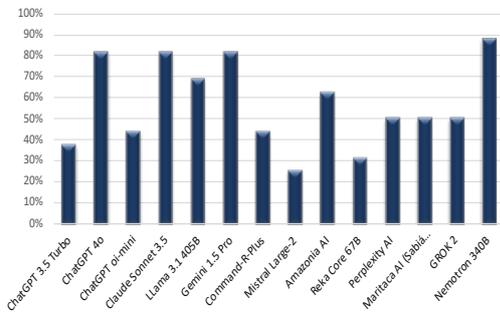

% Accuracy **(Civil Procedure Law)** 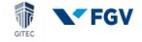

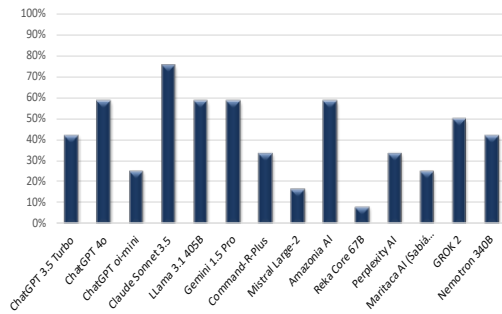

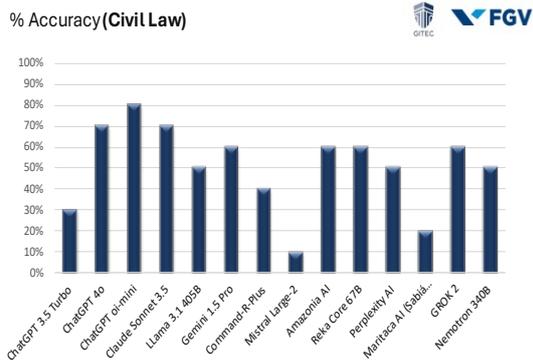

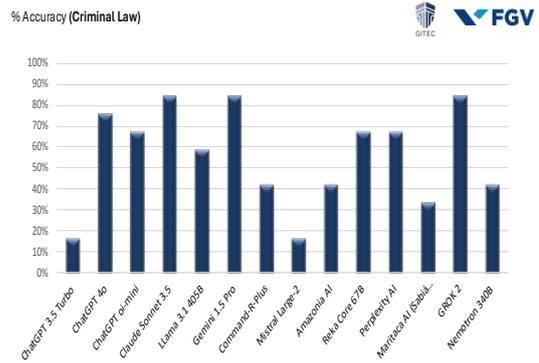

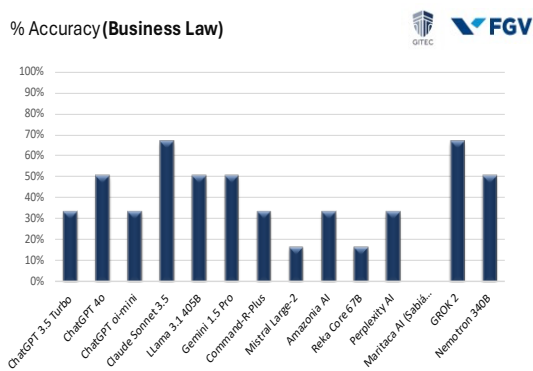

## Comparison Between The Two Selection Processes – ENAM

|  | ENAM 1 | ENAM 2 |
|---|---|---|
| **Model** | **Total Correct** | **Total Correct** |
| **ChatGPT 3.5 Turbo** | 33 | 33 |
| **ChatGPT 4o** | 46 | 62 |
| **ChatGPT oi-mini** | 30 | 47 |
| **Claude Sonnet 3.5** | 61 | 64 |
| **LLama 3.1 405B** | 41 | 54 |
| **Gemini 1.5 Pro** | 48 | 58 |
| **Command-R-Plus** | 34 | 40 |
| **Mistral Large-2** | 33 | 18 |
| **Amazonia AI** | 44 | 50 |
| **Reka Core 67B** | 35 | 36 |
| **Perplexity AI** | 37 | 41 |
| **Maritaca AI (Sabiá 3)** | 51 | 23 |
| **GROK 2** | 45 | 52 |
| **Nemotron 340B** | 43 | 51 |

### Brazilian Bar Association (OAB)

To qualify as a lawyer in Brazil, passing the Brazilian Bar Association Examination (OAB Exam) is mandatory. This exam determines who is eligible to enter the legal profession, assessing law graduates and final-year law students. It is important to note that only students from institutions accredited by the Ministry of Education

(MEC) are eligible to apply (OAB, 2022a). By subjecting candidates to exams that combine legal theory and practice, the OAB Exam evaluates whether they are prepared for the real-world challenges of the legal profession. The Statute of Advocacy (Law 8.906/94) is unequivocal: without passing the exam, one cannot register with the OAB or practice law.

This requirement also serves as a catalyst for raising the educational standards of Brazilian legal institutions. Law schools are compelled to improve their teaching methods, infrastructure, and curriculum. This highlights a key concern raised in our research regarding the extraordinarily high number of law programs in Brazil.

The exam acts as a filter to ensure that the legal market benefits from professionals capable of upholding the ethical and professional standards expected by society. Technically, the exam is divided into two phases. The first is a multiple-choice exam, consisting of 80 questions. For the purposes of this research, the exam was analyzed based on the following areas of knowledge:

- Professional Ethics
- Philosophy of Law
- Constitutional Law
- Human Rights
- Electoral Law
- Public International Law
- Financial Law
- Tax Law
- Administrative Law
- Environmental Law
- Civil Law
- Child and Adolescent Law
- Consumer Law
- Business Law
- Civil Procedure
- Criminal Law
- Criminal Procedure

- Social Security Law
- Labor Law
- Labor Procedure

The candidate must achieve a minimum score of 50%, which corresponds to at least 40 correct answers out of 80 questions.

In the subsequent stage of the Examination, candidates face the challenge of the practical-professional exam. At the time of registration, candidates must choose one of the following seven areas of specialization: Civil Law, Criminal Law, Labor Law, Tax Law, Administrative Law, Constitutional Law, or Business Law.

This stage consists of a written evaluation divided into two parts: a professional legal document related to the chosen area worth half of the total score, and four practical questions worth the other half. The format is based on real-life scenarios in legal practice, requiring candidates to demonstrate solid legal reasoning, well-founded argumentation, interpretative skills, mastery of professional techniques, and clear and organized expression. To pass the exam, candidates must achieve a minimum score of 6.00 and successfully pass both stages. This result demonstrates their aptitude to practice law competently.

The exam is conducted three times a year, approximately every 4 months. The Getulio Vargas Foundation (FGV) is the official organizer of the OAB Exam. By December 2024, 41 editions of the exam had been completed, with the 42nd exam currently in progress (only the first stage has been administered). Therefore, this most recent edition was not included in the empirical research. The study focused on the last 10 editions of the OAB exam, from the 33rd to the 41st, to test different artificial intelligence models.

The ranking of candidates is not published by the organizers. Only the names of those approved are disclosed, sorted by the location where they registered and took the exam.

We present the calculated approval rate index for the most recent editions of the exam, based on the data provided in spreadsheets made available by the Federal Council of the OAB (OAB, 2020), in which the performance of candidates was analyzed in relation to the number of registered applicants, regardless of their location or the name of their higher education institution (HEI):

| ANALYSIS OAB EXAMS | | | |
|---|---|---|---|
| EXAM | Candidates | Approved 2nd Phase | Percentage of Approval |
| 32 | 218838 | 36942 | 16,88% |
| 33 | 150044 | 39558 | 26,36% |
| 34 | 123558 | 15193 | 12,29% |
| 35 | 147652 | 29671 | 18,26% |
| 36 | 118439 | 14720 | 12,42% |
| 37 | 151654 | 30262 | 19,95% |
| 38 | 102828 | 10986 | 10,68% |
| 39 | 133777 | 17743 | 13,26% |
| 40 | 101287 | 12355 | 12,19% |

In addition, we present the following relevant data gathered during the empirical research, detailing the performance of the LLMs, or artificial intelligence models, for each exam analyzed, in descending order.

## Comparison between OAB Exams

| | OAB 32 | OAB 33 | OAB 34 | OAB 35 | OAB 36 | OAB 37 | OAB 38 | OAB 39 | OAB 40 | OAB 41 |
|---|---|---|---|---|---|---|---|---|---|---|
| | Total Correct | Total Correct | Total Correct | Total Correct | Total Correct | Total Correct | Total Correct | Total Correct | Total Correct | Total Correct |
| ChatGPT 3.5 Turbo | 41 | 47 | 41 | 43 | 41 | 46 | 34 | 48 | 45 | 51 |
| ChatGPT 4o | 62 | 64 | 71 | 63 | 67 | 65 | 67 | 68 | 68 | 69 |
| ChatGPT oi-mini | 47 | 43 | 41 | 43 | 49 | 53 | 50 | 50 | 62 | 52 |
| Claude Sonnet 3.5 | 63 | 64 | 67 | 70 | 75 | 66 | 69 | 66 | 68 | 68 |
| LLama 3.1 405B | 49 | 52 | 59 | 61 | 66 | 61 | 57 | 59 | 64 | 65 |
| Gemini 1.5 Pro | 60 | 58 | 66 | 63 | 70 | 61 | 58 | 58 | 59 | 58 |
| Command-R-Plus | 43 | 45 | 49 | 54 | 51 | 52 | 47 | 52 | 50 | 47 |
| Mistral Large-2 | 48 | 52 | 57 | 54 | 57 | 49 | 56 | 54 | 55 | 56 |
| Amazonia AI | 59 | 56 | 61 | 64 | 63 | 52 | 50 | 58 | 47 | 66 |
| Reka Core 67B | 48 | 40 | 37 | 50 | 52 | 56 | 49 | 45 | 55 | 50 |
| Perplexity AI | 51 | 53 | 58 | 53 | 53 | 55 | 52 | 57 | 58 | 53 |
| Sabiá 3 | 37 | 58 | 66 | 64 | 67 | 61 | 53 | 58 | 62 | 63 |
| GROK 2 | 51 | 58 | 60 | 43 | 67 | 55 | 59 | 56 | 58 | 57 |
| Nemotron 340B | 52 | 57 | 59 | 64 | 66 | 62 | 59 | 51 | 59 | 56 |

## Graphical Analysis of the Performance of Artificial Intelligence Models Tested in the 41st OAB Exam

| Model | Total Correct | Total Incorrect | % Accuracy (Total) | % Accuracy (Easy Questions) | % Accuracy (Difficult Questions) |
|---|---|---|---|---|---|
| ChatGPT 3.5 Turbo | 51 | 29 | 64% | 69% | 44% |
| ChatGPT 4o | 69 | 11 | 86% | 87% | 83% |
| ChatGPT oi-mini | 52 | 28 | 65% | 69% | 50% |
| Claude Sonnet 3.5 | 68 | 12 | 85% | 84% | 89% |
| LLama 3.1 405B | 65 | 15 | 81% | 84% | 72% |
| Gemini 1.5 Pro | 58 | 22 | 73% | 74% | 67% |
| Command-R-Plus | 47 | 33 | 59% | 63% | 44% |
| Mistral Large-2 | 56 | 24 | 70% | 73% | 61% |
| **Amazonia AI** | 66 | 14 | 83% | 82% | 83% |
| **Reka Core 67B** | 50 | 30 | 63% | 69% | 39% |
| **Perplexity AI** | 53 | 27 | 66% | 71% | 50% |
| **Maritaca AI (Sabiá 3)** | 63 | 17 | 79% | 77% | 83% |
| **GROK 2** | 57 | 23 | 71% | 73% | 67% |
| **Nemotron 340B** | 56 | 24 | 70% | 73% | 61% |

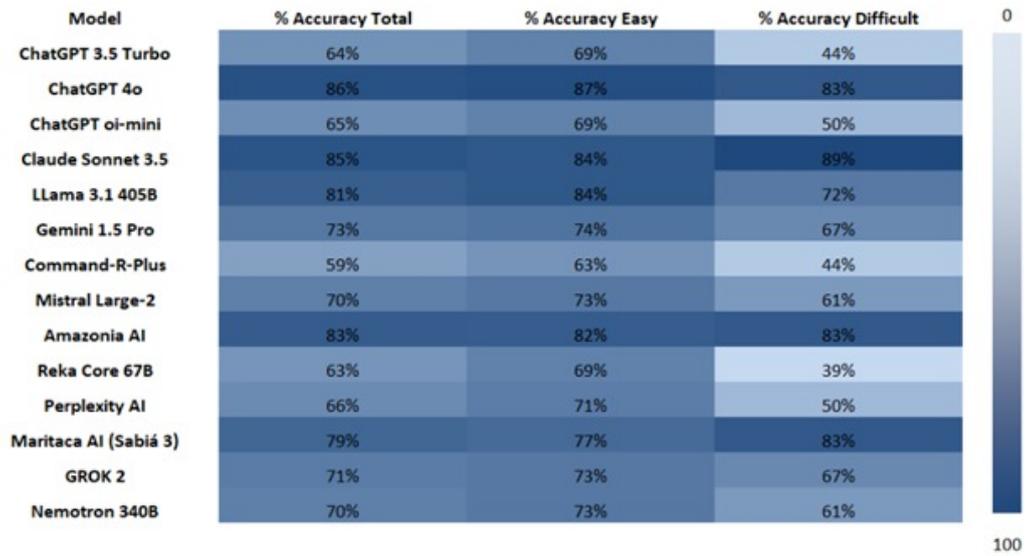

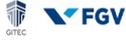

% Accuracy Exam OAB 41

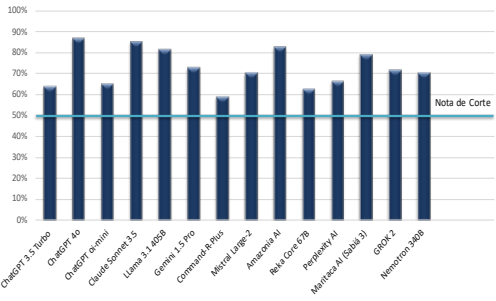

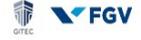

% Accuracy **(Philosophy of Law)**

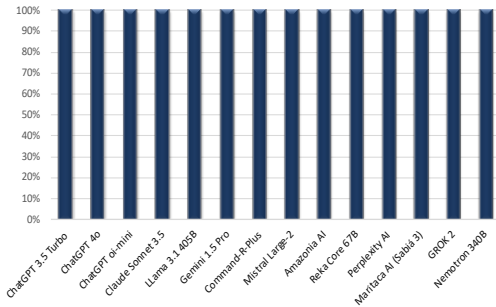

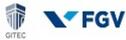

% Accuracy **(Easy Questions)**

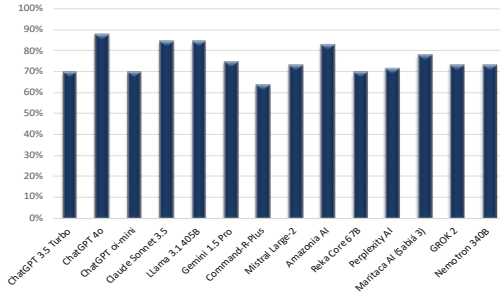

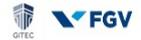

% Accuracy **(Constitutional Law)**

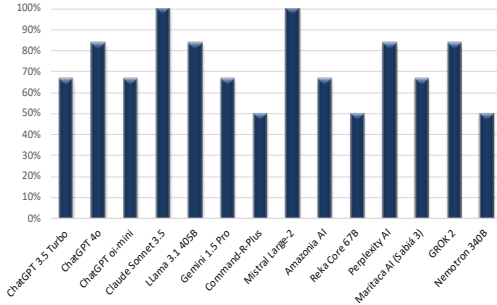

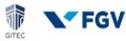

% Accuracy **(Difficult Questions)**

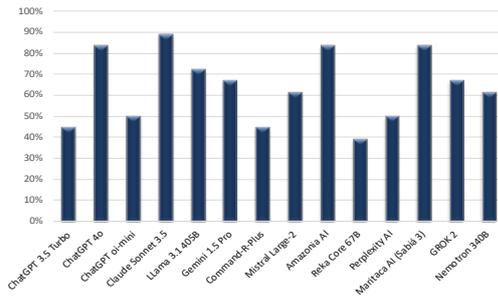

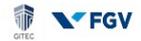

% Accuracy **(Human Rights)**

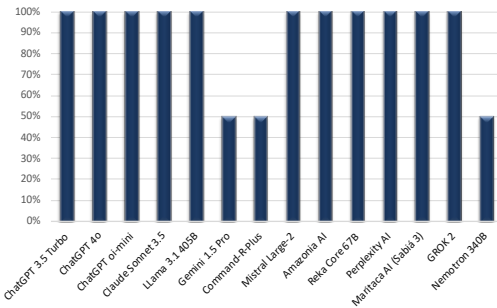

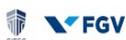

% Accuracy **(Ethics)**

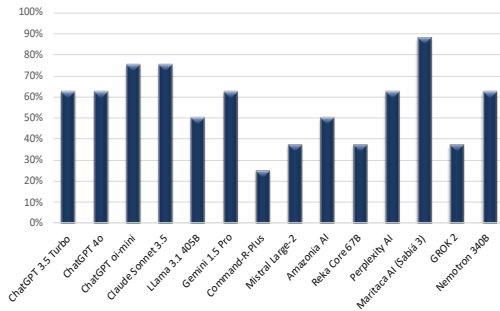

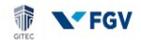

% Accuracy **(Electoral Law)**

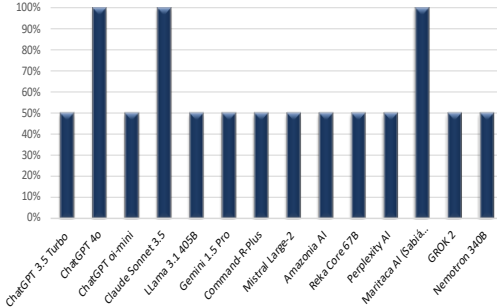

% Accuracy **(International Law)**
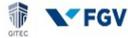
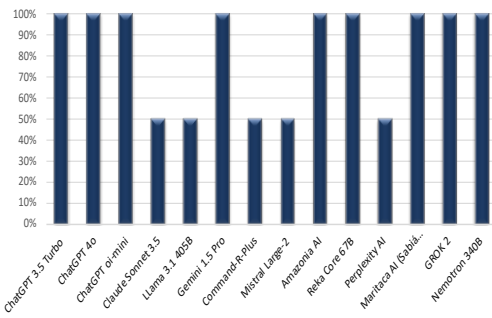

% Accuracy **(Environmental Law)**
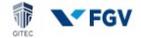
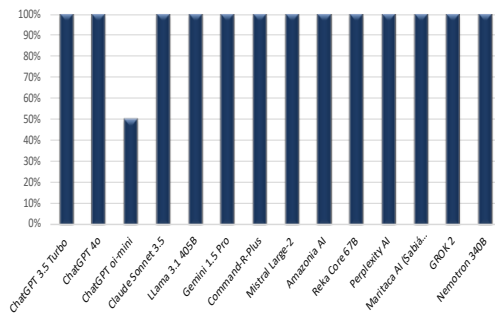

% Accuracy **(Financial Law)**
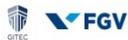
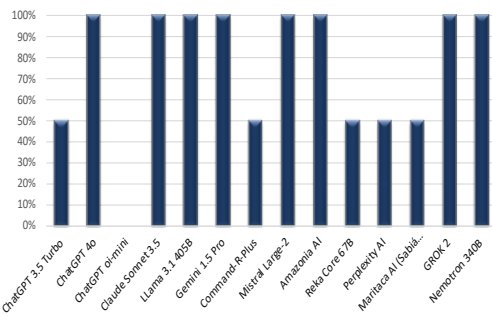

% Accuracy **(Civil Law)**
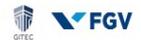
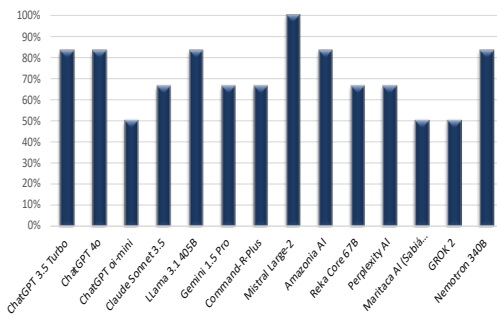

% Accuracy **(Tax Law)**
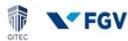
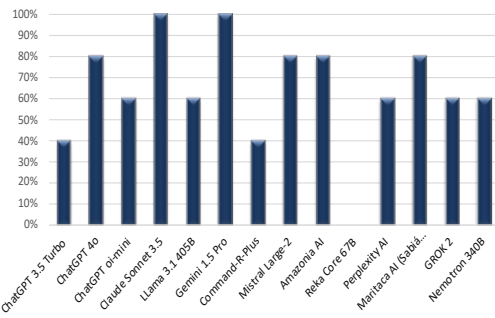

% Accuracy **(Child and Adolescent Law)**
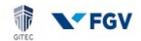
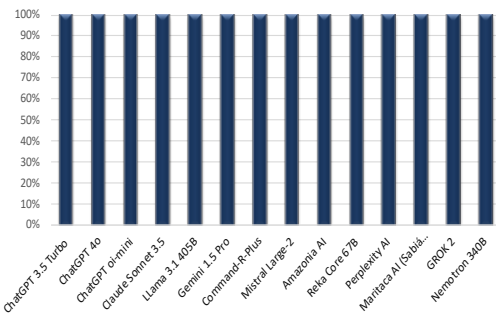

% Accuracy **(Administrative Law)**
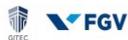
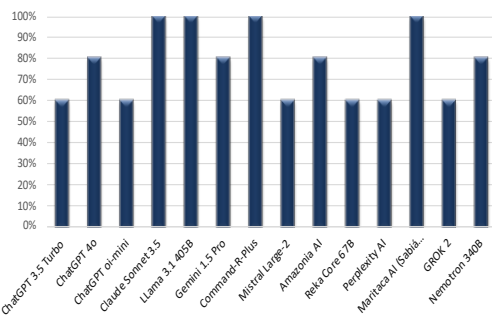

% Accuracy **(Consumer law)**
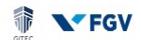
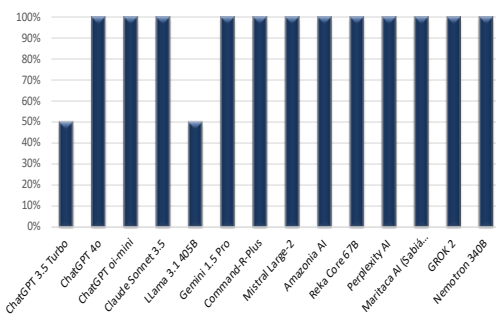

## % Accuracy (Business Law)

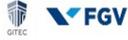
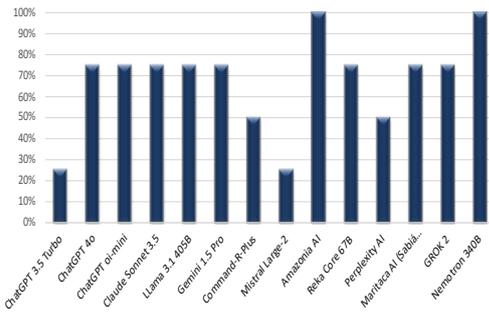

## % Accuracy (Social Security Law)

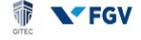
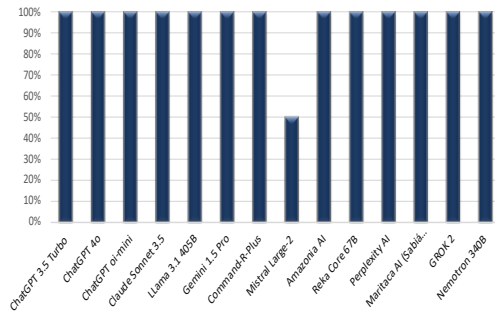

## % Accuracy (Civil Procedure Law)

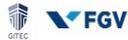
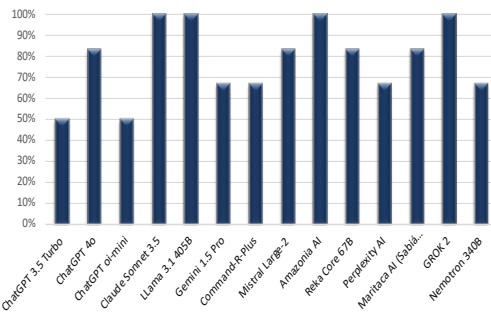

## % Accuracy (Labor Law)

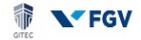
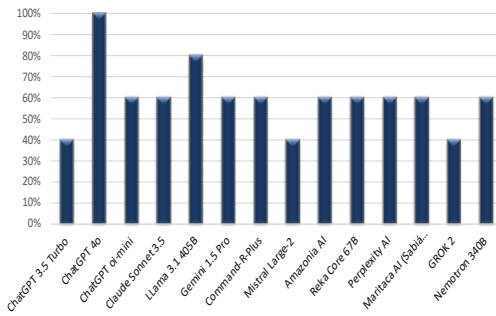

## % Accuracy (Criminal Law)

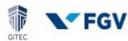
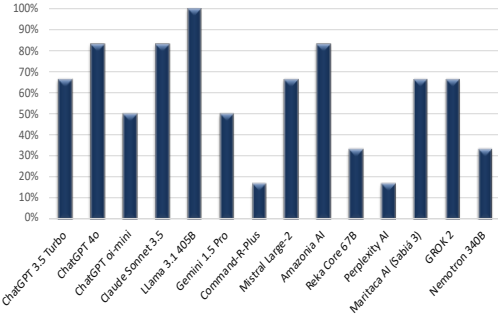

## % Accuracy (Labor Procedural Law)

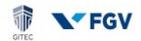
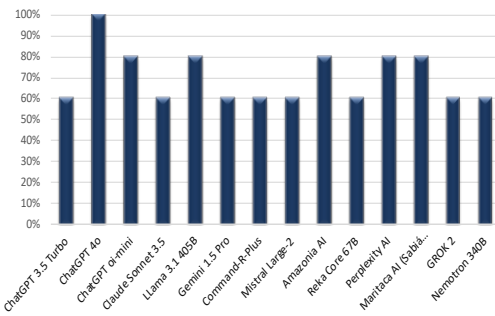

## % Accuracy (Criminal Procedure Law)

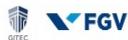
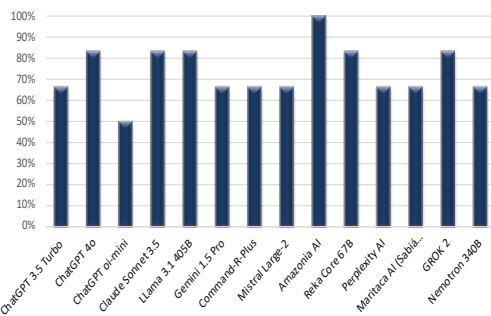

**Graphical Analysis of the Performance of Artificial Intelligence Models Tested in the 40th OAB Exam**

| Model | Total Correct | Total Incorrect | % Accuracy (Total) | % Accuracy (Easy Questions) | % Accuracy (Difficult Questions) |
|---|---|---|---|---|---|
| **ChatGPT 3.5 Turbo** | 45 | 35 | 56% | 65% | 24% |
| **ChatGPT 4o** | 68 | 12 | 85% | 87% | 76% |
| **ChatGPT oi-mini** | 62 | 18 | 78% | 81% | 65% |
| **Claude Sonnet 3.5** | 68 | 12 | 85% | 85% | 82% |
| **LLama 3.1 405B** | 64 | 16 | 80% | 82% | 71% |
| **Gemini 1.5 Pro** | 59 | 21 | 74% | 81% | 47% |
| **Command-R-Plus** | 50 | 30 | 63% | 74% | 18% |
| **Mistral Large-2** | 55 | 25 | 69% | 73% | 53% |
| **Amazonia AI** | 47 | 33 | 59% | 63% | 41% |
| **Reka Core 67B** | 55 | 25 | 69% | 73% | 53% |
| **Perplexity AI** | 58 | 22 | 73% | 74% | 65% |
| **Maritaca AI (Sabiá 3)** | 62 | 18 | 78% | 82% | 59% |
| **GROK 2** | 58 | 22 | 73% | 74% | 65% |
| **Nemotron 340B** | 59 | 21 | 74% | 79% | 53% |

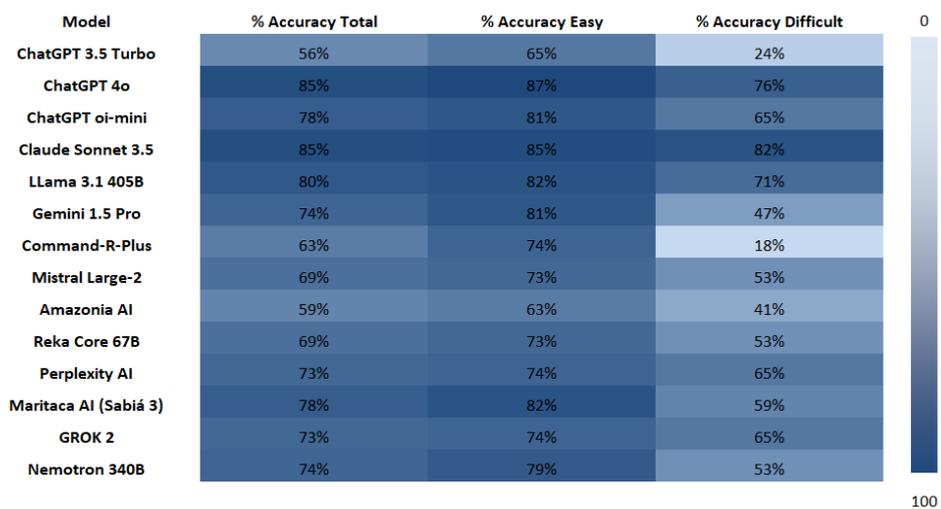

% Accuracy **Exam OAB 40** 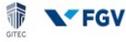

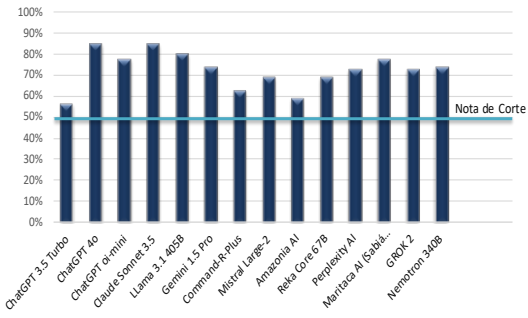

% Accuracy **(Philosophy of Law)** 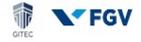

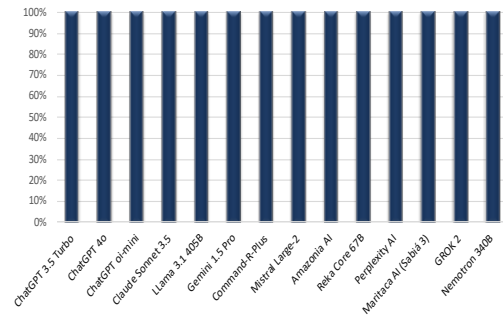

% Accuracy **(Easy Questions)** 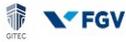

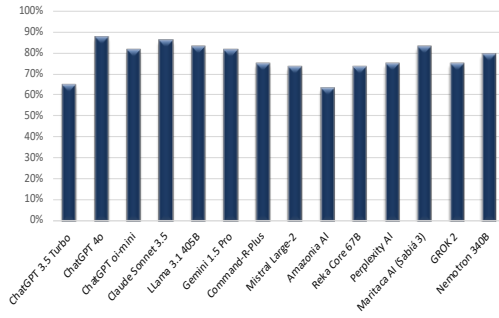

% Accuracy **(Constitutional Law)** 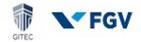

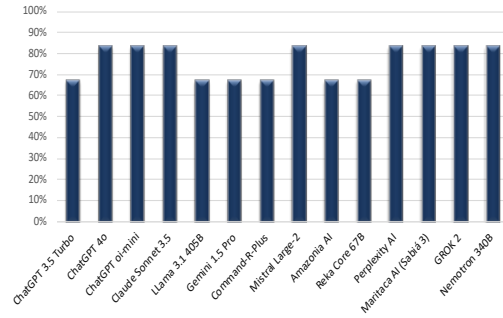

% Accuracy **(Difficult Questions)** 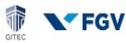

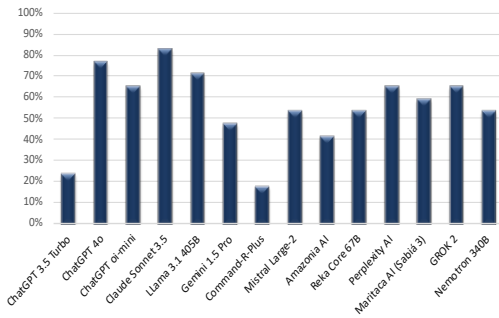

% Accuracy **(Human Rights)** 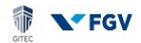

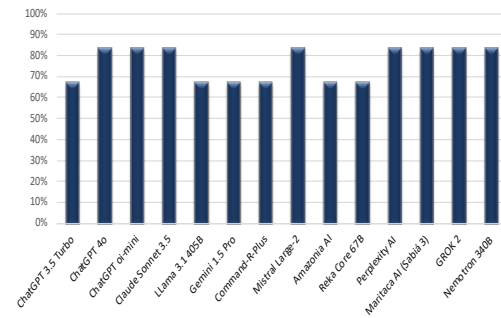

% Accuracy **(Ethics)** 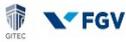

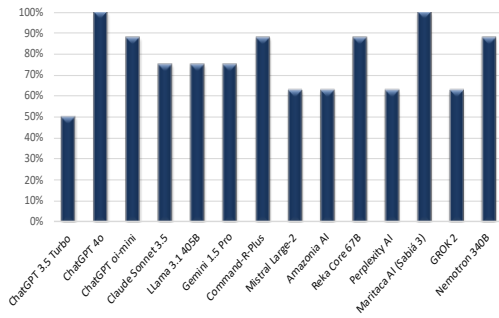

% Accuracy **(Electoral Law)** 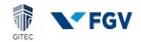

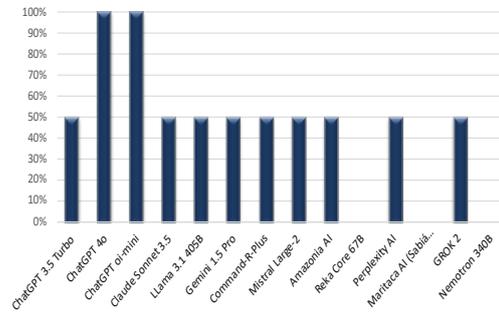

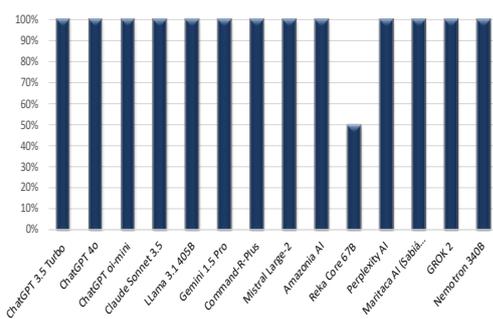

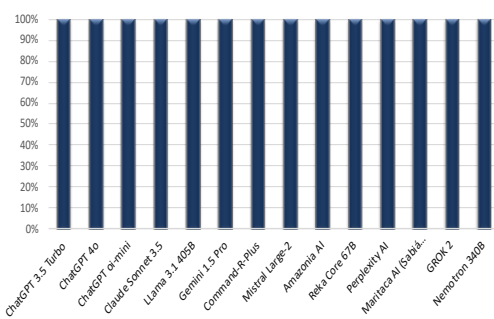

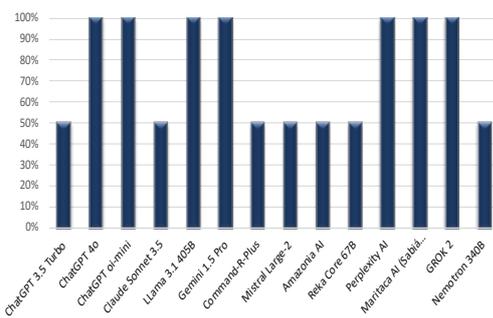

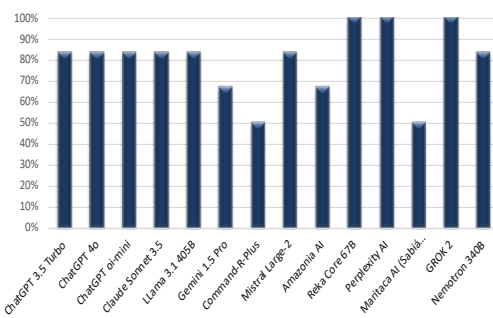

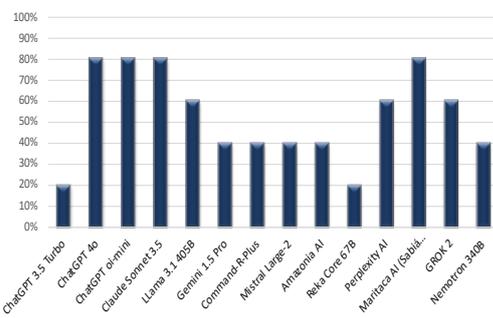

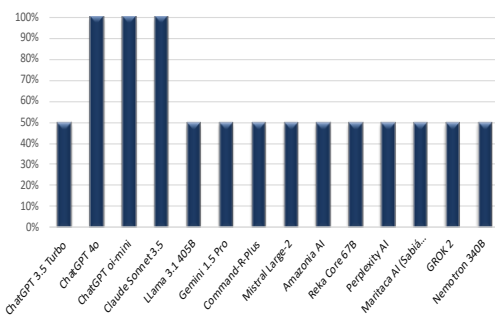

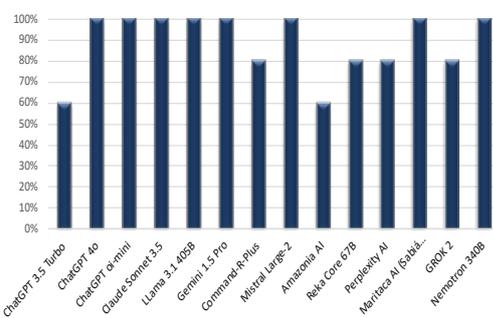

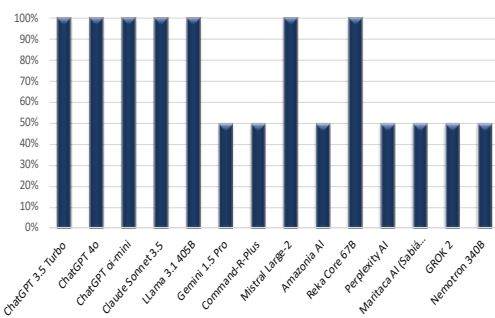

% Accuracy **(Business Law)** 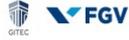

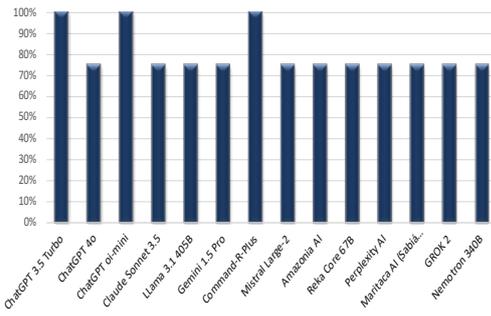

% Accuracy **(Social Security Law)** 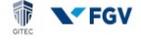

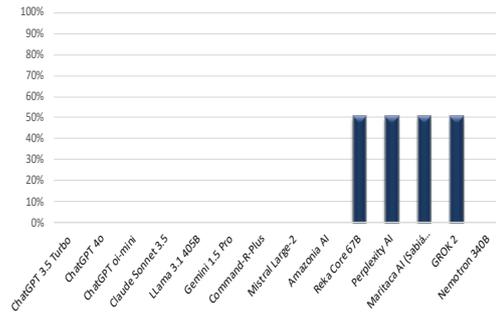

% Accuracy **(Civil Procedure Law)** 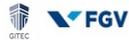

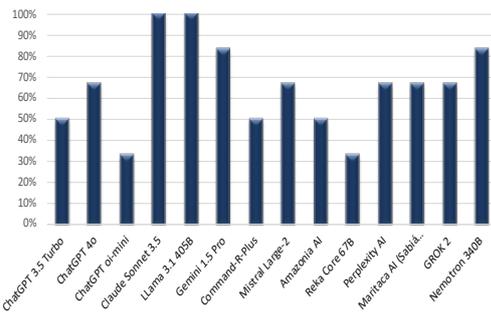

% Accuracy **(Labor Law)** 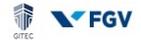

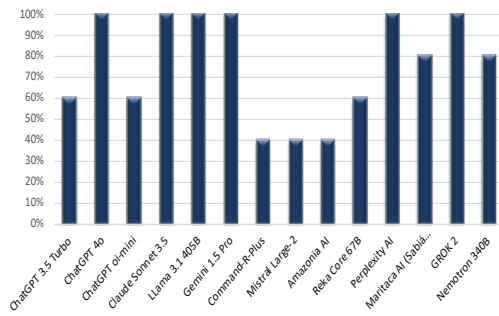

% Accuracy **(Criminal Law)** 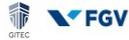

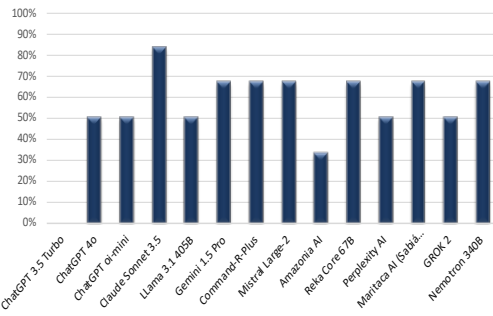

% Accuracy **(Labor Procedural Law)** 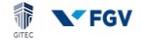

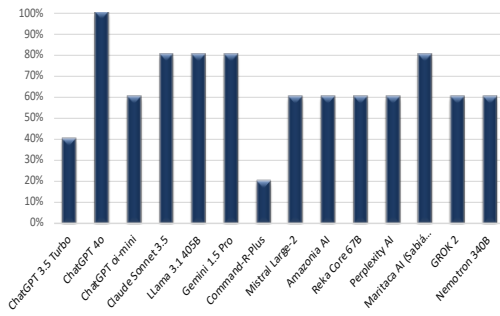

% Accuracy **(Criminal Procedure Law)** 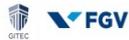

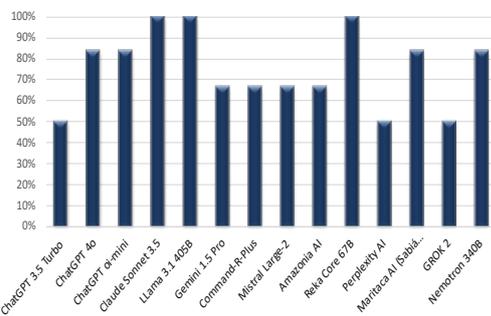

**Graphical Analysis of the Performance of Artificial Intelligence Models Tested in the 39[th] OAB Exam**

| Model | Total Correct | Total Incorrect | % Accuracy (Total) | % Accuracy (Easy Questions) | % Accuracy (Difficult Questions) |
|---|---|---|---|---|---|
| **ChatGPT 3.5 Turbo** | 48 | 32 | 60% | 62% | 44% |
| **ChatGPT 4o** | 68 | 12 | 85% | 89% | 69% |
| **ChatGPT oi-mini** | 50 | 30 | 63% | 70% | 25% |
| **Claude Sonnet 3.5** | 66 | 14 | 83% | 85% | 69% |
| **LLama 3.1 405B** | 59 | 21 | 74% | 79% | 50% |
| **Gemini 1.5 Pro** | 58 | 22 | 73% | 79% | 44% |
| **Command-R-Plus** | 52 | 28 | 65% | 72% | 31% |
| **Mistral Large-2** | 54 | 26 | 68% | 75% | 31% |
| **Amazonia AI** | 58 | 22 | 73% | 80% | 38% |
| **Reka Core 67B** | 45 | 35 | 56% | 61% | 31% |
| **Perplexity AI** | 57 | 23 | 71% | 75% | 50% |
| **Sabiá - 3** | 58 | 22 | 73% | 77% | 50% |
| **GROK 2** | 56 | 24 | 70% | 74% | 50% |
| **Nemotron 340B** | 51 | 29 | 64% | 67% | 44% |

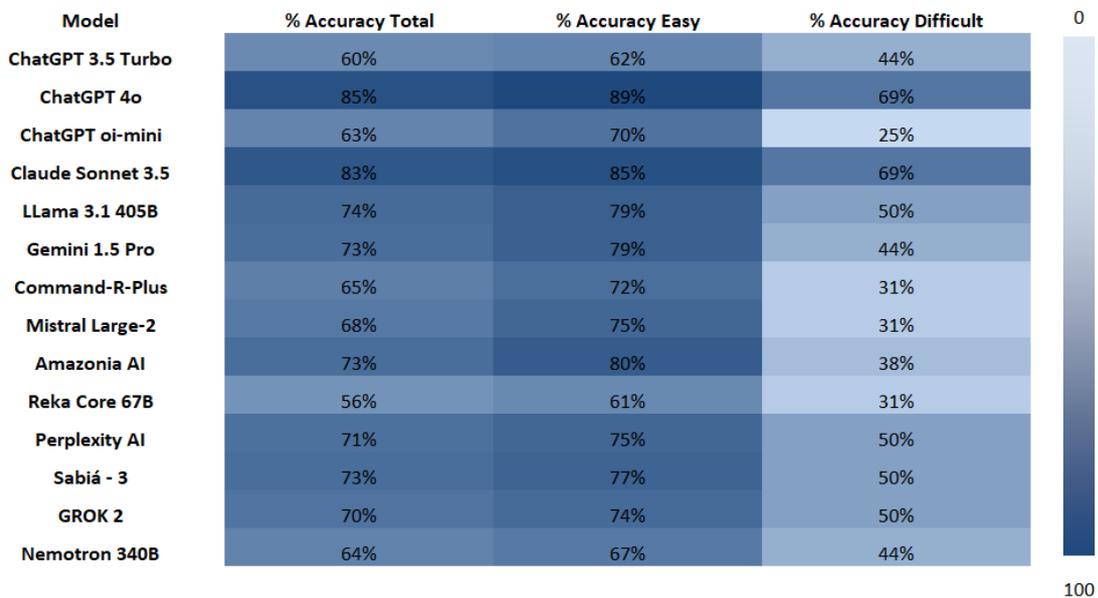

% Accuracy Exam OAB 39

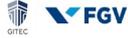
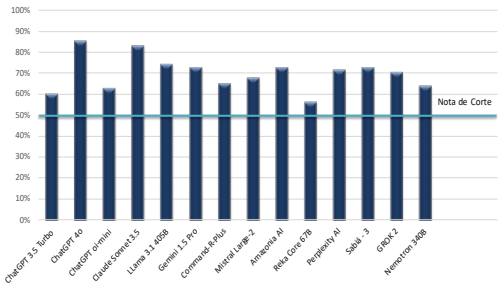

% Accuracy **(Easy Questions)**

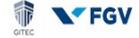
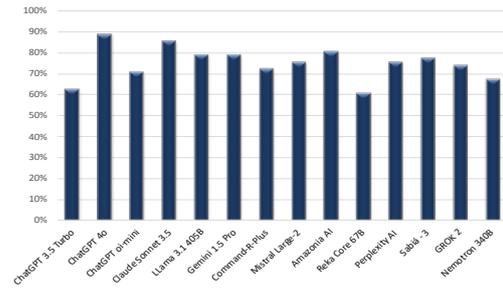

% Accuracy **(Difficult Questions)**

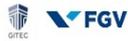
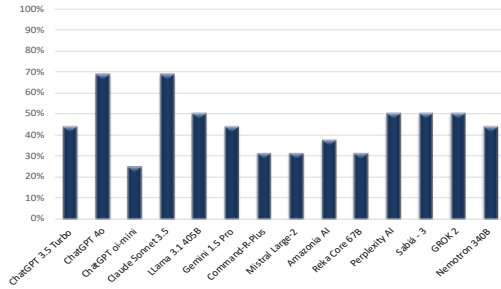

% Accuracy **(Ethics)**

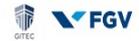
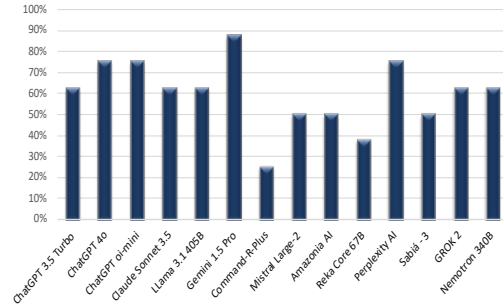

% Accuracy **(Philosophy of Law)**

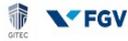
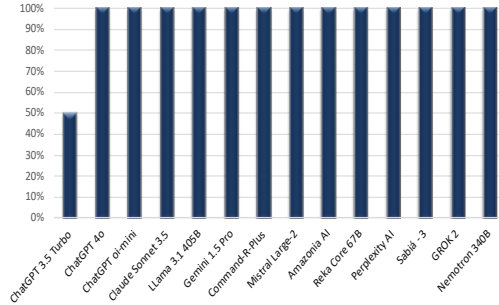

% Accuracy **(Constitutional Law)**

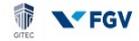
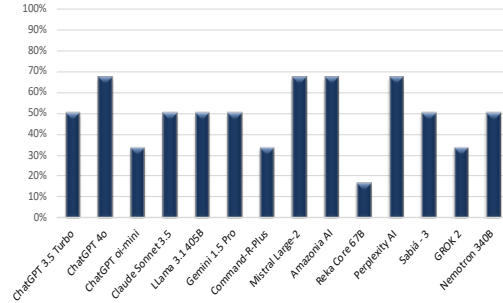

% Accuracy **(Human Rights)**

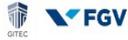
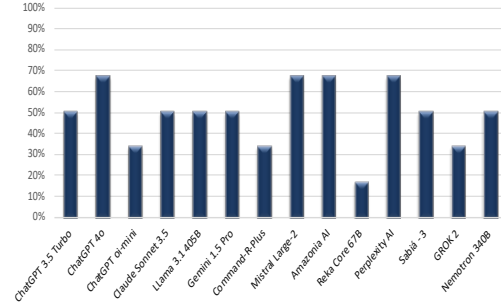

% Accuracy **(Electoral Law)**

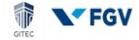
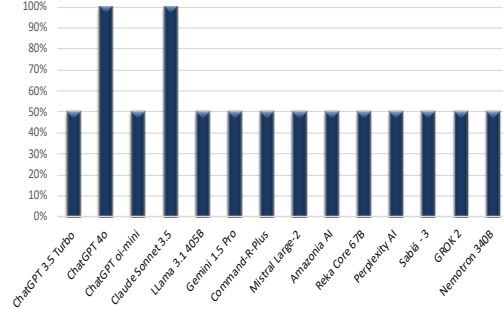

% Accuracy **(International Law)**
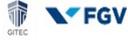
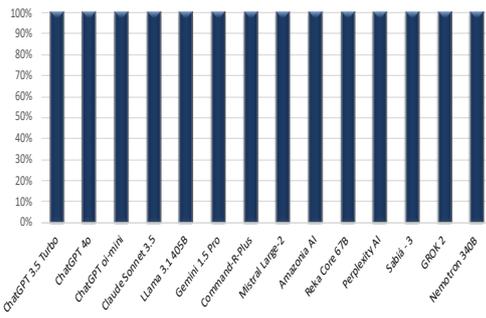

% Accuracy **(Financial Law)**
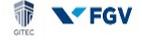
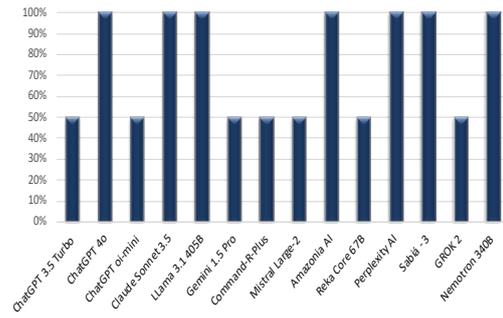

% Accuracy **(Tax Law)**
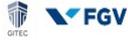
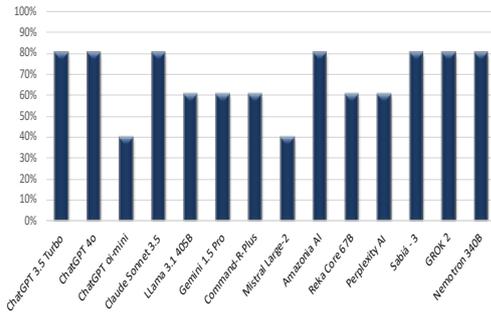

% Accuracy **(Administrative Law)**
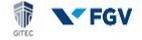
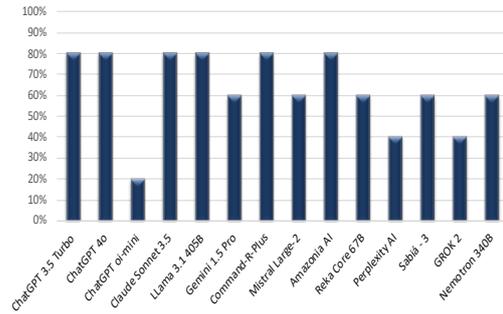

% Accuracy **(Environmental Law)**
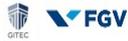
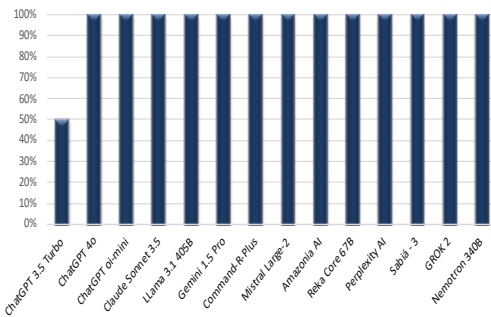

% Accuracy **(Civil Law)**
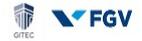
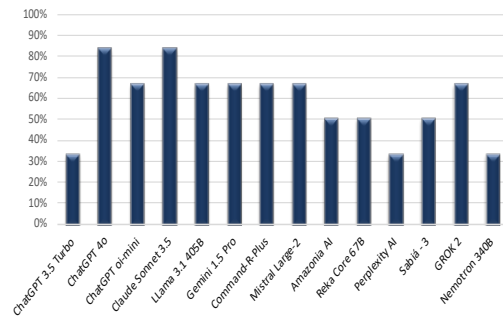

## % Accuracy **(Child and Adolescent Law)**

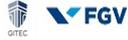
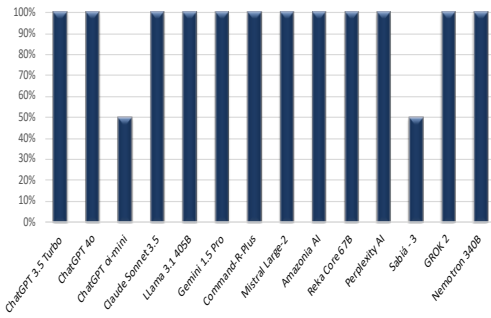

## % Accuracy **(Criminal Law)**

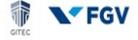
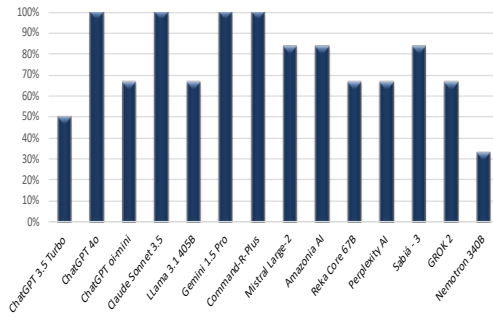

## % Accuracy **(Consumer law)**

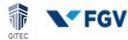
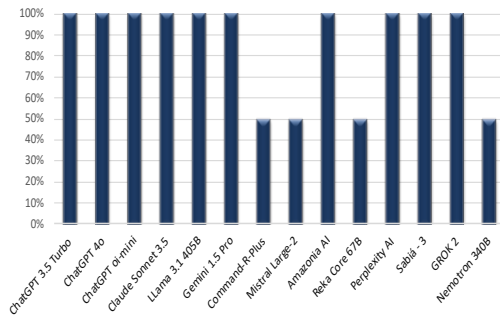

## % Accuracy **(Criminal Procedure Law)**

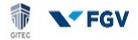
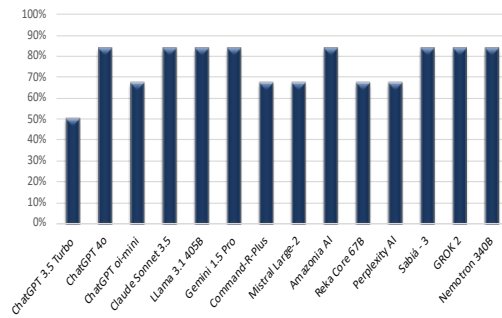

## % Accuracy **(Business Law)**

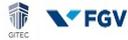
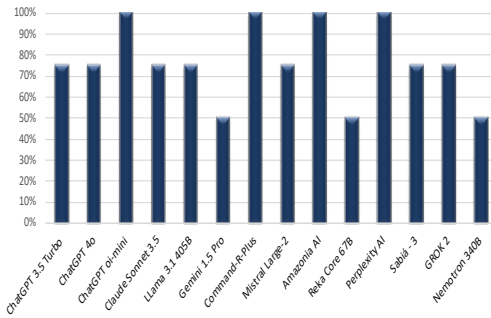

## % Accuracy **(Social Security Law)**

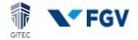
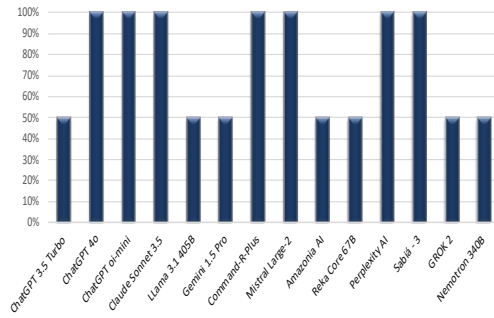

## % Accuracy **(Civil Procedure Law)**

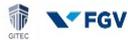
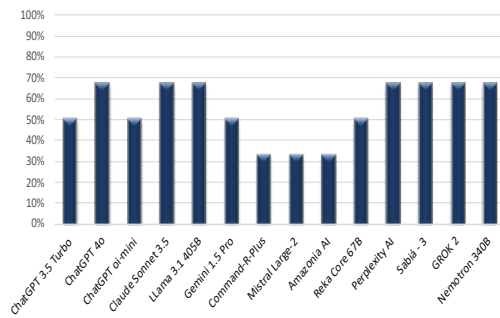

## % Accuracy **(Labor Law)**

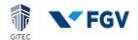
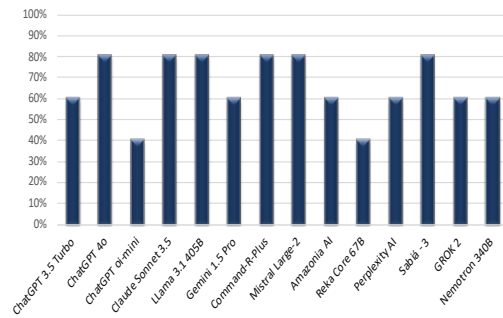

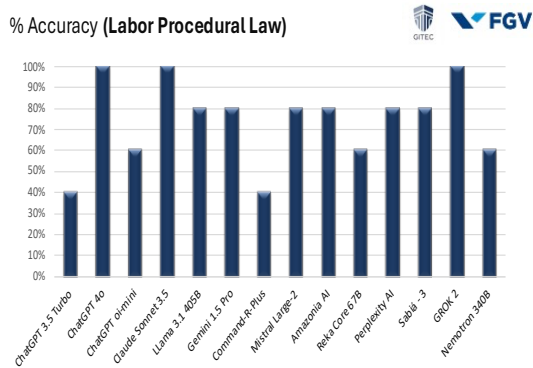

**Graphical Analysis of the Performance of Artificial Intelligence Models Tested in the 38th OAB Exam**

| Model | Total Correct | Total Incorrect | % Accuracy (Total) | % Accuracy (Easy Questions) | % Accuracy (Difficult Questions) |
|---|---|---|---|---|---|
| **ChatGPT 3.5 Turbo** | 34 | 46 | 43% | 44% | 38% |
| **ChatGPT 4o** | 67 | 13 | 84% | 81% | 88% |
| **ChatGPT oi-mini** | 50 | 30 | 63% | 61% | 65% |
| **Claude Sonnet 3.5** | 69 | 11 | 86% | 89% | 81% |
| **LLama 3.1 405B** | 57 | 23 | 71% | 76% | 62% |
| **Gemini 1.5 Pro** | 58 | 22 | 73% | 69% | 81% |
| **Command-R-Plus** | 47 | 33 | 59% | 61% | 54% |
| **Mistral Large-2** | 56 | 24 | 70% | 72% | 65% |
| **Amazonia AI** | 50 | 30 | 63% | 61% | 65% |
| **Reka Core 67B** | 49 | 31 | 61% | 69% | 46% |
| **Perplexity AI** | 52 | 28 | 65% | 70% | 54% |
| **Maritaca AI (Sabiá 3)** | 53 | 27 | 66% | 70% | 58% |
| **GROK 2** | 59 | 21 | 74% | 72% | 77% |
| **Nemotron 340B** | 59 | 21 | 74% | 72% | 77% |

| Model | % Accuracy Total | % Accuracy Easy | % Accuracy Difficult |
|---|---|---|---|
| ChatGPT 3.5 Turbo | 43% | 44% | 38% |
| ChatGPT 4o | 84% | 81% | 88% |
| ChatGPT oi-mini | 63% | 61% | 65% |
| Claude Sonnet 3.5 | 86% | 89% | 81% |
| LLama 3.1 405B | 71% | 76% | 62% |
| Gemini 1.5 Pro | 73% | 69% | 81% |
| Command-R-Plus | 59% | 61% | 54% |
| Mistral Large-2 | 70% | 72% | 65% |
| Amazonia AI | 63% | 61% | 65% |
| Reka Core 67B | 61% | 69% | 46% |
| Perplexity AI | 65% | 70% | 54% |
| Maritaca AI (Sabiá 3) | 66% | 70% | 58% |
| GROK 2 | 74% | 72% | 77% |
| Nemotron 340B | 74% | 72% | 77% |

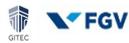
% Accuracy Exam OAB 38

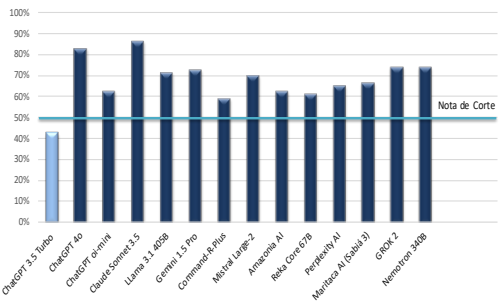

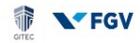
% Accuracy (Difficult Questions)

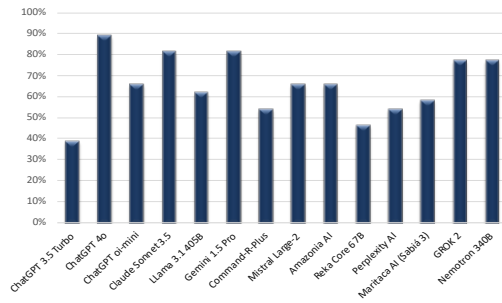

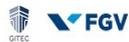
% Accuracy (Easy Questions)

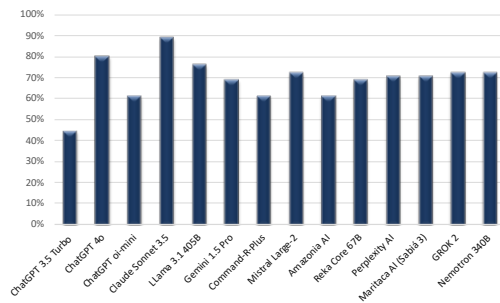

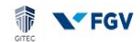
% Accuracy (Ethics)

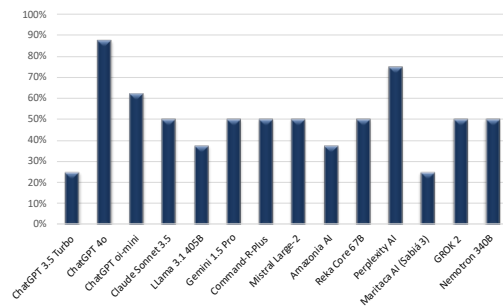

## % Accuracy (Philosophy of Law)

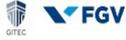
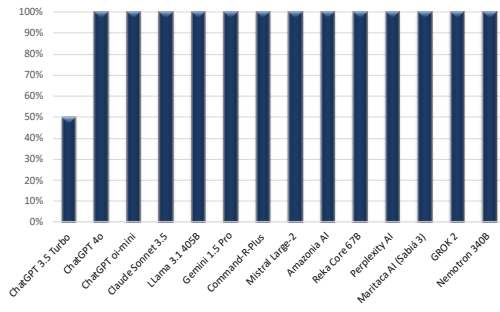

## % Accuracy (International Law)

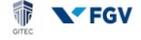
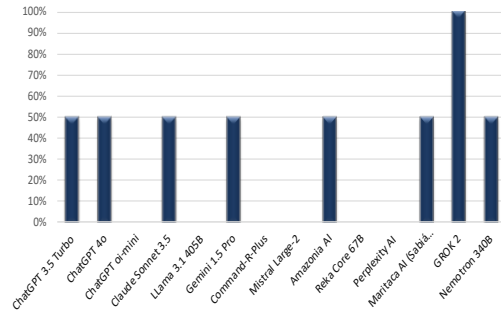

## % Accuracy (Constitutional Law)

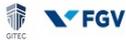
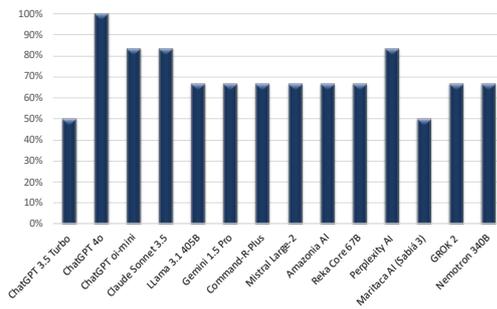

## % Accuracy (Financial Law)

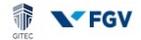
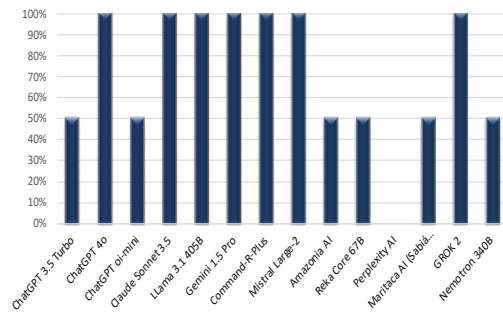

## % Accuracy (Human Rights)

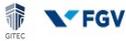
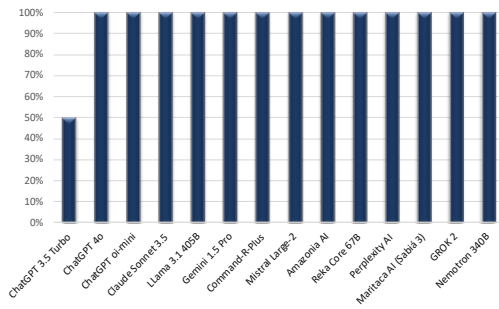

## % Accuracy (Tax Law)

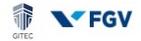
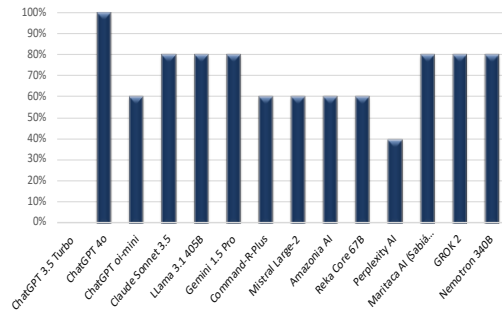

## % Accuracy (Electoral Law)

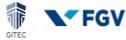
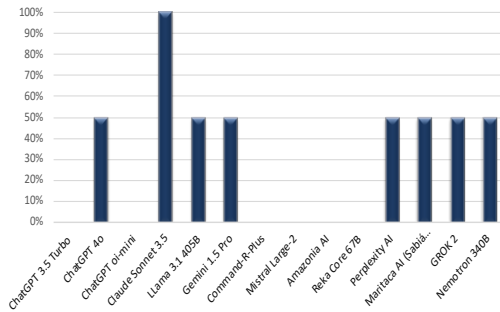

## % Accuracy (Administrative Law)

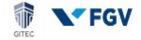
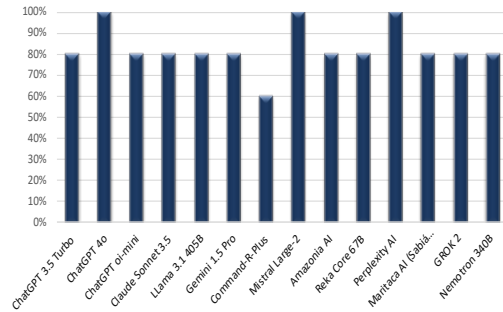

% Accuracy **(Environmental Law)** 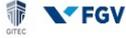

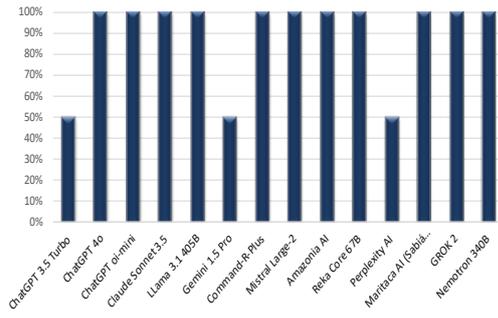

% Accuracy **(Business Law)** 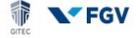

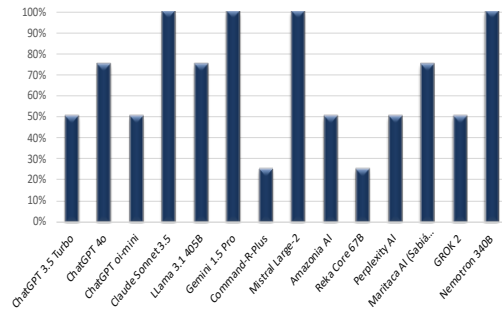

% Accuracy **(Civil Law)** 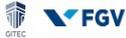

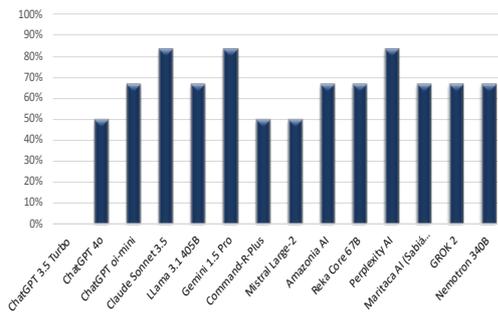

% Accuracy **(Civil Procedure Law)** 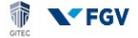

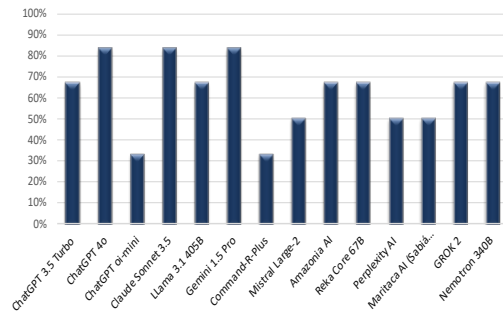

% Accuracy **(Child and Adolescent Law)** 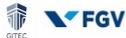

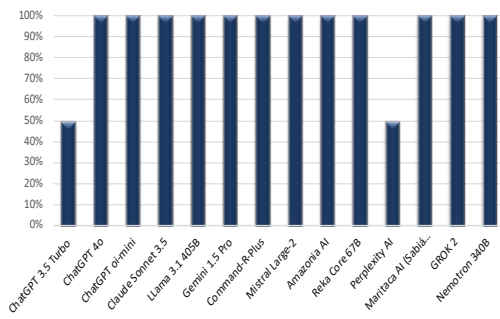

% Accuracy **(Criminal Law)** 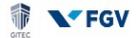

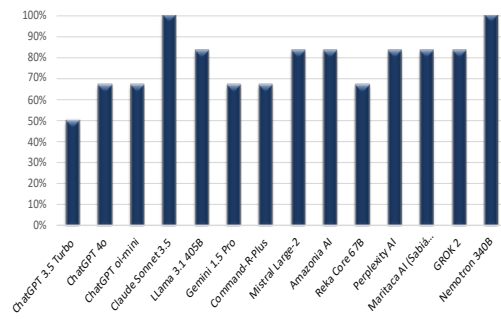

% Accuracy **(Consumer law)** 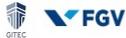

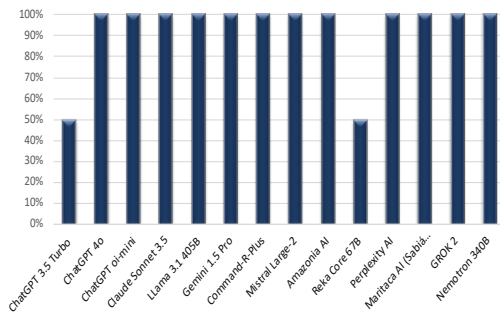

% Accuracy **(Criminal Procedure Law)** 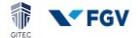

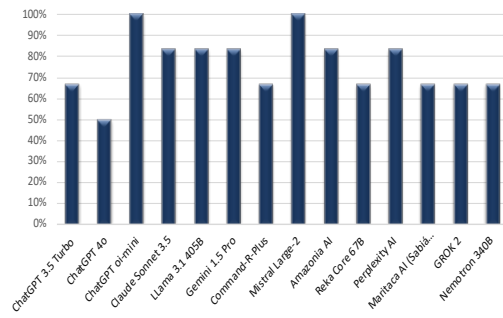

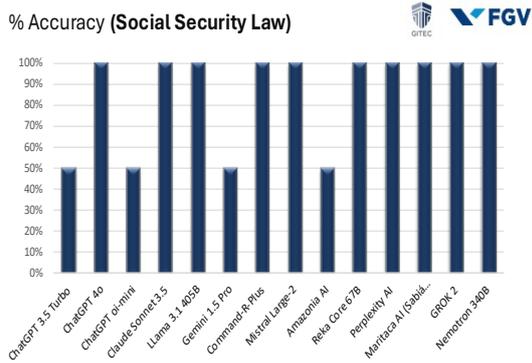

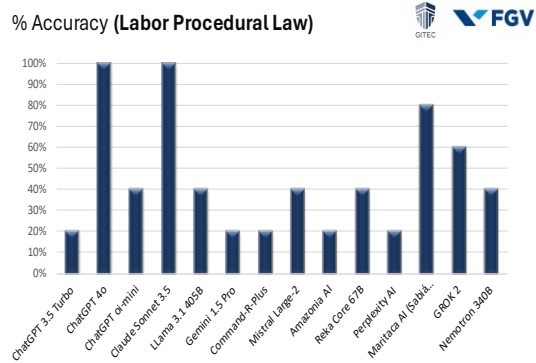

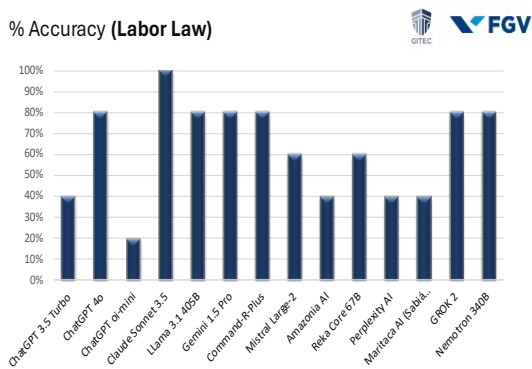

**Graphical Analysis of the Performance of Artificial Intelligence Models Tested in the 37th OAB Exam**

| Model | Total Correct | Total Incorrect | % Accuracy (Total) | % Accuracy (Easy Questions) | % Accuracy (Difficult Questions) |
|---|---|---|---|---|---|
| **ChatGPT 3.5 Turbo** | 46 | 34 | 58% | 56% | 63% |
| **ChatGPT 4o** | 65 | 15 | 81% | 84% | 69% |
| **ChatGPT oi-mini** | 53 | 27 | 66% | 70% | 50% |
| **Claude Sonnet 3.5** | 66 | 14 | 83% | 84% | 75% |
| **LLama 3.1 405B** | 61 | 19 | 76% | 75% | 81% |
| **Gemini 1.5 Pro** | 61 | 19 | 76% | 73% | 88% |
| **Command-R-Plus** | 52 | 28 | 65% | 67% | 56% |
| **Mistral Large-2** | 49 | 31 | 61% | 66% | 44% |
| **Amazonia AI** | 52 | 28 | 65% | 64% | 69% |
| **Reka Core 67B** | 56 | 24 | 70% | 70% | 69% |
| **Perplexity AI** | 55 | 25 | 69% | 70% | 63% |
| **Sabiá 3** | 61 | 19 | 76% | 78% | 69% |
| **GROK 2** | 55 | 25 | 69% | 70% | 63% |
| **Nemotron 340B** | 62 | 18 | 78% | 81% | 63% |

| Model | % Accuracy Total | % Accuracy Easy | % Accuracy Difficult |
|---|---|---|---|
| ChatGPT 3.5 Turbo | 58% | 56% | 63% |
| ChatGPT 4o | 81% | 84% | 69% |
| ChatGPT o1-mini | 66% | 70% | 50% |
| Claude Sonnet 3.5 | 83% | 84% | 75% |
| LLama 3.1 405B | 76% | 75% | 81% |
| Gemini 1.5 Pro | 76% | 73% | 88% |
| Command-R-Plus | 65% | 67% | 56% |
| Mistral Large-2 | 61% | 66% | 44% |
| Amazonia AI | 65% | 64% | 69% |
| Reka Core 67B | 70% | 70% | 69% |
| Perplexity AI | 69% | 70% | 63% |
| Sabiá 3 | 76% | 78% | 69% |
| GROK 2 | 69% | 70% | 63% |
| Nemotron 340B | 78% | 81% | 63% |

% Accuracy Exam OAB 37

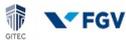
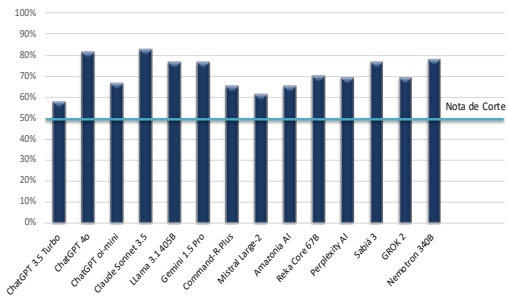

% Accuracy (Difficult Questions)

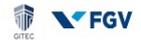
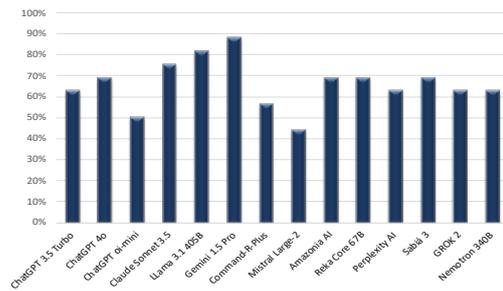

% Accuracy (Easy Questions)

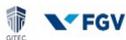
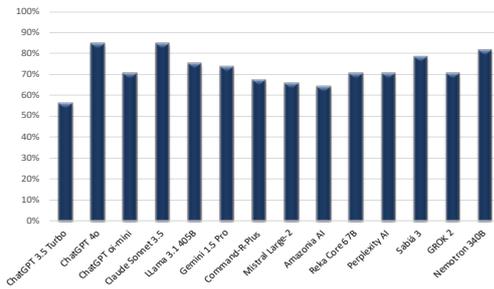

% Accuracy (Ethics)

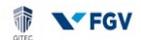
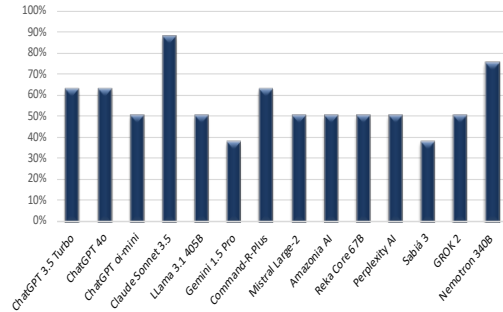

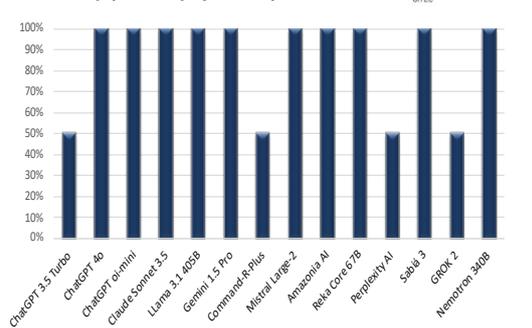

% Accuracy (Philosophy of Law)

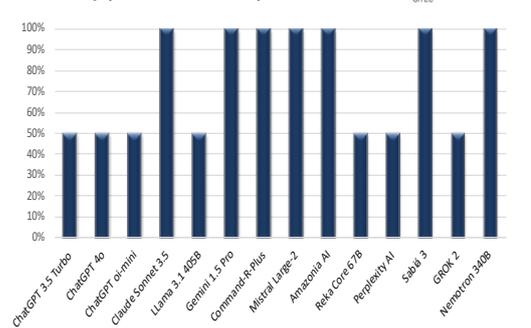

% Accuracy (International Law)

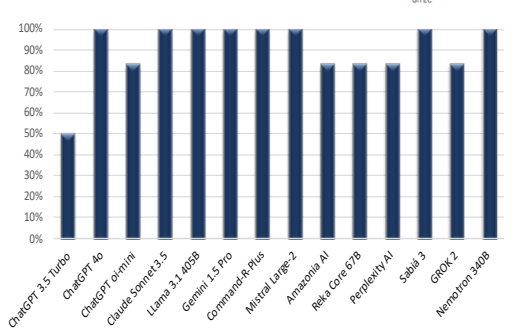

% Accuracy (Constitutional Law)

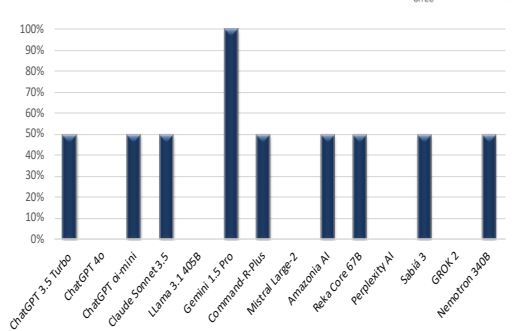

% Accuracy (Financial Law)

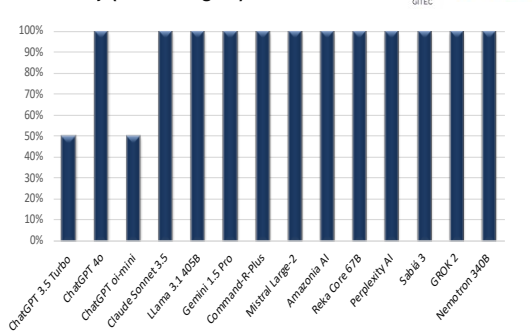

% Accuracy (Human Rights)

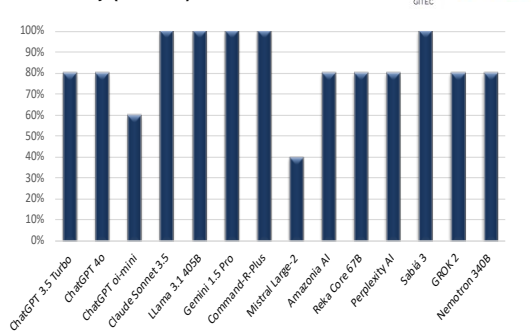

% Accuracy (Tax Law)

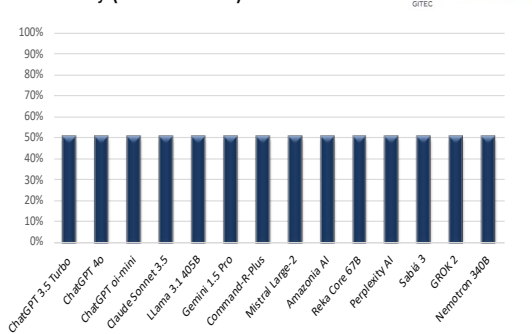

% Accuracy (Electoral Law)

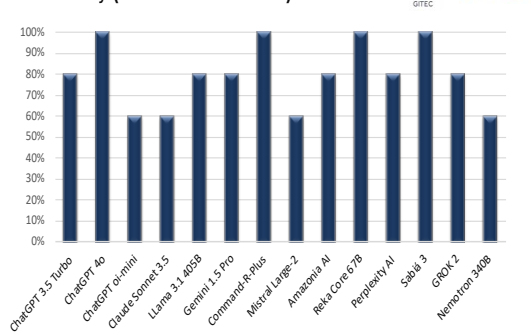

% Accuracy (Administrative Law)

% Accuracy **(Environmental Law)** 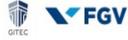

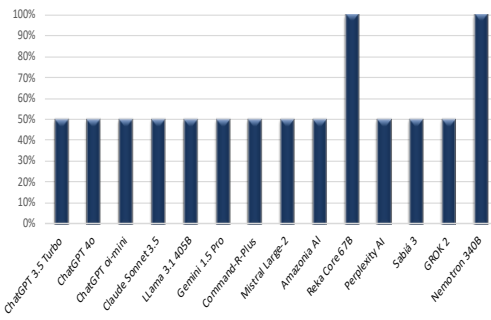

% Accuracy **(Business Law)** 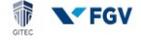

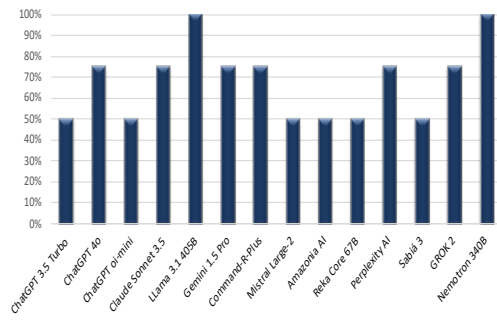

% Accuracy **(Civil Law)** 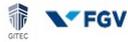

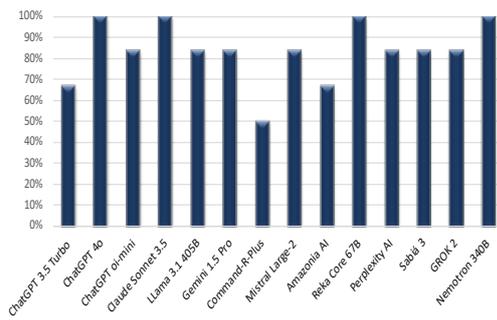

% Accuracy **(Civil Procedure Law)** 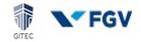

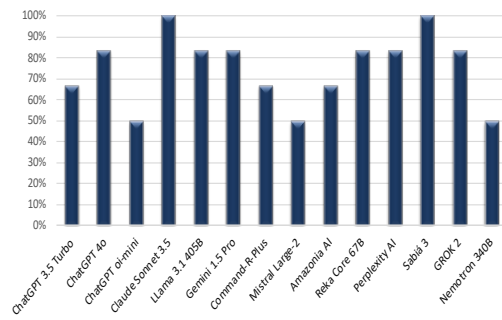

% Accuracy **(Child and Adolescent Law)** 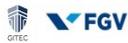

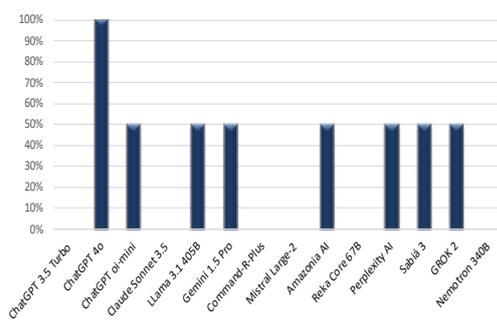

% Accuracy **(Criminal Law)** 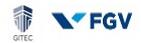

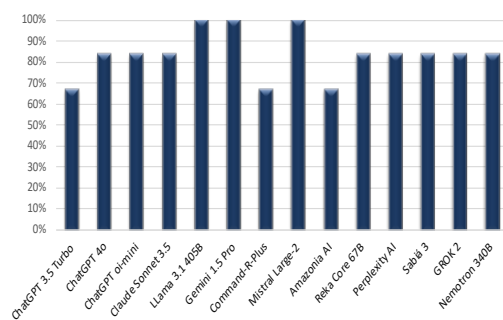

% Accuracy **(Consumer law)** 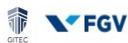

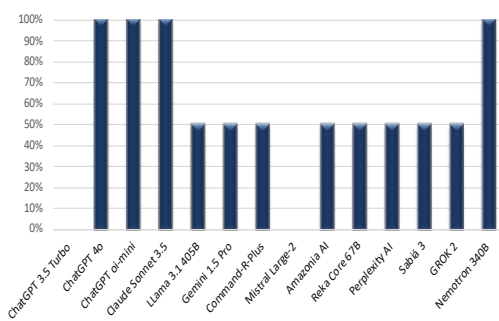

% Accuracy **(Criminal Procedure Law)** 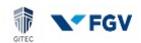

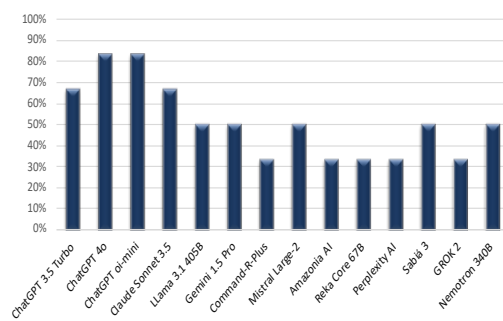

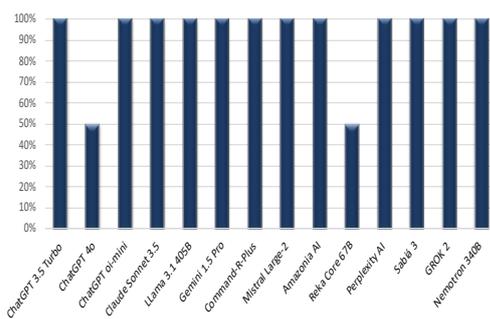

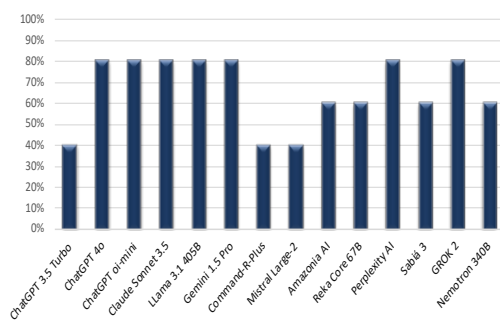

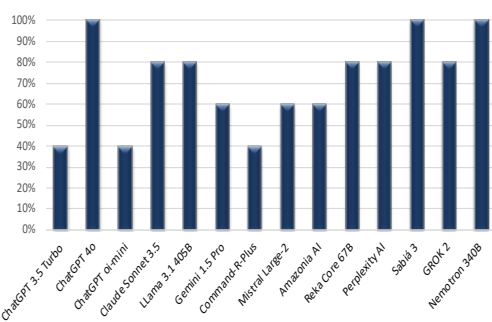

**Graphical Analysis of the Performance of Artificial Intelligence Models Tested in the 36[th] OAB Exam**

| Model | Total Correct | Total Incorrect | % Accuracy (Total) | % Accuracy (Easy Questions) | % Accuracy (Difficult Questions) |
|---|---|---|---|---|---|
| **ChatGPT 3.5 Turbo** | 41 | 39 | 51% | 51% | 54% |
| **ChatGPT 4o** | 67 | 13 | 84% | 87% | 69% |
| **ChatGPT oi-mini** | 49 | 31 | 61% | 64% | 46% |
| **Claude Sonnet 3.5** | 75 | 5 | 94% | 96% | 85% |
| **LLama 3.1 405B** | 66 | 14 | 83% | 84% | 77% |
| **Gemini 1.5 Pro** | 70 | 10 | 88% | 90% | 77% |
| **Command-R-Plus** | 51 | 29 | 64% | 64% | 62% |
| **Mistral Large-2** | 57 | 23 | 71% | 72% | 69% |
| **Amazonia AI** | 63 | 17 | 79% | 81% | 69% |
| **Reka Core 67B** | 52 | 28 | 65% | 64% | 69% |
| **Perplexity AI** | 53 | 27 | 66% | 64% | 77% |
| **Sabiá 3** | 67 | 13 | 84% | 85% | 77% |
| **GROK 2** | 67 | 13 | 84% | 88% | 62% |
| **Nemotron 340B** | 66 | 14 | 83% | 81% | 92% |

| Modelo | % Acerto Total | % Acerto Facil | % Acerto Difícil |
|---|---|---|---|
| ChatGPT 3.5 Turbo | 51% | 51% | 54% |
| ChatGPT 4o | 84% | 87% | 69% |
| ChatGPT oi-mini | 61% | 64% | 46% |
| Claude Sonnet 3.5 | 94% | 96% | 85% |
| LLama 3.1 405B | 83% | 84% | 77% |
| Gemini 1.5 Pro | 88% | 90% | 77% |
| Command-R-Plus | 64% | 64% | 62% |
| Mistral Large-2 | 71% | 72% | 69% |
| Amazonia AI | 79% | 81% | 69% |
| Reka Core 67B | 65% | 64% | 69% |
| Perplexity AI | 66% | 64% | 77% |
| Sabiá 3 | 84% | 85% | 77% |
| GROK 2 | 84% | 88% | 62% |
| Nemotron 340B | 83% | 81% | 92% |

**% Acerto Prova OAB 36**

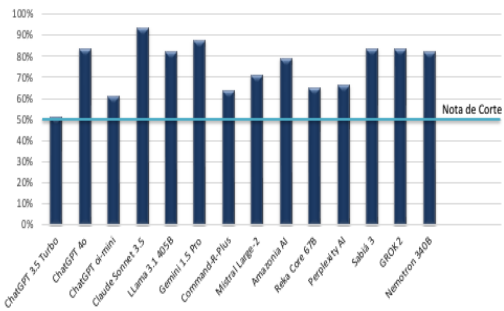

**% Acerto (Questões Difíceis)**

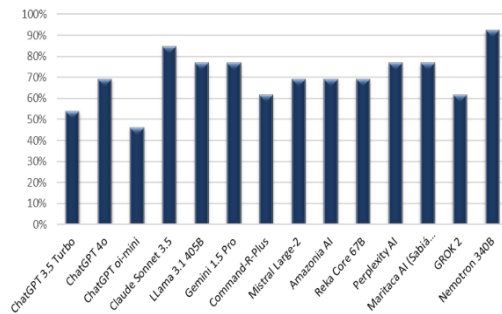

**% Acerto (Questões Fáceis)**

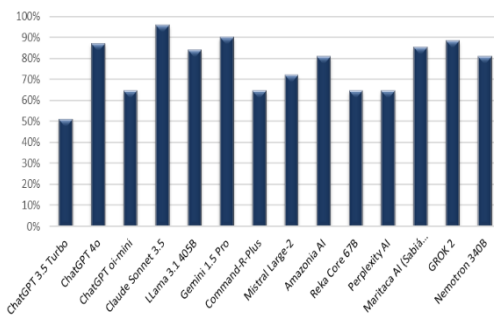

**% Acerto (Ética)**

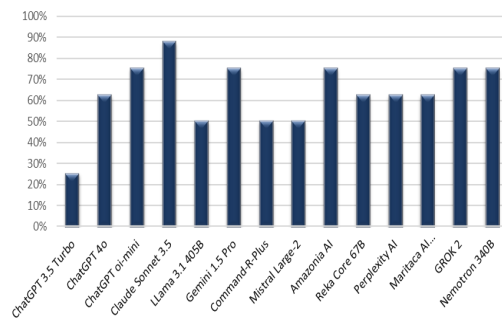

## % Acerto (Filosofia do Direito)

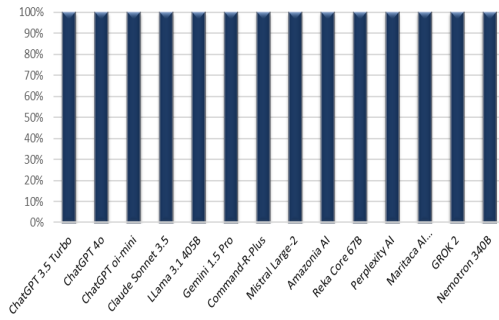

## % Acerto (Direitos Internacional)

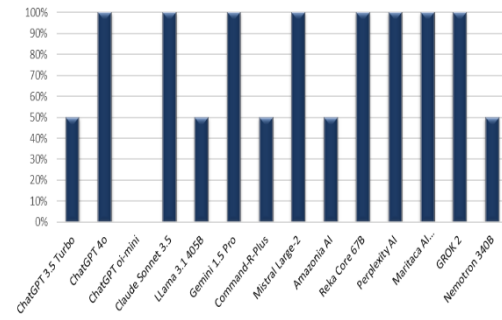

## % Acerto (Direito Constitucional)

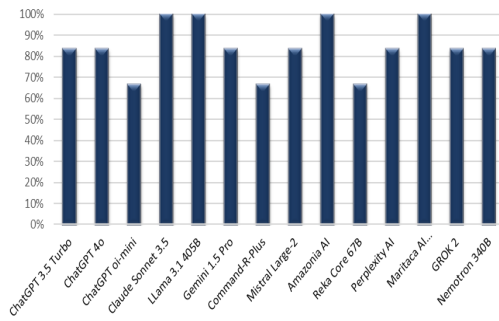

## % Acerto (Direito Financeiro)

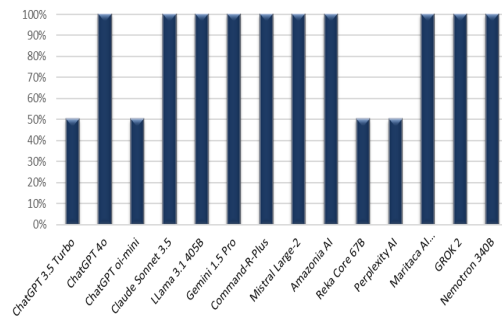

## % Acerto (Direitos Humanos)

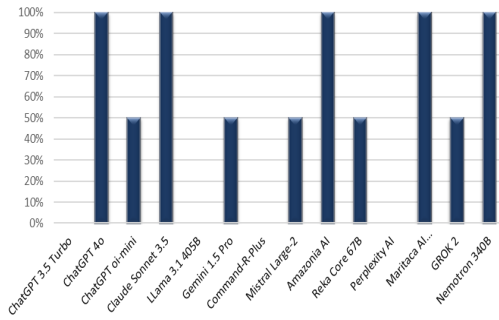

## % Acerto (Direito Tributário)

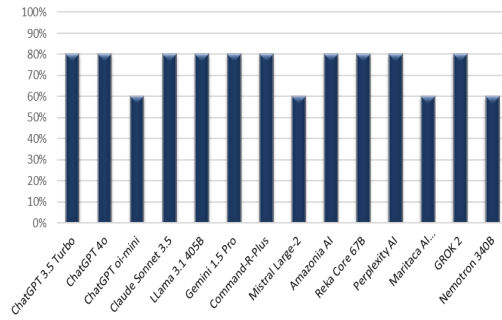

## % Acerto (Direitos Eleitoral)

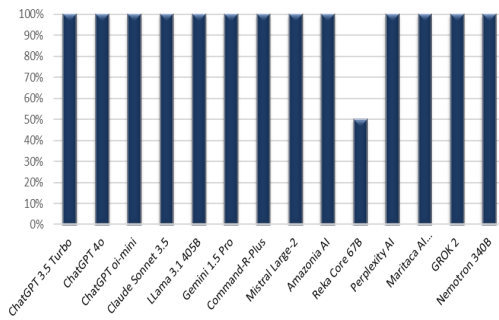

## % Acerto (Direito Administrativo)

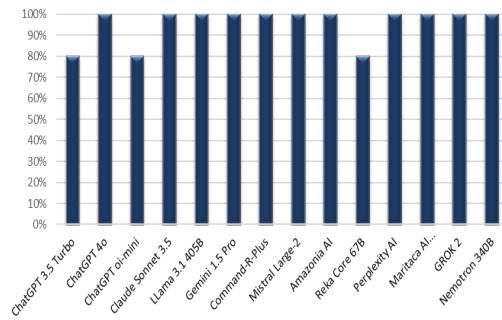

## % Acerto **(Direito Ambiental)**

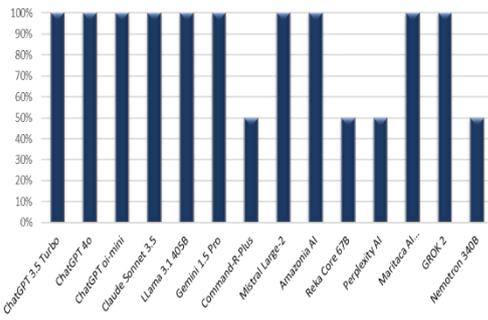

## % Acerto **(Direito Empresarial)**

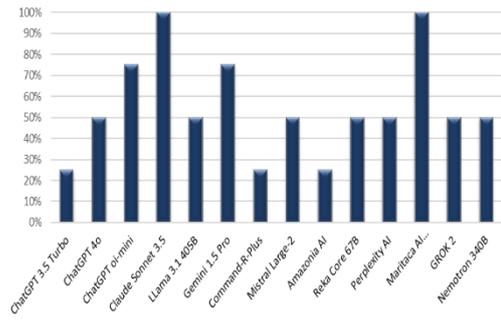

## % Acerto **(Direito Civil)**

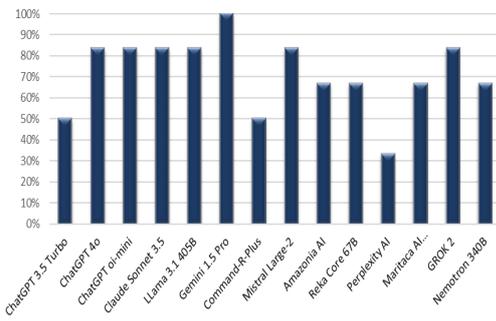

## % Acerto **(Direito Processual Civil)**

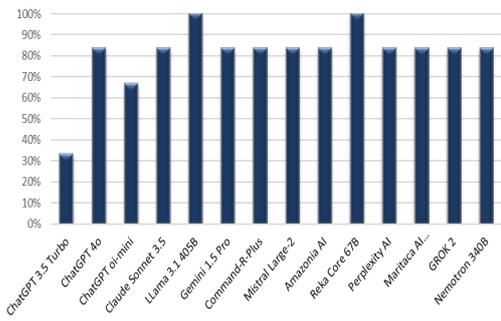

## % Acerto **(Direito da Criança e do Adolescente)**

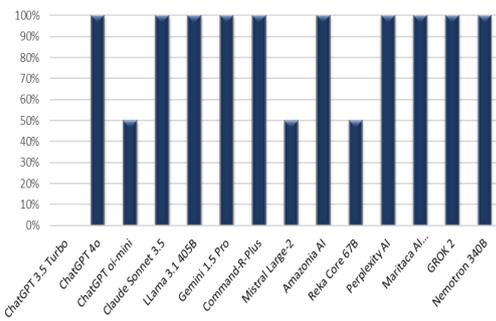

## % Acerto **(Direito Penal)**

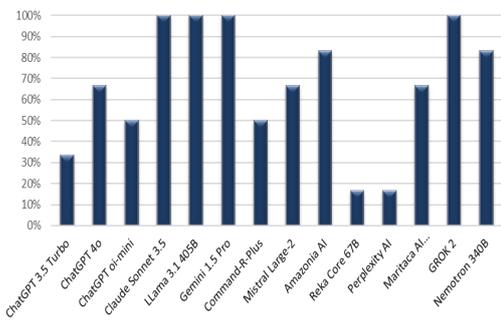

## % Acerto **(Direito do Consumidor)**

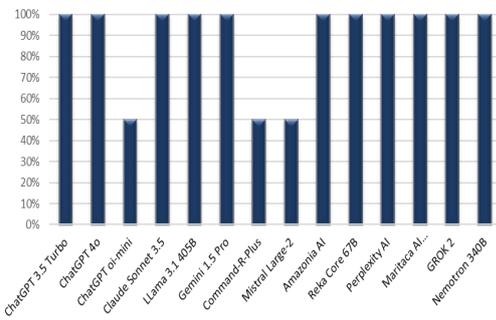

## % Acerto **(Direito Processual Penal)**

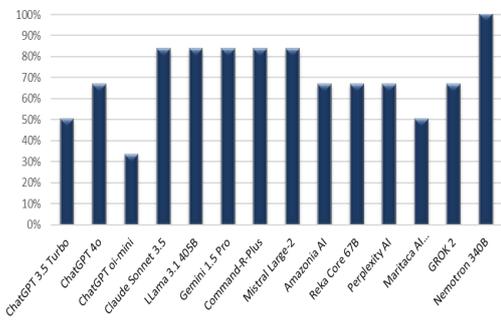

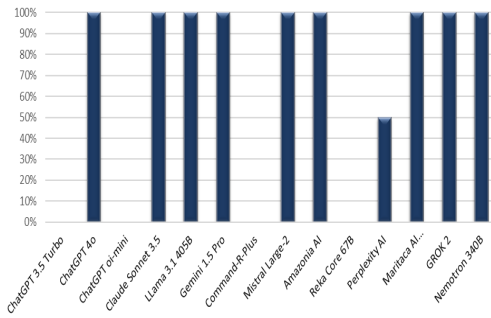

% Acerto (Direito Previdenciário)

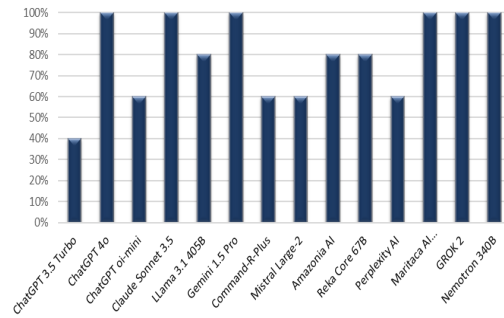

% Acerto (Direito Processual do Trabalho)

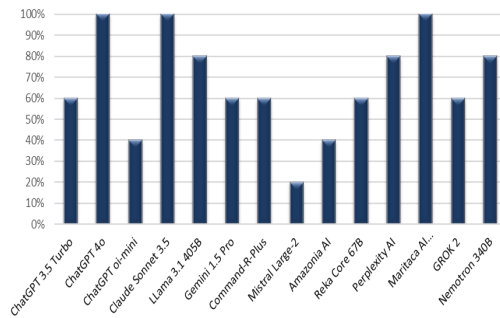

% Acerto (Direito do Trabalho)

**Graphical Analysis of the Performance of Artificial Intelligence Models Tested in the 35th OAB Exam**

| Model | Total Correct | Total Incorrect | % Accuracy (Total) | % Accuracy (Easy Questions) | % Accuracy (Difficult Questions) |
|---|---|---|---|---|---|
| **ChatGPT 3.5 Turbo** | 43 | 37 | 54% | 56% | 40% |
| **ChatGPT 4o** | 63 | 17 | 79% | 84% | 40% |
| **ChatGPT oi-mini** | 43 | 37 | 54% | 54% | 50% |
| **Claude Sonnet 3.5** | 70 | 10 | 88% | 91% | 60% |
| **LLama 3.1 405B** | 61 | 19 | 76% | 79% | 60% |
| **Gemini 1.5 Pro** | 63 | 17 | 79% | 86% | 30% |
| **Command-R-Plus** | 54 | 26 | 68% | 70% | 50% |
| **Mistral Large-2** | 54 | 26 | 68% | 70% | 50% |
| **Amazonia AI** | 64 | 16 | 80% | 83% | 60% |
| **Reka Core 67B** | 50 | 30 | 63% | 66% | 40% |
| **Perplexity AI** | 53 | 27 | 66% | 70% | 40% |
| **Sabiá 3** | 64 | 16 | 80% | 83% | 60% |
| **GROK 2** | 43 | 37 | 54% | 57% | 30% |
| **Nemotron 340B** | 64 | 16 | 80% | 86% | 40% |

| Model | % Accuracy Total | % Accuracy Easy | % Accuracy Difficult |
|---|---|---|---|
| ChatGPT 3.5 Turbo | 54% | 56% | 40% |
| ChatGPT 4o | 79% | 84% | 40% |
| ChatGPT oi-mini | 54% | 54% | 50% |
| Claude Sonnet 3.5 | 88% | 91% | 60% |
| LLama 3.1 405B | 76% | 79% | 60% |
| Gemini 1.5 Pro | 79% | 86% | 30% |
| Command-R-Plus | 68% | 70% | 50% |
| Mistral Large-2 | 68% | 70% | 50% |
| Amazonia AI | 80% | 83% | 60% |
| Reka Core 67B | 63% | 66% | 40% |
| Perplexity AI | 66% | 70% | 40% |
| Sabiá 3 | 80% | 83% | 60% |
| GROK 2 | 54% | 57% | 30% |
| Nemotron 340B | 80% | 86% | 40% |

0

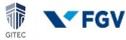

% Accuracy ExamOAB 35

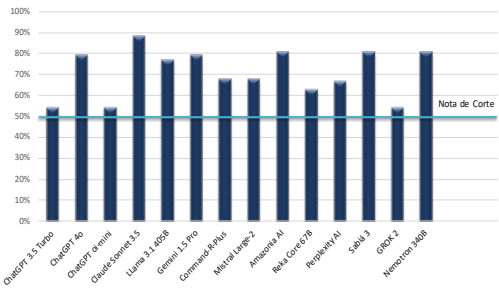

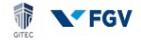

% Accuracy (Difficult Questions)

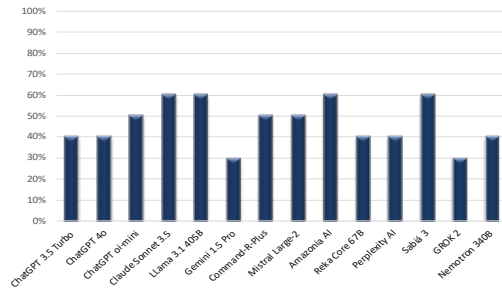

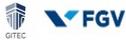

% Accuracy (Easy Questions)

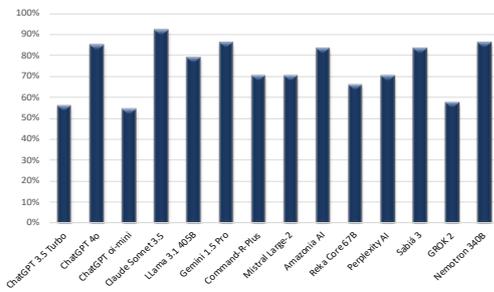

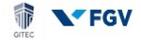

% Accuracy (Ethics)

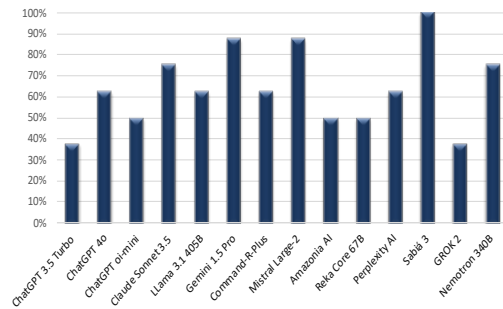

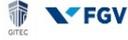

% Accuracy (**Philosophy of Law**)

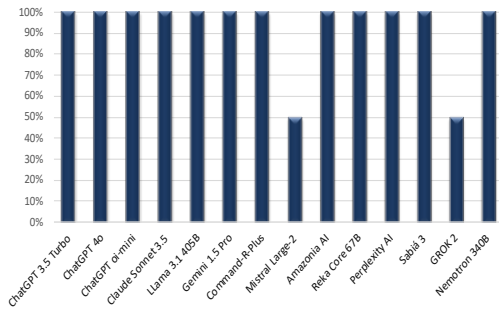

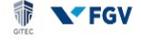

% Accuracy (**International Law**)

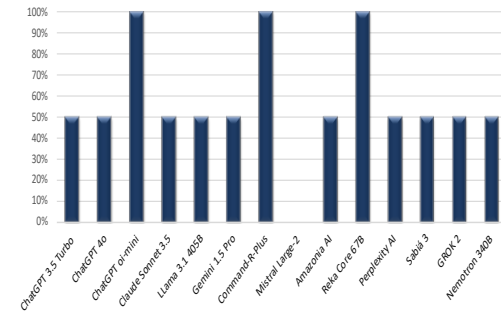

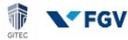

% Accuracy (**Constitutional Law**)

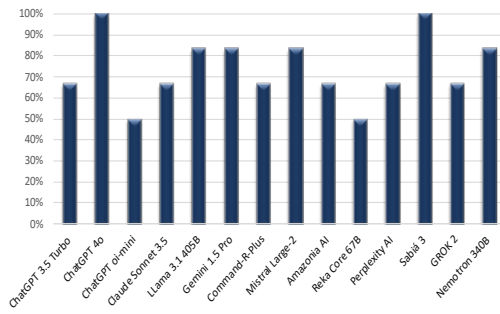

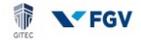

% Accuracy (**Financial Law**)

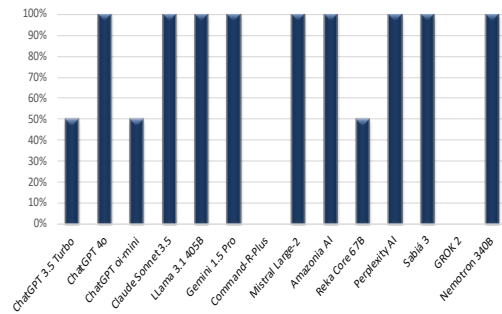

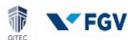

% Accuracy (**Human Rights**)

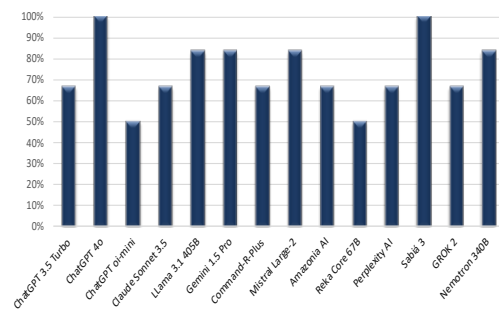

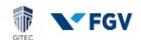

% Accuracy (**Tax Law**)

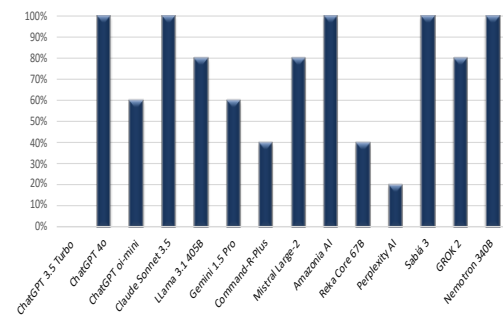

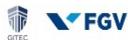

% Accuracy (**Electoral Law**)

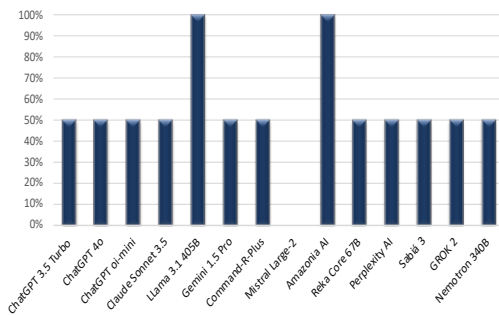

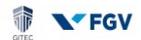

% Accuracy (**Administrative Law**)

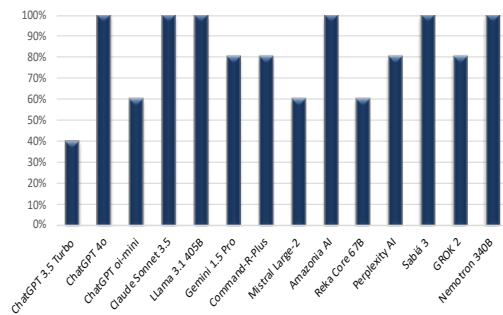

% Accuracy **(Environmental Law)** 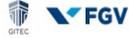

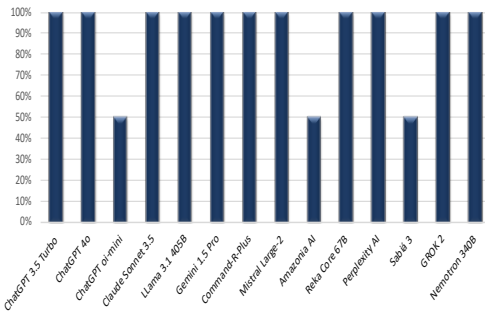

% Accuracy **(Business Law)** 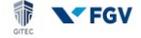

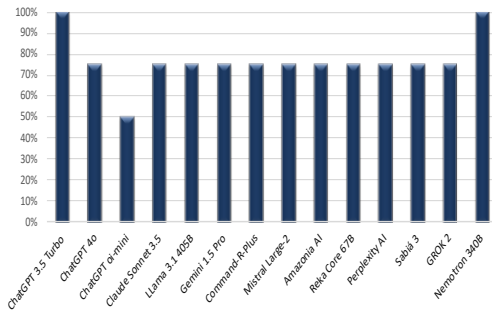

% Accuracy **(Civil Law)** 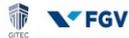

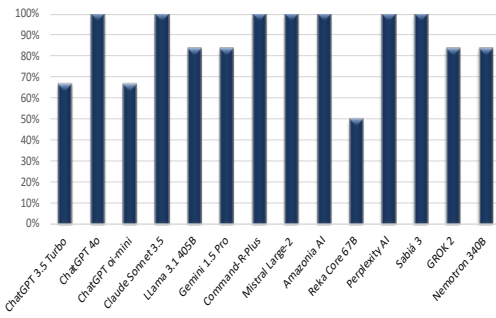

% Accuracy **(Civil Procedure Law)** 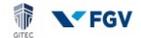

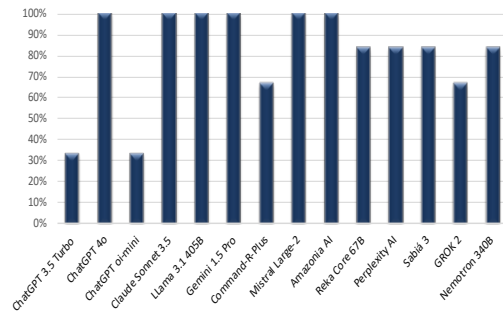

% Accuracy **(Child and Adolescent Law)** 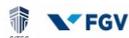

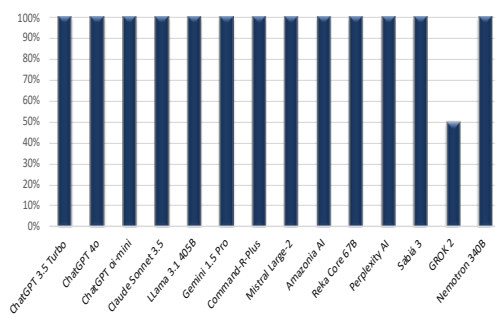

% Accuracy **(Criminal Law)** 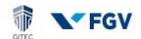

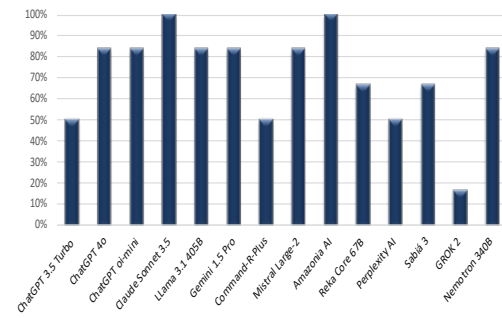

% Accuracy **(Consumer law)** 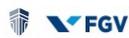

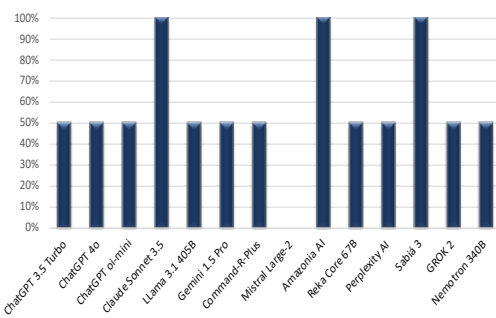

% Accuracy **(Criminal Procedure Law)** 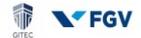

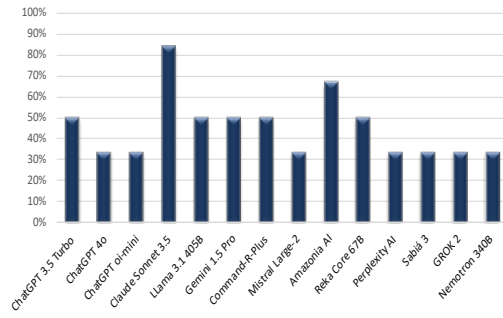

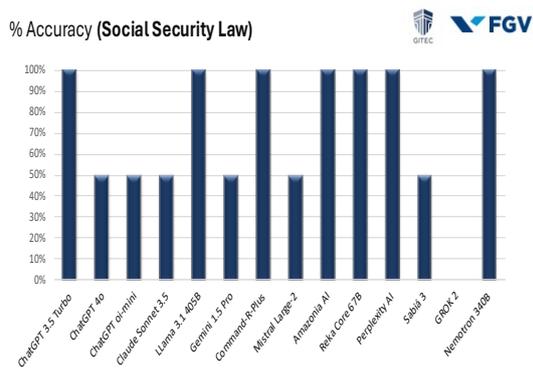

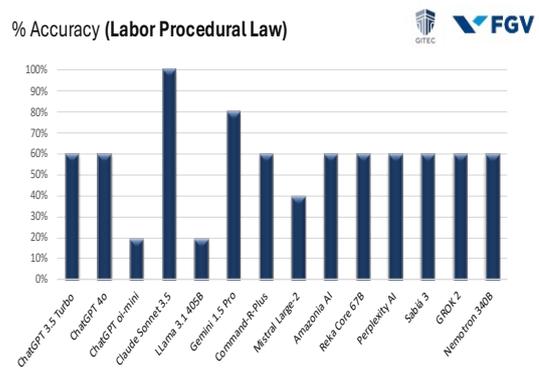

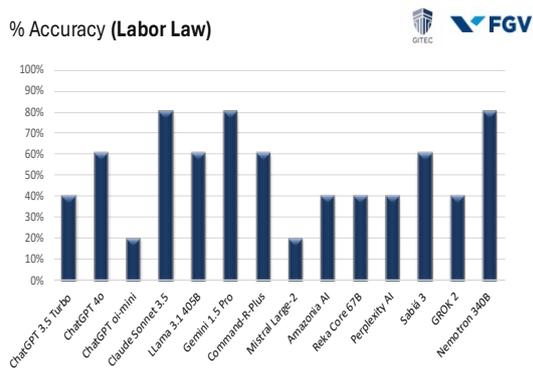

**Graphical** Analysis **of the Performance of Artificial Intelligence Models Tested in the 34th OAB Exam**

| Model | Total Correct | Total Incorrect | % Accuracy (Total) | % Accuracy (Easy Questions) | % Accuracy (Difficult Questions) |
|---|---|---|---|---|---|
| **ChatGPT 3.5 Turbo** | 41 | 39 | 51% | 54% | 38% |
| **ChatGPT 4o** | 71 | 9 | 89% | 88% | 92% |
| **ChatGPT oi-mini** | 41 | 39 | 51% | 52% | 46% |
| **Claude Sonnet 3.5** | 67 | 13 | 84% | 85% | 77% |
| **LLama 3.1 405B** | 59 | 21 | 74% | 76% | 62% |
| **Gemini 1.5 Pro** | 66 | 14 | 83% | 84% | 77% |
| **Command-R-Plus** | 49 | 31 | 61% | 64% | 46% |
| **Mistral Large-2** | 57 | 23 | 71% | 73% | 62% |
| **Amazonia AI** | 61 | 19 | 76% | 79% | 62% |
| **Reka Core 67B** | 37 | 43 | 46% | 52% | 15% |
| **Perplexity AI** | 58 | 22 | 73% | 78% | 46% |
| **Sabiá 3** | 66 | 14 | 83% | 87% | 62% |
| **GROK 2** | 60 | 20 | 75% | 79% | 54% |
| **Nemotron 340B** | 59 | 21 | 74% | 79% | 46% |

| Model | % Accuracy Total | % Accuracy Easy | % Accuracy Difficult |
|---|---|---|---|
| ChatGPT 3.5 Turbo | 51% | 54% | 38% |
| ChatGPT 4o | 89% | 88% | 92% |
| ChatGPT oi-mini | 51% | 52% | 46% |
| Claude Sonnet 3.5 | 84% | 85% | 77% |
| LLama 3.1 405B | 74% | 76% | 62% |
| Gemini 1.5 Pro | 83% | 84% | 77% |
| Command-R-Plus | 61% | 64% | 46% |
| Mistral Large-2 | 71% | 73% | 62% |
| Amazonia AI | 76% | 79% | 62% |
| Reka Core 67B | 46% | 52% | 15% |
| Perplexity AI | 73% | 78% | 46% |
| Sabiá 3 | 83% | 87% | 62% |
| GROK 2 | 75% | 79% | 54% |
| Nemotron 340B | 74% | 79% | 46% |

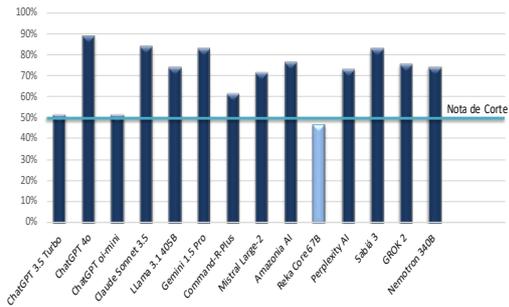

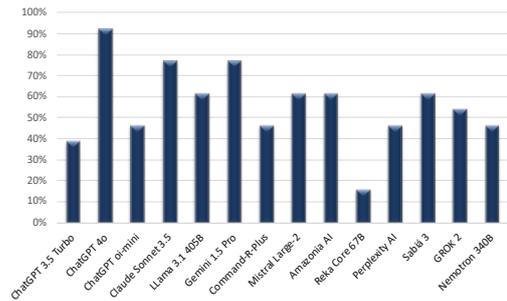

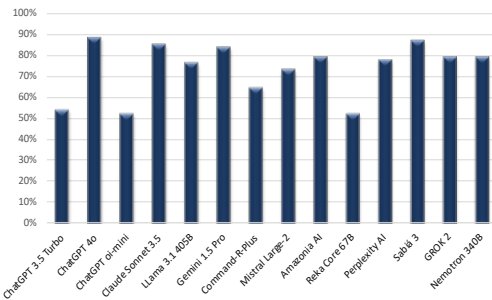

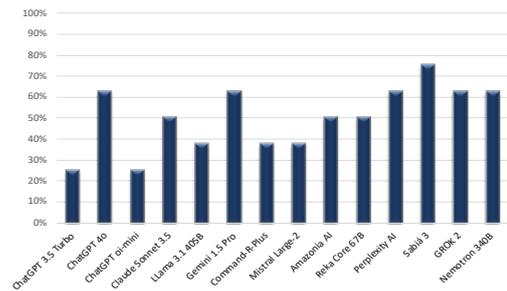

**% Accuracy (Philosophy of Law)**

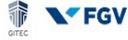
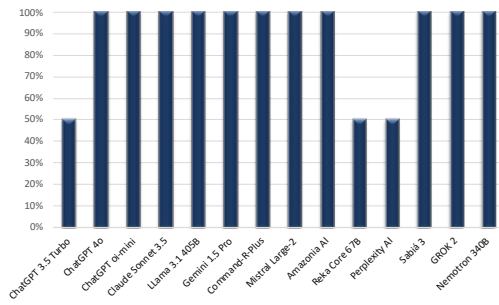

**% Accuracy (International Law)**

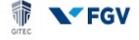
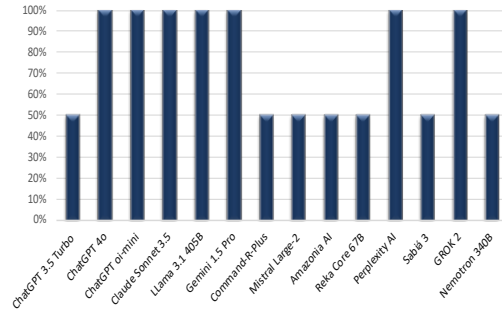

**% Accuracy (Constitutional Law)**

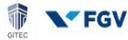
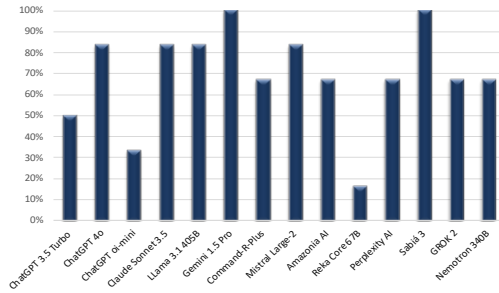

**% Accuracy (Financial Law)**

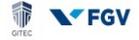
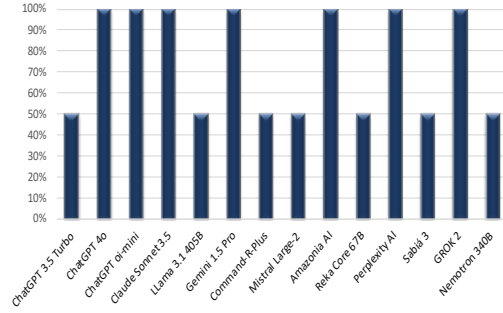

**% Accuracy (Human Rights)**

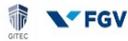
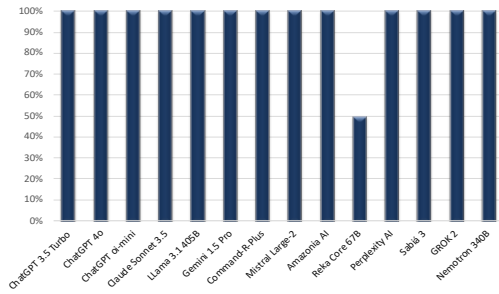

**% Accuracy (Tax Law)**

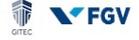
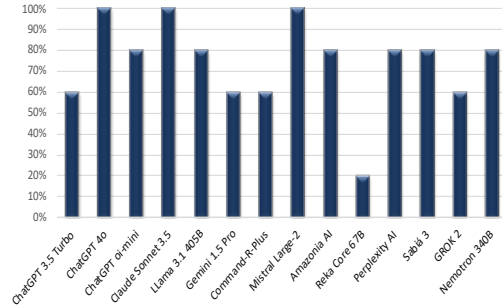

**% Accuracy (Electoral Law)**

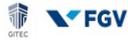
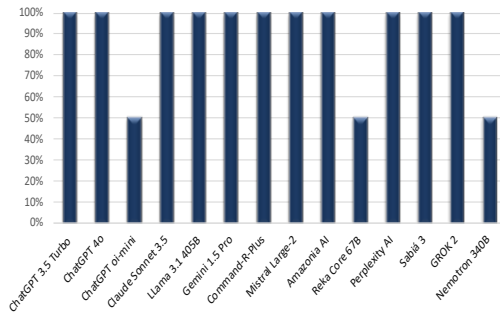

**% Accuracy (Administrative Law)**

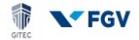
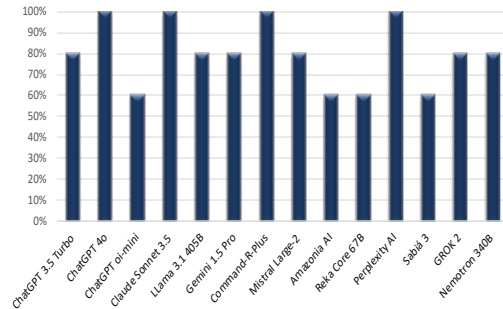

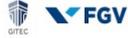
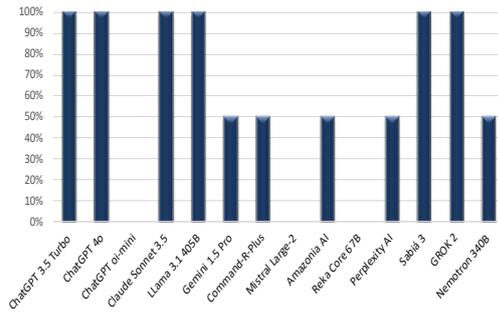

% Accuracy **(Environmental Law)**

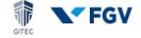
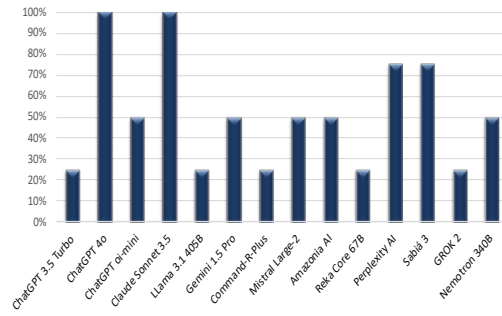

% Accuracy **(Business Law)**

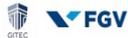
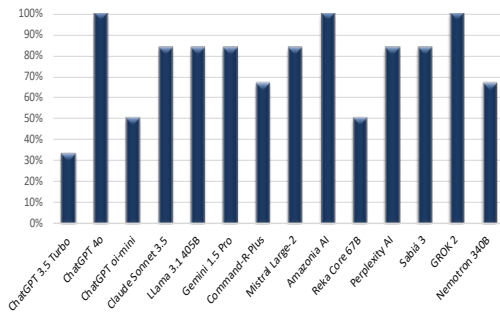

% Accuracy **(Civil Law)**

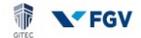
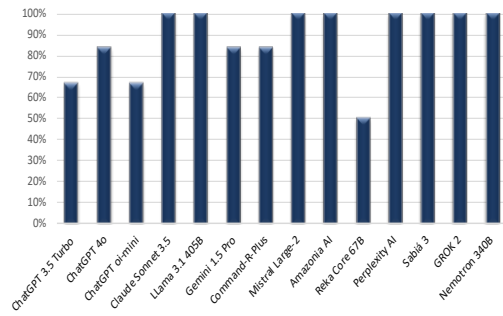

% Accuracy **(Civil Procedure Law)**

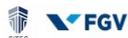
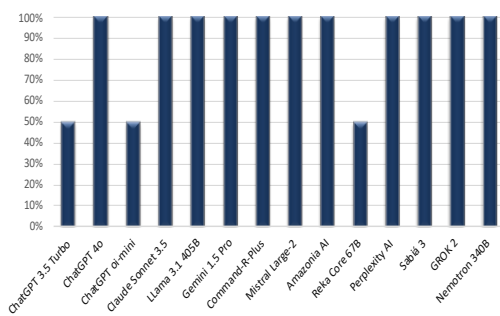

% Accuracy **(Child and Adolescent Law)**

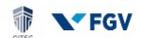
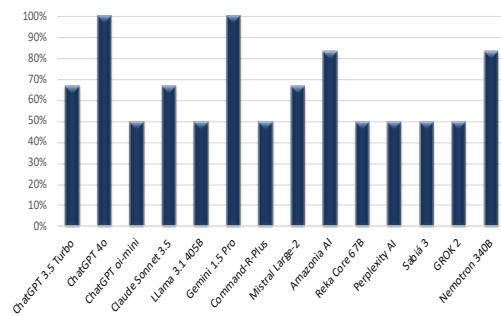

% Accuracy **(Criminal Law)**

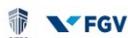
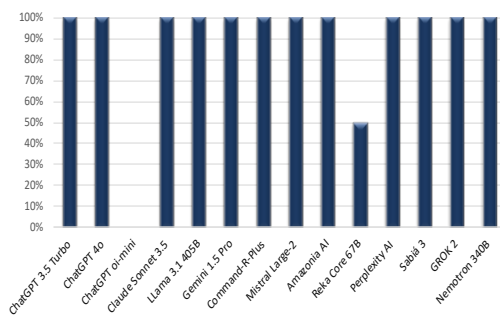

% Accuracy **(Consumer law)**

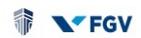
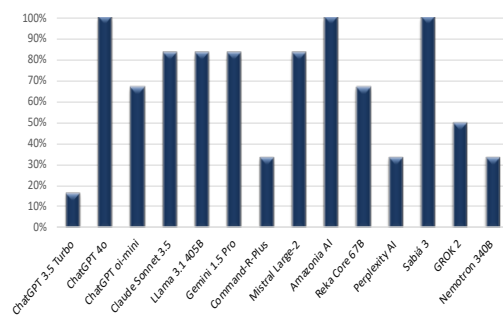

% Accuracy **(Criminal Procedure Law)**

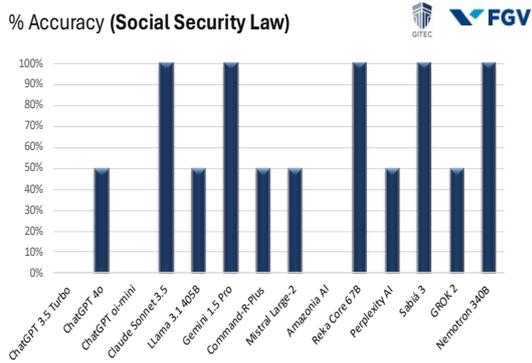

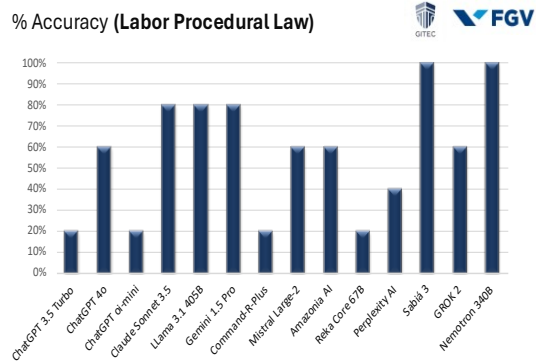

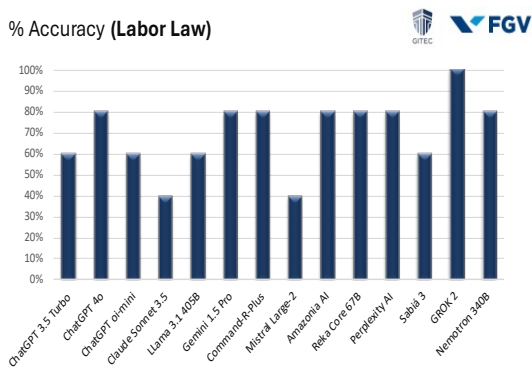

**Graphical Analysis of the Performance of Artificial Intelligence Models Tested in the 33rd OAB Exam**

| Model | Total Correct | Total Incorrect | % Accuracy (Total) | % Accuracy (Easy Questions) | % Accuracy (Difficult Questions) |
|---|---|---|---|---|---|
| ChatGPT 3.5 Turbo | 47 | 33 | 59% | 61% | 42% |
| ChatGPT 4o | 64 | 16 | 80% | 82% | 67% |
| ChatGPT oi-mini | 43 | 37 | 54% | 61% | 8% |
| Claude Sonnet 3.5 | 64 | 16 | 80% | 84% | 58% |
| LLama 3.1 405B | 52 | 28 | 65% | 67% | 50% |
| Gemini 1.5 Pro | 58 | 22 | 73% | 78% | 42% |
| Command-R-Plus | 45 | 35 | 56% | 63% | 17% |
| Mistral Large-2 | 52 | 28 | 65% | 69% | 42% |
| Amazonia AI | 56 | 24 | 70% | 69% | 75% |
| Reka Core 67B | 40 | 40 | 50% | 51% | 42% |
| Perplexity AI | 53 | 27 | 66% | 69% | 50% |
| Sabiá 3 | 58 | 22 | 73% | 75% | 58% |
| GROK 2 | 58 | 22 | 73% | 75% | 58% |
| Nemotron 340B | 57 | 23 | 71% | 73% | 58% |

| Model | % Accuracy Total | % Acurracy Easy | % Accuracy Difficult | 0 |
|---|---|---|---|---|
| ChatGPT 3.5 Turbo | 59% | 61% | 42% | |
| ChatGPT 4o | 80% | 82% | 67% | |
| ChatGPT oi-mini | 54% | 61% | 8% | |
| Claude Sonnet 3.5 | 80% | 84% | 58% | |
| LLama 3.1 405B | 65% | 67% | 50% | |
| Gemini 1.5 Pro | 73% | 78% | 42% | |
| Command-R-Plus | 56% | 63% | 17% | |
| Mistral Large-2 | 65% | 69% | 42% | |
| Amazonia AI | 70% | 69% | 75% | |
| Reka Core 67B | 50% | 51% | 42% | |
| Perplexity AI | 66% | 69% | 50% | |
| Sabiá 3 | 73% | 75% | 58% | |
| GROK 2 | 73% | 75% | 58% | |
| Nemotron 340B | 71% | 73% | 58% | 100 |

% Accuracy Exam OAB 33 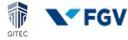

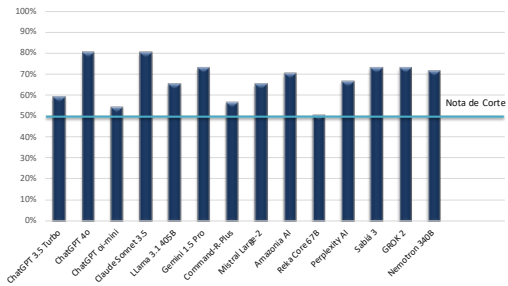

% Accuracy (Difficult Questions) 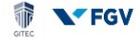

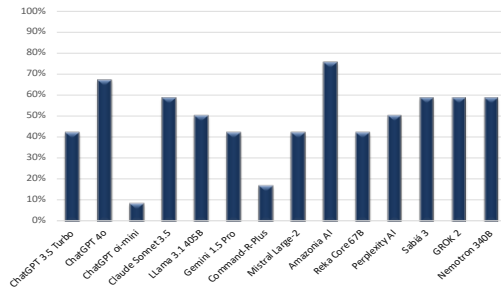

% Accuracy (Easy Questions) 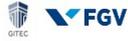

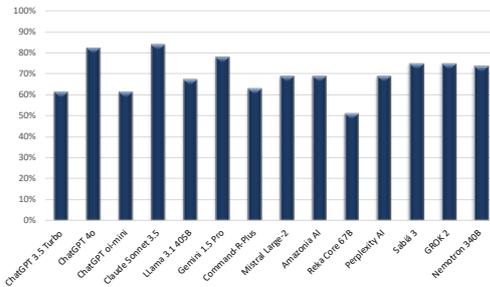

% Accuracy (Ethics) 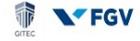

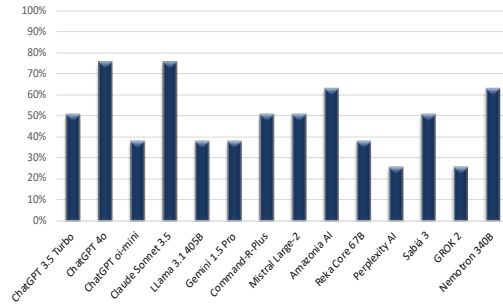

% Accuracy (Philosophy of Law)
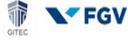
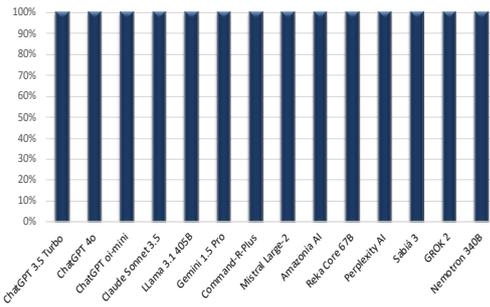

% Accuracy (International Law)
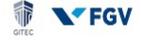
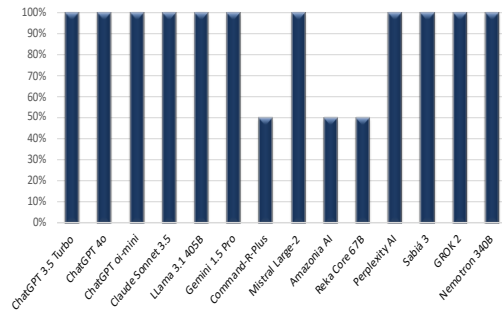

% Accuracy (Constitutional Law)
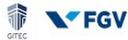
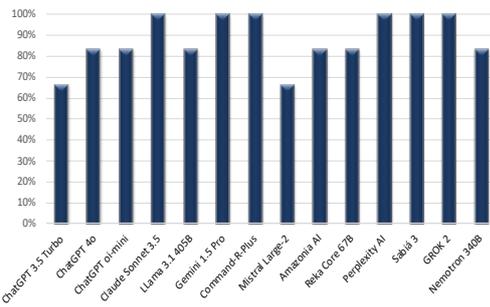

% Accuracy (Financial Law)
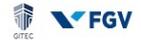
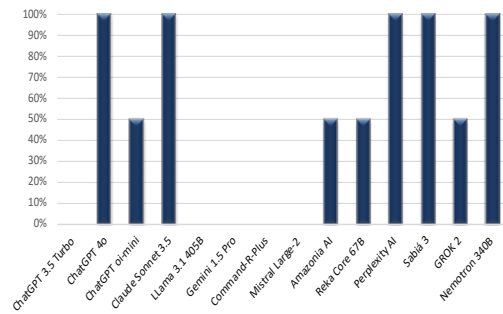

% Accuracy (Human Rights)
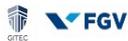
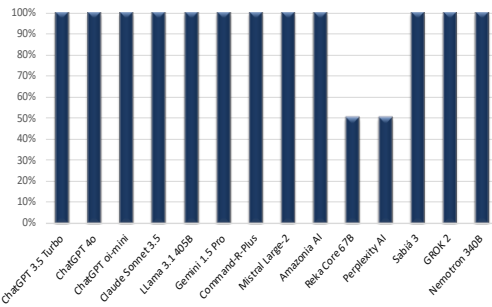

% Accuracy (Administrative Law)
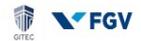
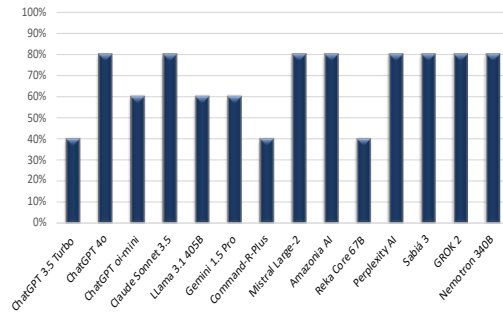

% Accuracy (Electoral Law)
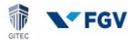
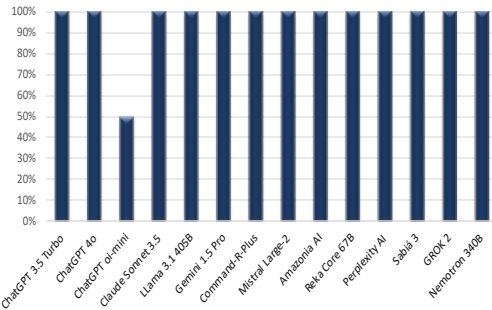

% Accuracy (Environmental Law)
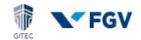
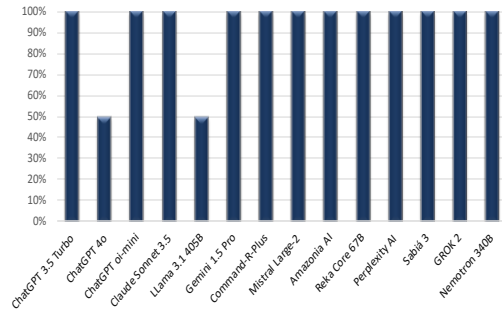

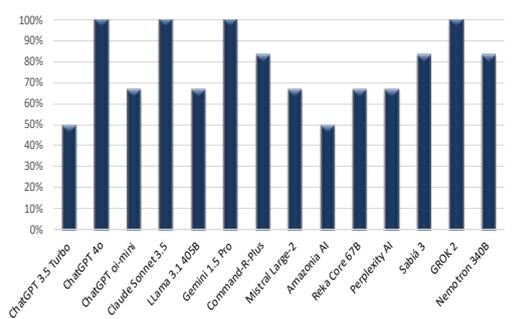

% Accuracy **(Civil Law)**

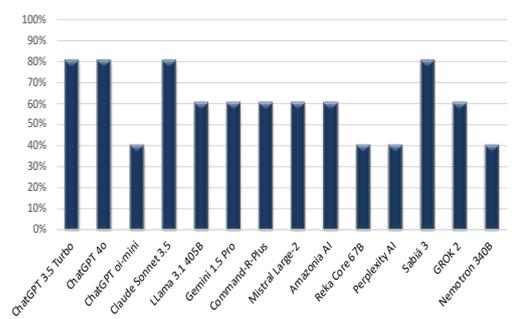

% Accuracy **(Tax Law)**

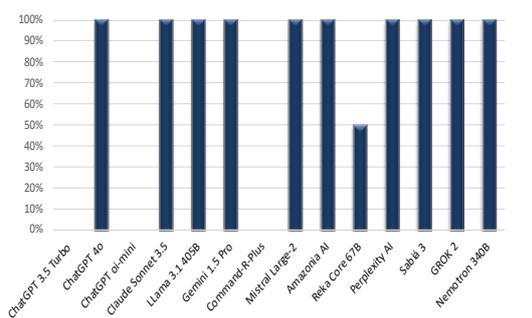

% Accuracy **(Consumer law)**

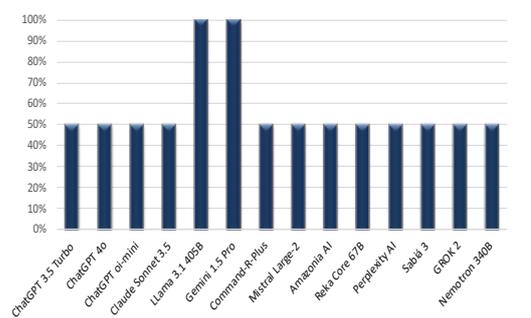

% Accuracy **(Child and Adolescent Law)**

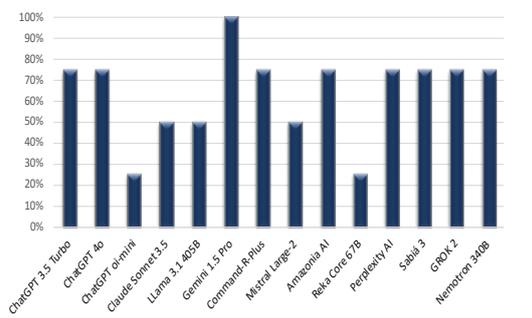

% Accuracy **(Business Law)**

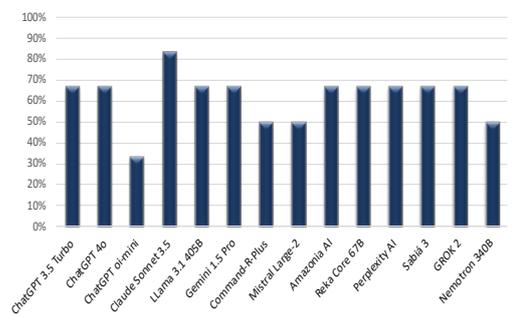

% Accuracy **(Criminal Law)**

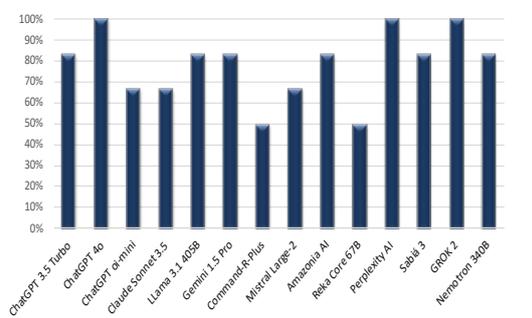

% Accuracy **(Civil Procedure Law)**

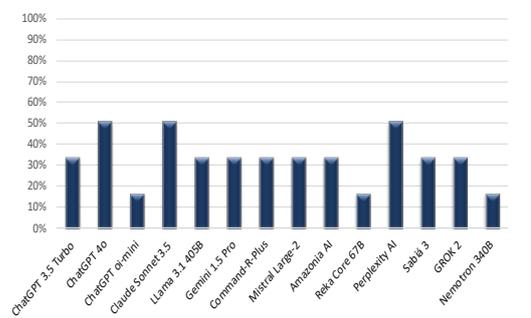

% Accuracy **(Criminal Procedure Law)**

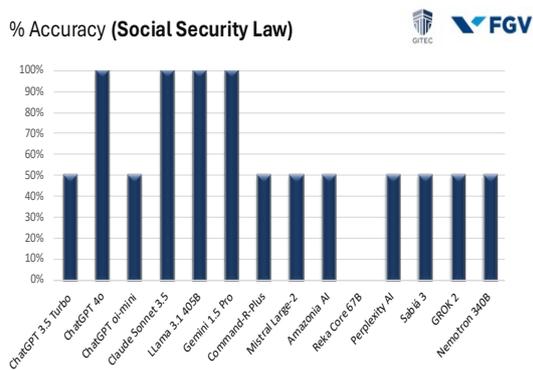

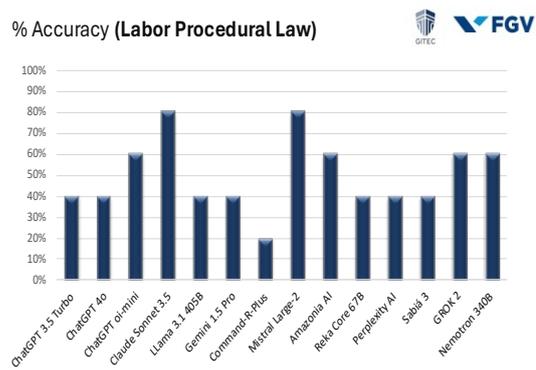

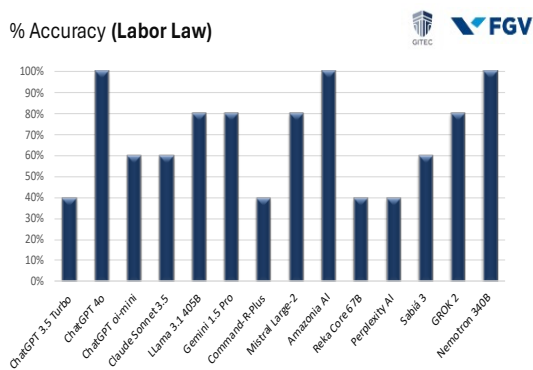

**Graphical Analysis of the Performance of Artificial Intelligence Models Tested in the 32[nd] OAB Exam**

| Model | Total Correct | Total Incorrect | % Accuracy (Total) | % Accuracy (Easy Questions) | % Accuracy (Difficult Questions) |
|---|---|---|---|---|---|
| ChatGPT 3.5 Turbo | 41 | 39 | 51% | 55% | 42% |
| ChatGPT 4o | 62 | 18 | 78% | 82% | 67% |
| ChatGPT oi-mini | 47 | 33 | 59% | 64% | 46% |
| Claude Sonnet 3.5 | 63 | 17 | 79% | 80% | 75% |
| LLama 3.1 405B | 49 | 31 | 61% | 63% | 58% |
| Gemini 1.5 Pro | 60 | 20 | 75% | 79% | 67% |
| Command-R-Plus | 43 | 37 | 54% | 61% | 38% |
| Mistral Large-2 | 48 | 32 | 60% | 70% | 38% |
| Amazonia AI | 59 | 21 | 74% | 79% | 63% |
| Reka Core 67B | 48 | 32 | 60% | 63% | 54% |
| Perplexity AI | 51 | 29 | 64% | 71% | 46% |
| Sabiá 3 | 37 | 43 | 46% | 45% | 50% |
| GROK 2 | 51 | 29 | 64% | 73% | 42% |
| Nemotron 340B | 52 | 28 | 65% | 71% | 50% |

| Model | % Accuracy Total | % Accuracy Easy | % Accuracy Difficult | 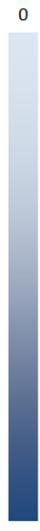 |
|---|---|---|---|---|
| ChatGPT 3.5 Turbo | 51% | 55% | 42% | |
| ChatGPT 4o | 78% | 82% | 67% | |
| ChatGPT oi-mini | 59% | 64% | 46% | |
| Claude Sonnet 3.5 | 79% | 80% | 75% | |
| LLama 3.1 405B | 61% | 63% | 58% | |
| Gemini 1.5 Pro | 75% | 79% | 67% | |
| Command-R-Plus | 54% | 61% | 38% | |
| Mistral Large-2 | 60% | 70% | 38% | |
| Amazonia AI | 74% | 79% | 63% | |
| Reka Core 67B | 60% | 63% | 54% | |
| Perplexity AI | 64% | 71% | 46% | |
| Sabiá 3 | 46% | 45% | 50% | |
| GROK 2 | 64% | 73% | 42% | |
| Nemotron 340B | 65% | 71% | 50% | |

% Accuracy Exam OAB 32 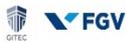

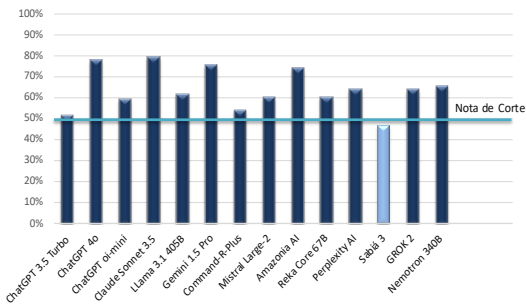

% Accuracy (Difficult Questions)

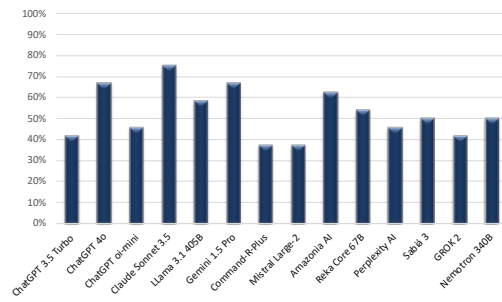

% Accuracy (Easy Questions)

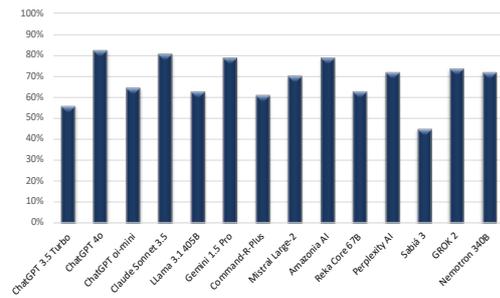

% Accuracy (Ethics)

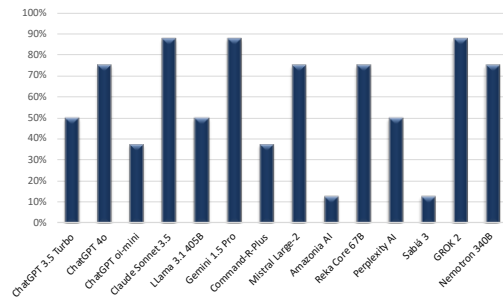

**% Accuracy (Philosophy of Law)**

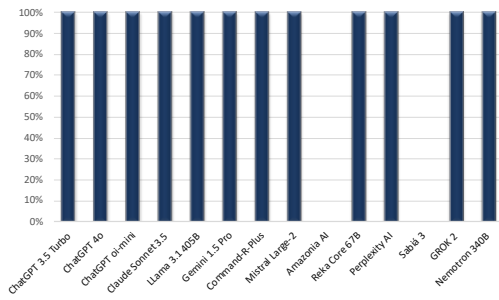

**% Accuracy (International Law)**

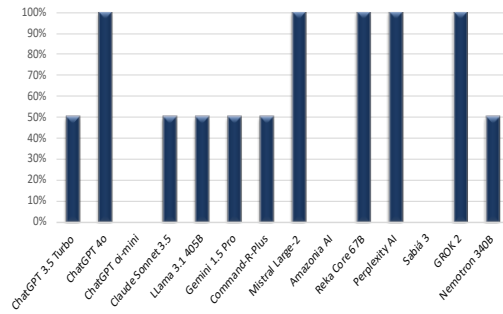

**% Accuracy (Constitutional Law)**

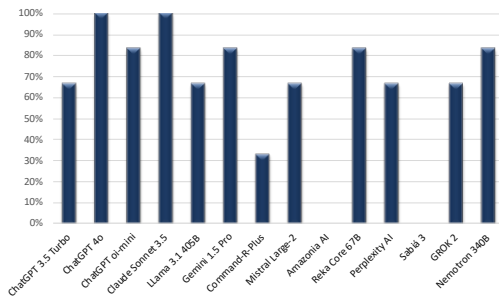

**% Accuracy (Financial Law)**

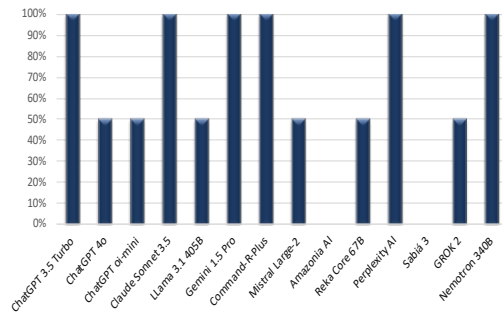

**% Accuracy (Human Rights)**

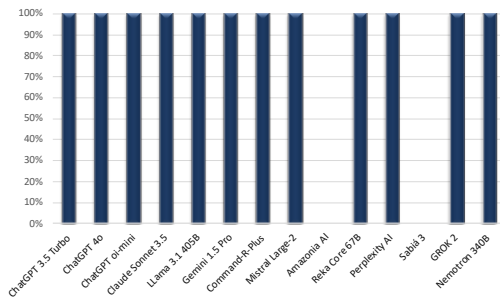

**% Accuracy (Tax Law)**

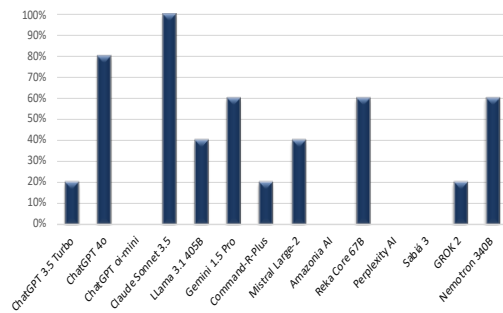

**% Accuracy (Electoral Law)**

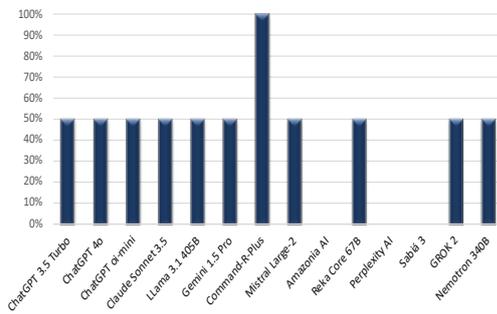

**% Accuracy (Administrative Law)**

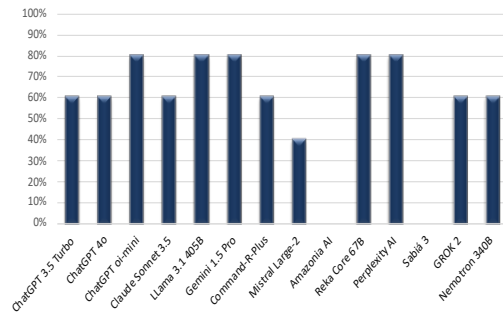

% Accuracy **(Environmental Law)**

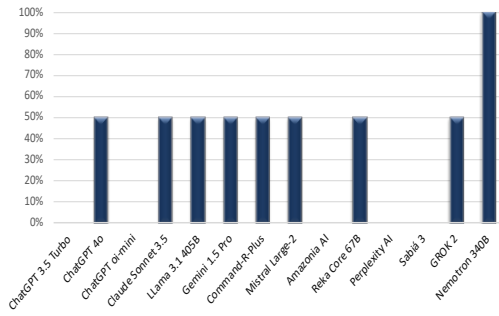

% Accuracy **(Business Law)**

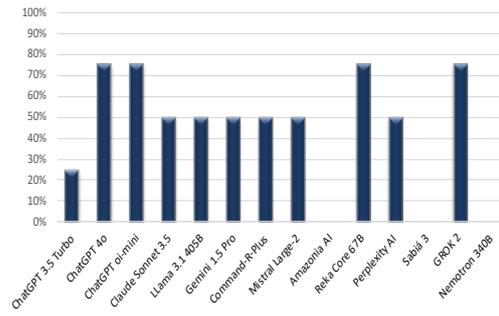

% Accuracy **(Civil Law)**

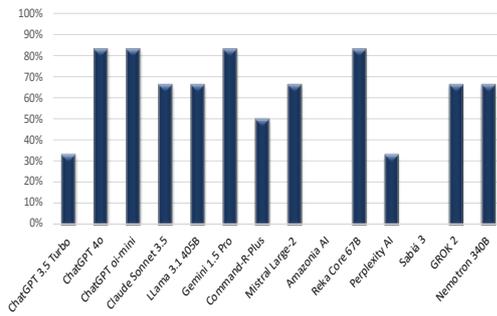

% Accuracy **(Civil Procedure Law)**

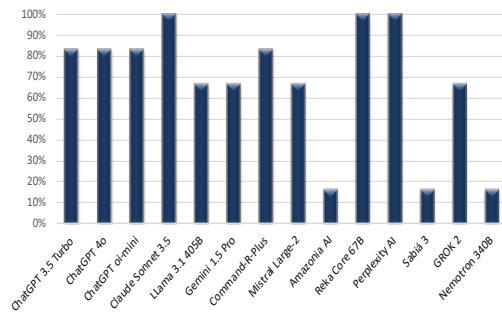

% Accuracy**(Child and Adolescent Law)**

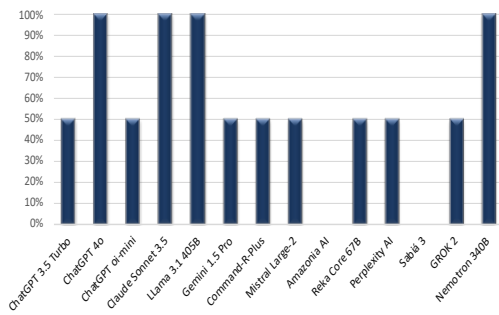

% Accuracy **(Criminal Law)**

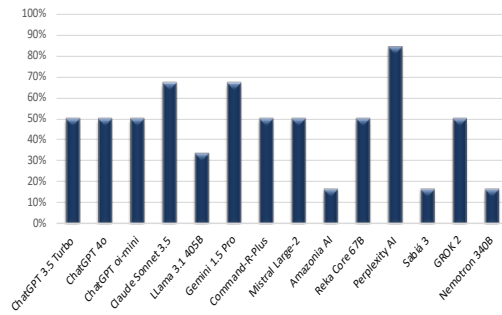

% Accuracy **(Consumer law)**

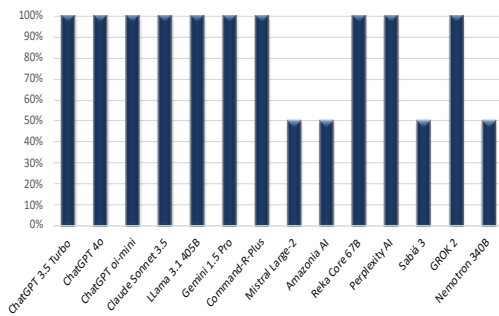

% Accuracy **(Criminal Procedure Law)**

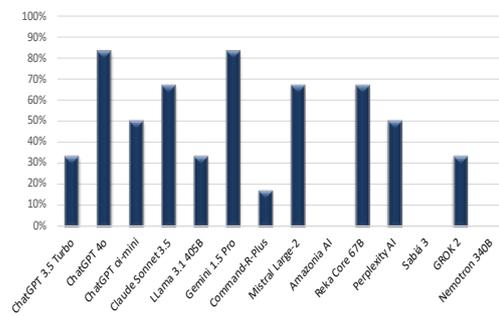

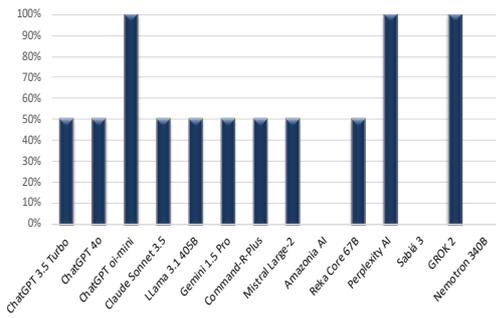

% Accuracy (**Social Security Law**)

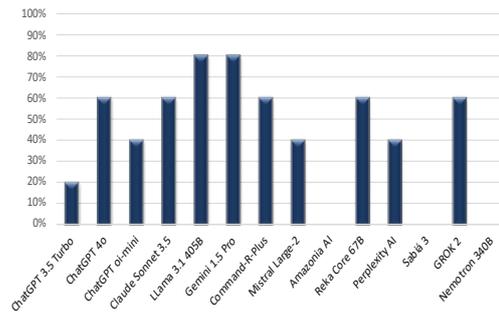

% Accuracy (**Labor Procedural Law**)

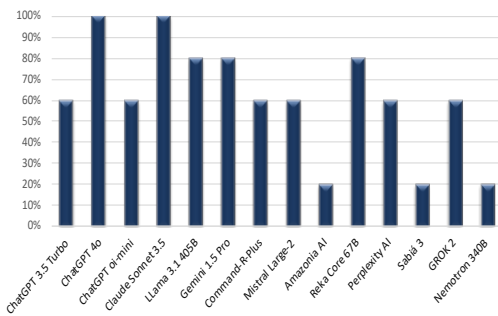

% Accuracy (**Labor Law**)

## Selection Process for Federal Labor Prosecutors

The selection process for Federal Labor Prosecutors, currently linked to the Office of the Attorney General of Brazil (AGU), is a public examination for entry into the career of Labor Prosecutor, a position within the Labor Public Prosecutor's Office (MPT). This career focuses on defending labor rights as established in the Federal Constitution and Brazilian labor laws.

Labor Prosecutors primarily work to oversee and protect collective and individual labor rights. To join this career, candidates must meet the following requirements: hold a law degree and demonstrate at least three years of legal experience after completing their law degree.

Since 2022, the selection process for Labor Prosecutors has been unified, with the implementation of the 1st Unified National Public Selection Process for Entry into the Career of the Labor Public Prosecutor's Office (MPT), centralizing the selection process to ensure uniformity and efficiency (Edital nº 01, de 09 de agosto de 2022, 2022)).

Beyond its attractive compensation, the career of a Labor Prosecutor is highly prestigious due to the significance of its role in overseeing and protecting collective and individual labor rights. Key focus areas include combating slave labor and child labor, promoting decent working conditions, defending union freedom, and enforcing compliance with workplace safety and health regulations.

## Considerations on the 23rd Selection Process for Labor Prosecutors

The public selection process for Labor Prosecutors, according to Notice No. 169/2023 (EDITAL Nº 169, 2023), offers vacancies distributed as follows:

- General Category: 3 positions

- Persons with Disabilities (PwD): 5% of the total vacancies
- Quota for Black Candidates: 20% of the total vacancies
- The reservation of positions applies whenever the number of vacancies offered in the selection process, or that arise during its validity period, is equal to or greater than 3.

The selection process is structured into five stages:

I. Multiple-Choice Exam: Eliminatory and classificatory

II. Discursive Exam: Eliminatory and classificatory

III. Practical Exam: Eliminatory and classificatory

IV. Oral Exams: Eliminatory and classificatory

V. Credentials Assessment: Classificatory

The selection process for Labor Prosecutors covers a variety of subjects, divided into Basic Knowledge and Specific Knowledge.

Basic Knowledge:

- Administrative Law
- Environmental Labor Law
- Civil Law
- Collective Labor Law
- Constitutional Law
- Individual Labor Law
- International Law
- Criminal Law
- Social Security Law
- Civil Procedure
- Labor Procedure
- Human Rights

The criteria for approval are:

- Minimum score:
- A minimum final score of 60 is required on the multiple-choice exam to be considered approved.

- Additionally, candidates may be required to achieve at least 50% accuracy in the multiple-choice, discursive, and practical exams. For each subject in the oral exams, a minimum score of 50 (on a scale of 0 to 100) is required.
- The final score for the written exams will be calculated as the arithmetic mean of the scores obtained in the multiple-choice, discursive, and practical exams.

- Multiple-Choice Exam:
- The top 300 candidates with the highest scores will be classified, excluding from this limit candidates registered as beneficiaries of reserved vacancies, candidates benefiting from an appeal due to a material error, and candidates granted classification through a judicial decision unrelated to preliminary registration.

The exam has a maximum score of 100 points. The grading criteria limited the approval for the discursive phase to 300 candidates in the general category and 100 candidates in the reserved quota category.

The cutoff scores for this selection process were not disclosed, preventing a deeper analysis of the results and a comparative assessment of the performance of the LLM tools.

**Observed Results and Analysis of LLM Model Performance in the 23rd Selection Process for Labor Prosecutors**

**Overall Accuracy Rate:** The highest accuracy rate was achieved by Claude Sonnet 3.5 with 64%, followed by ChatGPT 4.0 with 56%. The lowest performance was observed in ChatGPT 3.5 Turbo, which scored 34%.

**Performance on Easy Questions:** Models performed better on easy questions, with Claude Sonnet 3.5 standing out with 73.24% accuracy.

**Performance on Difficult Questions:** Performance was significantly lower, with Claude Sonnet 3.5 still leading at 48% accuracy.

**Analysis by Legal Area**

**Constitutional Law:** Claude Sonnet 3.5 achieved the highest accuracy rate at 80%.

**Human Rights:** Claude Sonnet 3.5 also excelled with 80%, followed closely by ChatGPT 4.0 with 70%.

**Individual Labor Law:** Claude Sonnet 3.5 stood out with 64.7% accuracy.

**Collective Labor Law:** Claude Sonnet 3.5 led with 62.5% accuracy.

**Environmental Labor Law:** ChatGPT 4.0, ChatGPT oi-mini, and Claude Sonnet 3.5 achieved 100% accuracy.

**Labor Procedure:** Claude Sonnet 3.5 attained 100% accuracy, with LLama 3.1 405B trailing at 83.3%.

**Conclusions**

Claude Sonnet 3.5 demonstrated the most consistent performance across nearly all legal areas and difficulty levels, proving to be the most efficient model in this context. ChatGPT 3.5 Turbo and ChatGPT oi-mini performed significantly worse, particularly on more complex questions and specific areas such as Labor Procedure and Administrative Law.

**Analysis**

Based on the collected data, we found that accuracy rates vary significantly across different areas of law. Some models demonstrated high performance in more specific fields, such as Environmental Law and Labor Procedure (for instance, Claude Sonnet 3.5 achieved 100% in Labor Procedure). This observation suggests that specific models excel in interpreting and applying structured rules and repetitive procedural patterns, as with Labor Procedure, regulated by well-defined and often standardized norms and procedures.

In Administrative and Criminal Law questions, models like ChatGPT 3.5 Turbo showed difficulties (0% and 66.67%, respectively), highlighting gaps in training data relevant to the Brazilian legal framework.

Considering the overall exam results, the areas where the models achieved the highest average accuracy were Environmental Labor Law (64.29%), International Law (62.05%), and Social Security Law (57.14%). On the other hand, the areas with the worst performance were Collective Labor Law (30.36%), Civil Law (32.14%), and Civil Procedure (35.16%)[3]. From these results, we can deduce that subjects with greater international alignment, such as Environmental Law and Human Rights, show better performance due to the likely presence of global data in the training process. Conversely, areas that require Brazilian legal contextualization, such as Collective Labor Law and Civil Law, exhibited the poorest performance, likely due to the absence of these data in the training sets. This highlights the need to adapt the models with datasets specific to Brazil. Lastly, subjects with strong interdependence, such as Constitutional Law and Human Rights, appear to have positive correlations, indicating that the models can transfer knowledge between related areas.

---

[3] In the scope of Labor Procedure, Civil Procedure Law is applied subsidiarily, meaning its application relies on the interpretation of specific and often contextual procedural norms.

This discrepancy reinforces the hypothesis that LLMs, especially commercial ones, are not perfectly aligned with the peculiarities of Brazilian law. The datasets used in their training, predominantly from foreign legal systems, limit their performance in highly complex local areas. This suggests that the more specific the Brazilian legal field being analyzed, the poorer the performance of the commercial models tested.

From a linguistic perspective, despite the gaps, models such as Claude Sonnet 3.5 and ChatGPT 4.0 demonstrated notable performance, with overall accuracy rates of 64% and 56%, respectively. The most significant challenges were contextual interpretations and areas requiring practical and up-to-date knowledge of Brazilian legislation (e.g., Administrative and Criminal Law). Furthermore, the impact of the original training language is evident in models like ChatGPT 3.5 Turbo, whose lower performance (34% overall accuracy) can be attributed to difficulties in interpreting the nuances of legal Portuguese, as well as cultural and systemic gaps.

LLMs still face difficulties in interpreting scenarios requiring practical knowledge of Brazilian law, particularly in more specialized contexts (such as the selection process for Labor Prosecutors). Additionally, many models lack updates on recent legislative changes or relevant judicial decisions.

Given this, it is possible to conclude that specific adaptations are necessary, such as additional training with Brazilian legal datasets, to mitigate the observed limitations and thereby enhance the efficiency of these models.

## 22nd Public Selection Process for Filling Positions of Labor Prosecutor

| Model | Total Correct | Total Incorrect | % Accuracy (Total) | % Accuracy (Easy Questions) | % Accuracy (Difficult Questions) |
|---|---|---|---|---|---|
| ChatGPT 3.5 Turbo | 27 | 73 | 27% | 32% | 13% |
| ChatGPT 4o | 50 | 50 | 50% | 54% | 42% |
| ChatGPT oi-mini | 43 | 57 | 43% | 49% | 32% |
| Claude Sonnet 3.5 | 63 | 37 | 63% | 71% | 48% |
| LLama 3.1 405B | 50 | 50 | 50% | 55% | 39% |
| Gemini 1.5 Pro | 42 | 58 | 42% | 45% | 39% |
| Command-R-Plus | 37 | 63 | 37% | 42% | 26% |
| Mistral Large-2 | 42 | 58 | 42% | 43% | 42% |
| Amazonia AI | 44 | 56 | 44% | 48% | 35% |
| Reka Core 67B | 39 | 61 | 39% | 46% | 23% |
| Perplexity AI | 39 | 61 | 39% | 45% | 26% |
| Sabiá 3 | 59 | 41 | 59% | 60% | 58% |
| GROK 2 | 49 | 51 | 49% | 51% | 45% |
| Nemotron 340B | 50 | 50 | 50% | 49% | 55% |

| Model | % Accuracy Total | % Accuracy Easy | % Accuracy Difficult | |
|---|---|---|---|---|
| ChatGPT 3.5 Turbo | 27% | 32% | 13% | 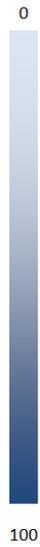 |
| ChatGPT 4o | 50% | 54% | 42% | |
| ChatGPT oi-mini | 43% | 49% | 32% | |
| Claude Sonnet 3.5 | 63% | 71% | 48% | |
| LLama 3.1 405B | 50% | 55% | 39% | |
| Gemini 1.5 Pro | 42% | 45% | 39% | |
| Command-R-Plus | 37% | 42% | 26% | |
| Mistral Large-2 | 42% | 43% | 42% | |
| Amazonia AI | 44% | 48% | 35% | |
| Reka Core 67B | 39% | 46% | 23% | |
| Perplexity AI | 39% | 45% | 26% | |
| Maritaca AI (Sabiá 3) | 59% | 60% | 58% | |
| GROK 2 | 49% | 51% | 45% | |
| Nemotron 340B | 50% | 49% | 55% | |

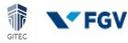

% Accuracy **MPT 2022**

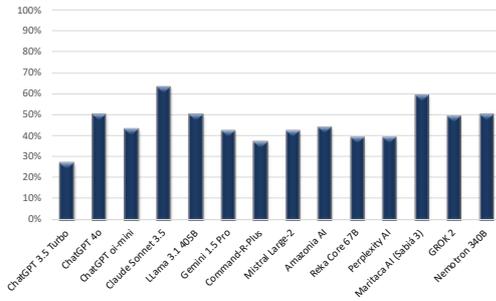

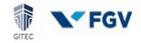

% Accuracy **(Difficult Questions)**

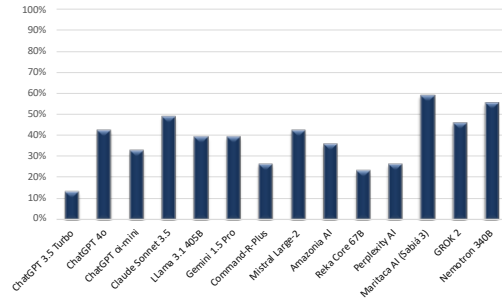

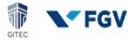

% Accuracy **(Easy Questions)**

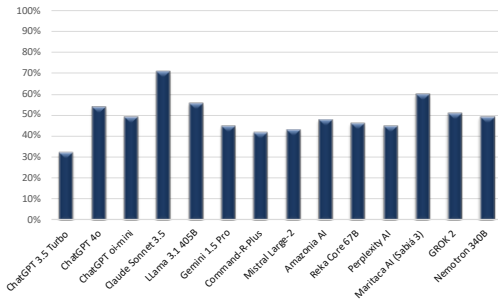

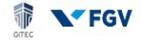

% Accuracy **(Constitutional Law)**

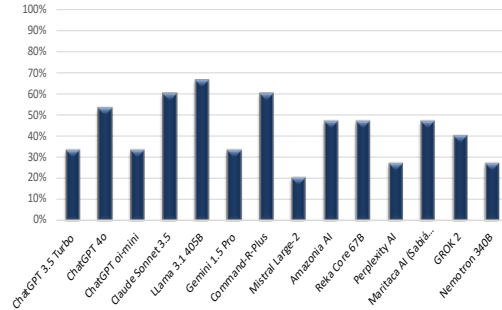

% Accuracy **(Human Rights)**
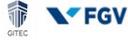
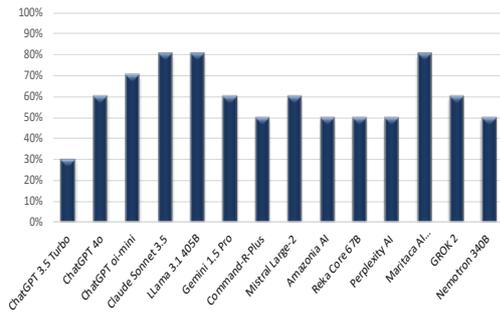

% Accuracy **(Labor Procedural Law)**
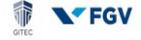
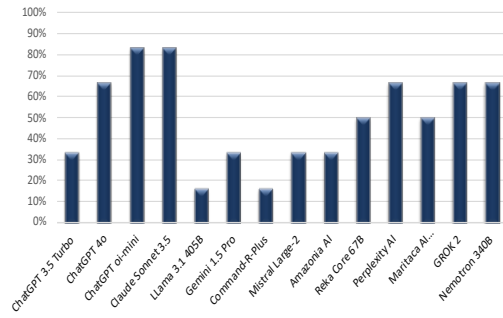

% Accuracy **(Individual Labor Law)**
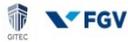
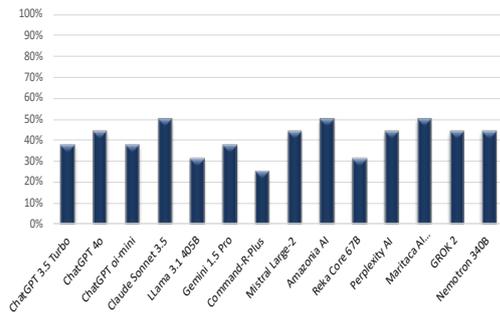

% Accuracy **(Civil Procedure Law)**
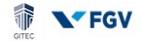
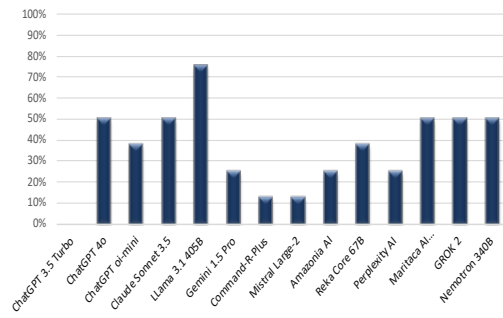

% Accuracy **(Collective Labor Law)**
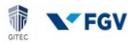
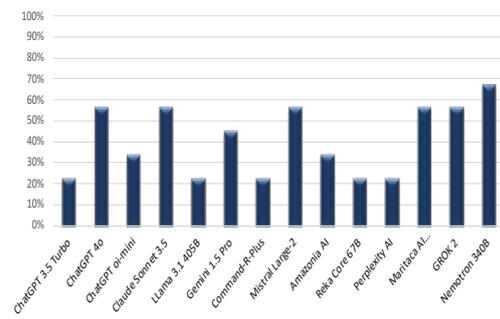

% Accuracy **(Administrative Law)**
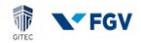
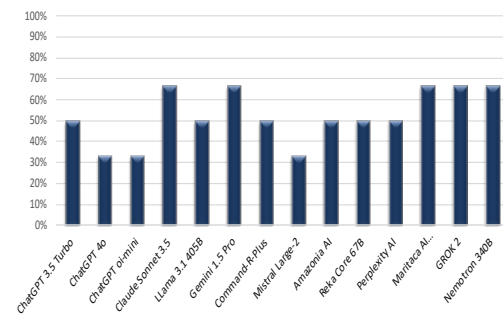

% Accuracy **(Direito Ambiental do Trabalho)**
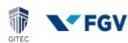
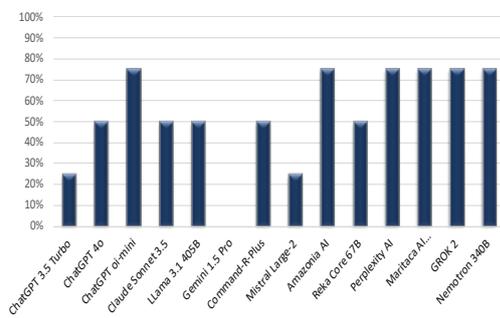

% Accuracy **(Civil Law)**
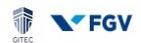
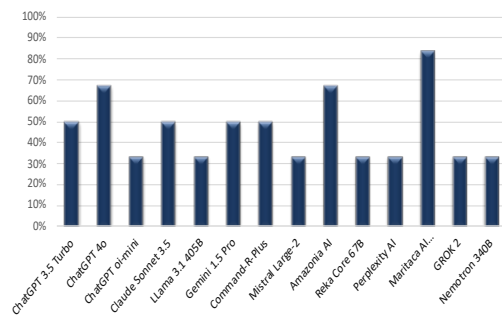

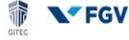
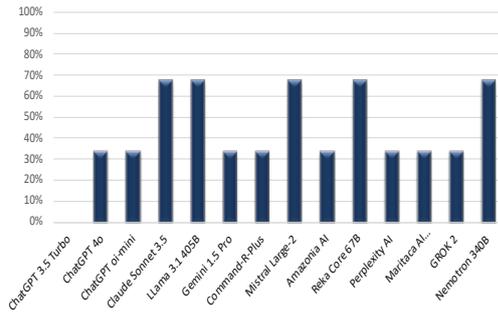

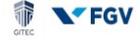
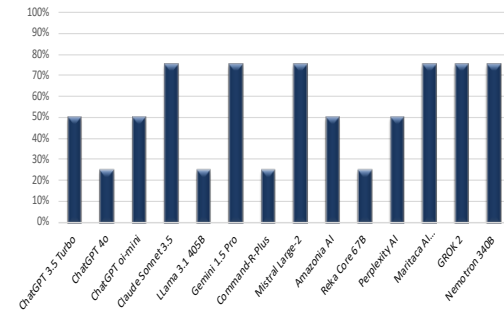

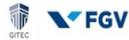
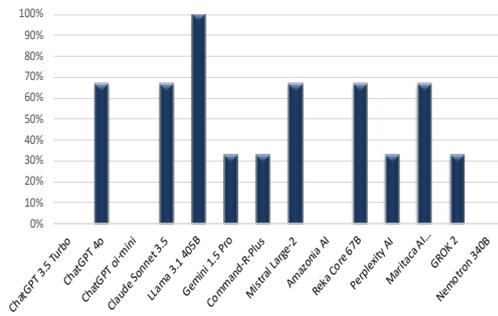

## 23rd Public Selection Process for Filling Positions of Labor Prosecutor

| Model | Total Correct | Total Incorrect | % Accuracy (Total) | % Accuracy (Easy Questions) | % Accuracy (Difficult Questions) |
|---|---|---|---|---|---|
| **ChatGPT 3.5 Turbo** | 36 | 64 | 36% | 42% | 16% |
| **ChatGPT 4o** | 58 | 42 | 58% | 68% | 32% |
| **ChatGPT oi-mini** | 39 | 61 | 39% | 46% | 16% |
| **Claude Sonnet 3.5** | 66 | 34 | 66% | 73% | 48% |
| **Llama 3.1 405B** | 56 | 44 | 56% | 63% | 36% |
| **Gemini 1.5 Pro** | 55 | 45 | 55% | 62% | 36% |
| **Command-R-Plus** | 34 | 66 | 34% | 39% | 16% |
| **Mistral Large-2** | 46 | 54 | 46% | 51% | 32% |
| **Amazonia AI** | 60 | 40 | 60% | 73% | 24% |
| **Reka Core 67B** | 48 | 52 | 48% | 54% | 32% |
| **Perplexity AI** | 45 | 55 | 45% | 55% | 16% |
| **Sabiá 3** | 65 | 35 | 65% | 75% | 40% |
| **GROK 2** | 59 | 41 | 59% | 62% | 52% |
| **Nemotron 340B** | 60 | 40 | 60% | 72% | 28% |

| Model | % Accuracy Total | % Accuracy Easy | % Accuracy Difficult |
|---|---|---|---|
| ChatGPT 3.5 Turbo | 36% | 42% | 16% |
| ChatGPT 4o | 58% | 68% | 32% |
| ChatGPT o1-mini | 39% | 46% | 16% |
| Claude Sonnet 3.5 | 66% | 73% | 48% |
| LLama 3.1 405B | 56% | 63% | 36% |
| Gemini 1.5 Pro | 55% | 62% | 36% |
| Command-R-Plus | 34% | 39% | 16% |
| Mistral Large-2 | 46% | 51% | 32% |
| Amazonia AI | 60% | 73% | 24% |
| Reka Core 67B | 48% | 54% | 32% |
| Perplexity AI | 45% | 55% | 16% |
| Maritaca AI (Sabiá 3) | 65% | 75% | 40% |
| GROK 2 | 59% | 62% | 52% |
| Nemotron 340B | 60% | 72% | 28% |

0

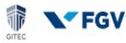

% Accuracy **MPT 2023**

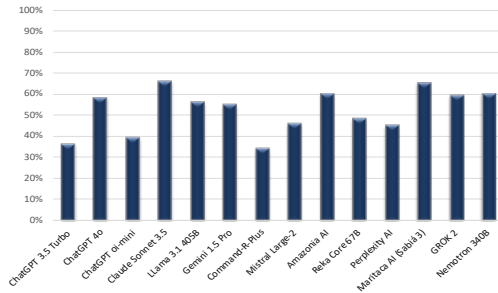

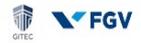

% Accuracy **(Difficult Questions)**

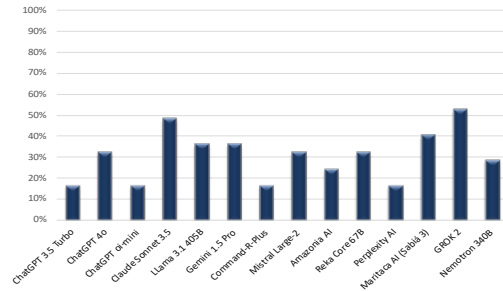

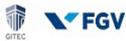

% Accuracy **(Easy Questions)**

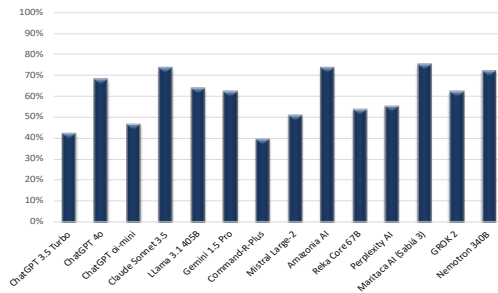

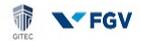

% Accuracy **(Constitutional Law)**

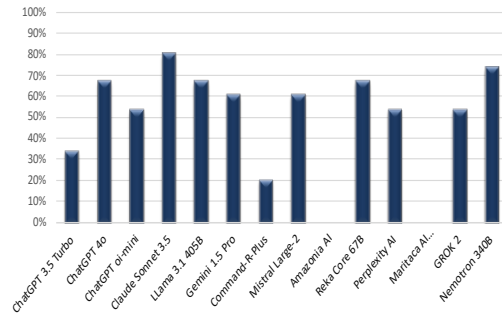

% Accuracy **(Human Rights)** 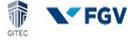

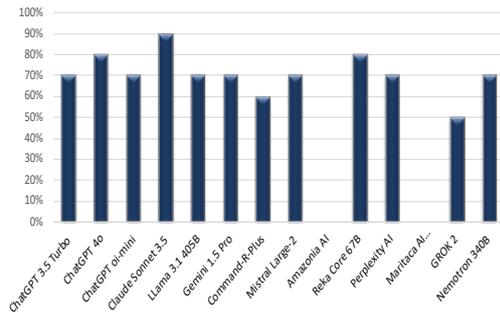

% Accuracy **(Labor Procedural Law)** 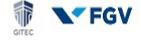

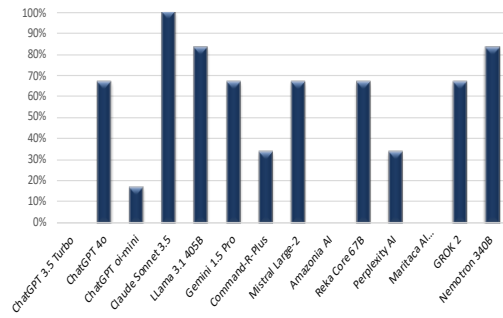

% Accuracy **(Individual Employment Law)** 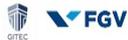

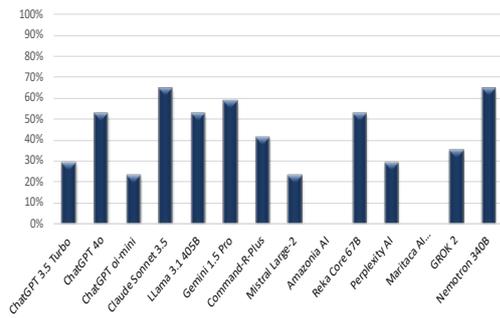

% Accuracy **(Civil Procedure Law)** 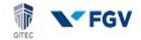

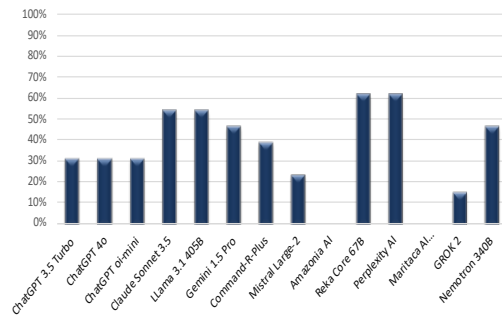

% Accuracy **(Collective Labor Law)** 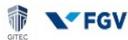

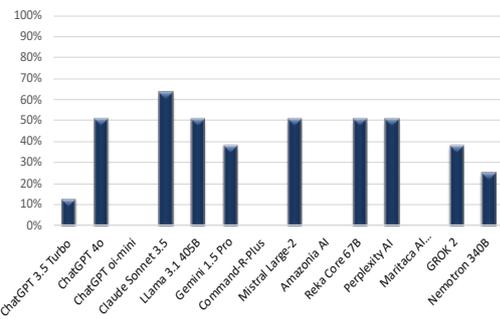

% Accuracy **(Administrative Law)** 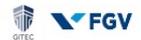

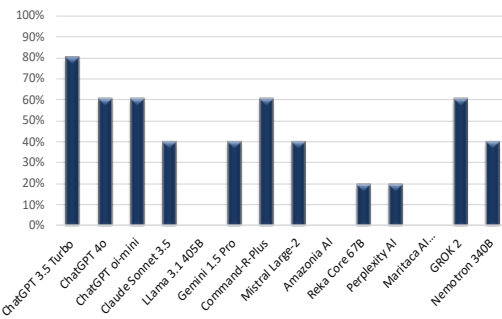

% Accuracy **(Environmental Labor Law)** 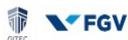

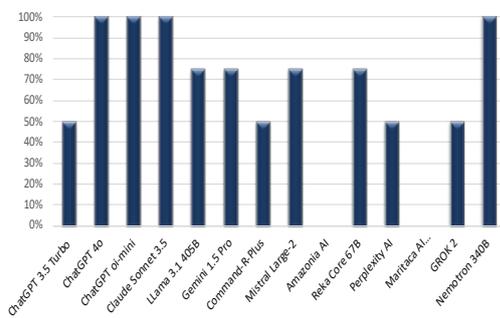

% Accuracy **(Civil Law)** 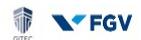

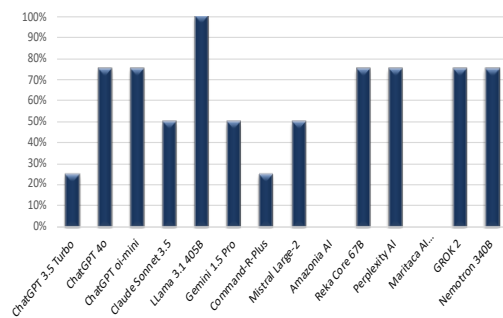

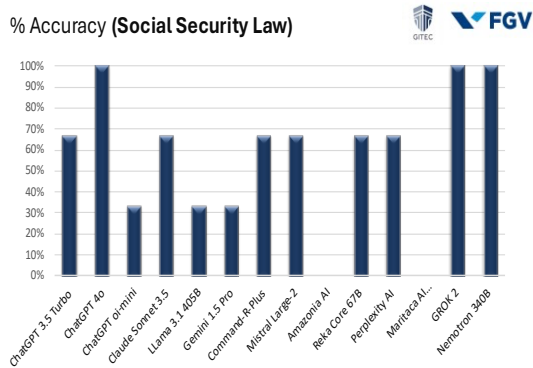

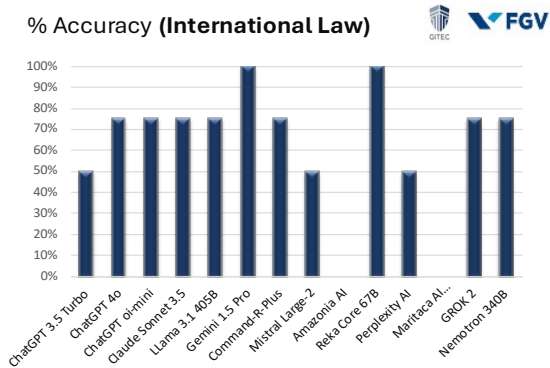

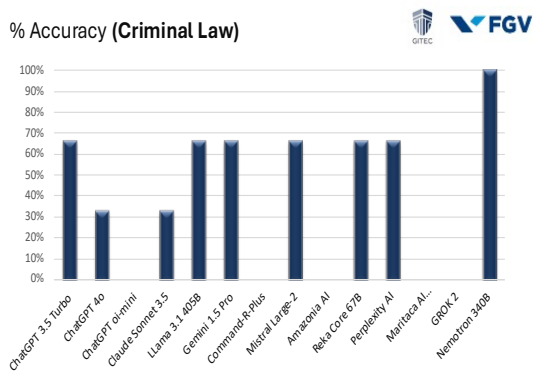

## Comparison Between the Two MPT Selection Processes

| Model | 22nd Exam Total Correct | 23rd Exam Total Correct |
|---|---|---|
| ChatGPT 3.5 Turbo | 27 | 36 |
| ChatGPT 4o | 50 | 58 |
| ChatGPT oi-mini | 43 | 39 |
| Claude Sonnet 3.5 | 63 | 66 |
| LLama 3.1 405B | 50 | 56 |
| Gemini 1.5 Pro | 42 | 55 |
| Command-R-Plus | 37 | 34 |
| Mistral Large-2 | 42 | 46 |
| Amazonia AI | 44 | 60 |
| Reka Core 67B | 39 | 48 |
| Perplexity AI | 39 | 45 |
| Maritaca AI (Sabiá 3) | 59 | 65 |
| GROK 2 | 49 | 59 |
| Nemotron 340B | 50 | 60 |